\documentclass[final,3p,times,authoryear]{elsarticle}
\usepackage{graphicx}
\usepackage{bm}%
\usepackage{subfigure}
\usepackage{amsmath,bm}
\usepackage{amsbsy}
\usepackage{amstext}
\usepackage{tabularx}
\usepackage{float}
\usepackage{amsmath}
\usepackage{xcolor}
\usepackage{hyperref}

\hypersetup{
    colorlinks = true,
    citecolor = blue,
    urlcolor = blue,
    linkcolor = blue
}

\DeclareMathOperator{\erf}{erf}

\numberwithin{equation}{section}

\begin{document}

\begin{frontmatter}

\title{Control of axial dispersion in polymeric flows: magnetic effect versus electroosmotic effect}

\author[1]{Xiaoping Wang}

\author[2,3]{Mengqi Zhang\corref{cor1}}

\author[4]{Haitao Qi\corref{cor2}}

\cortext[cor1]{Corresponding author. Email: ma.zmq@cityu.edu.hk}
\cortext[cor2]{Corresponding author. Email: htqi@sdu.edu.cn}

\address[1]{
Department of Mathematics, Harbin Institute of Technology, Weihai 264209, PR China
}
\address[2]{
Department of Mathematics, City University of Hong Kong, 83 Tat Chee Avenue, Kowloon Tong, Hong Kong, PR China
}
\address[3]{
Department of Mechanical Engineering, National University of Singapore, 9 Engineering Drive 1, 117575, Singapore
}
\address[4]{
School of Mathematics and Statistics, Shandong University, Weihai 264209, PR China
}

\begin{abstract}
Since Taylor's seminal work on the dispersion of soluble matter, the modeling and control of solute dispersion have received considerable attention. This study investigates the solute dispersion in polymer solutions driven by unsteady magnetohydrodynamic-oscillatory electroosmotic flow within a microtube, aiming to reveal the transport mechanisms in complex rheological media through microscale flow. Firstly, based on the Debye-H\"{u}ckel approximation, analytical solutions for the electric potential and velocity distributions of mixed magnetohydrodynamic-electrohydrodynamic flows are rigorously derived. Subsequently, using Sankarasubramanian \& Gill's generalized dispersion model, mathematical models are developed to systematically investigate the temporal evolution of hydrodynamic dispersion. By investigating the synergistic modulation of electrokinetic and magnetic fields, we identify three distinct flow regimes, namely the electric-field-dominant, competitive transition, and magnetic-field-limited regimes, and quantitatively define their boundaries and evolution with oscillatory Reynolds number $Re$ and Deborah number $De$, thus allowing accurate flow regime classification. Moreover, a non-monotonic critical response governing solute dispersion is observed over a wide range of oscillatory Reynolds numbers, which provides a novel control paradigm for microfluidic mixing and separation. Additionally, through a comparative analysis of fractional Maxwell, classical Maxwell, and Newtonian fluids, we show that microscopic structural relaxation plays a crucial role in macroscopic solute transport under multi-field coupling. Our findings help to reveal flow mechanisms for precise mass transport control in electro-magnetically coupled systems, advancing the development of microfluidic devices and biomedical detection technologies.
\end{abstract}

\end{frontmatter}

\section{Introduction}\label{sec:1}

Axial solute dispersion in micro/nano-fluidic systems holds significant theoretical importance and practical applications, with its regulation directly influencing the efficiency of critical processes such as micro-mixing~\citep{DZWK14,LDSL15}, separation analysis~\citep{Ghosal12,Dhellemmes24}, and electrophoresis~\citep{GhosalS06}. Over recent decades, alongside the rapid advancement of microfluidic technology, this field has demonstrated broad application prospects in scenarios such as biomedicine, chemical synthesis, and point-of-care diagnostics. Meanwhile, active control strategies for axial dispersion have continuously evolved, including: Modulating flow characteristics via applied pressure gradients~\citep{RanaM16,AzariS22}; Regulating solute diffusion behavior using electric fields~\citep{RIZB13}; Introducing magnetic fields to induce Lorentz forces and alter solute diffusions~\citep{Vargas2017}; Engineering interfacial interactions through surface modification techniques~\citep{SSMS2020}.

In microfluidic systems, electroosmosis (EO) attracts significant research interest due to its unique capability for pump-free fluid propulsion. Introducing magnetic fields to establish an additional magnetohydrodynamic (MHD) effect creates a novel fluid manipulation mechanism, enabling fine-scale regulation of hydrodynamic characteristics through the interaction between electromagnetic fields and fluids. Additionally, the synergistic combination of hydrophobic wall slip characteristics and slip-dependent zeta potential further enhances fluid behavior control. Therefore, this study investigates how interface slip and slip-dependent zeta potential modulate solute dispersion in viscoelastic solutions under oscillatory electroosmotic flow (EOF) coupled with MHD, developing enhanced technology for precise control of microscale mass transport.

\subsection{Theoretical models for solute dispersion}

Theoretical research on dispersion models has long attracted the attention of researchers. Taylor's pioneering work~\citep{TAYLOR53} laid the foundation for solute dispersion modeling in pressure-driven pipe flows. This model relies on transverse concentration averaging, which requires an initial phase for radial diffusion to homogenize cross-sectional concentration profiles. This process is characterized by the dimensionless time scale $D_m t / R^2$ (where $t$, $D_m$, and $R$ denote time, molecular diffusivity, and pipe radius, respectively), restricting the model's validity to long-time regimes. To overcome these limitations, Aris extended Taylor's dispersion theory using the moment method for an infinitely long pipe~\citep{ARIS56}. He derived a second-moment equation to describe the average concentration distribution under steady advection and demonstrated that, over sufficiently long times, solute tracers evolve into a Gaussian distribution due to the mean flow field. This research further refined the applicability and descriptive capability of Taylor dispersion theory. Subsequently, the methodologies based on the models proposed by Taylor and Aris were improved by~\citep{Chatwin70,Barton83,CJPJ84,BSSK92,ZWGQ14}.

Over the years, numerous studies have examined solute transport behavior under different conditions, such as porous layers~\citep{DIJJ18,DMPW24}, MHD flows~\citep{Vargas2017,Poddar24,DPKR2024,SMRR2025}, pollutant emissions in floating wetlandss~\citep{ADZCH22}, by improving Taylor's model. However, these studies have only analyzed solute diffusion at long times after injection into the fluid and have not provided any information on solute diffusion during the short time immediately following injection. Notably,~\citet{CCMS25} recently extended the Taylor dispersion in coupled electroosmotic and pressure-driven flows across all time regimes under the assumptions of a low zeta potential, the Debye-H\"{u}ckel approximation, and a finite electric double layer (EDL) thickness. Initial studies sought to characterize band broadening dynamics immediately following injection, \cite{Gill1970} put forth a general model, which articulates the transport coefficients as functions of time within steady-state flow. Subsequently, a series of advancements have been made. These include the consideration of time-variable flow~\citep{Gill1972,SANKAR1972,VEDEL2011}, oscillating flow~\citep{RanaM16,AzariS22}, the non-uniformity of the injected solute band~\citep{Gill1971,Gill1972}, and interfacial/boundary mass transfer processes~\citep{SANKAR1973,RanaM16}. In the present study, we are set to carry out a comprehensive exploration of the solute dispersion driven by the combined effects of MHD and oscillating EOF. Our research builds upon the generalized dispersion model that was introduced by~\citet{Gill1970}.

\subsection{Solute dispersion in microchannels driven by EOF and MHD}

The manipulation of fluid flow and dispersion in microchannels via EOF represents a foundational area of inquiry in microfluidics owing to its attributes of operational simplicity, high controllability, uniform flow velocity profiles, and reliance on non-mechanical actuation~\citep{YZZC2004,GDCS24,SKPK24}. For nearly a century, the literature has systematically investigated various aspects of hydrodynamic dispersion in direct-current (DC) EO.~\citet{HDRW24} showed that in transient Taylor-Gill dispersion during EOF, the peak concentration attenuates with increasing reaction intensity. Through multiscale analysis, ~\citet{SJSB24} found that stronger coupling parameters significantly slow solute transport in micropolar EOF within rectangular microchannels.~\citet{PRMD23} revealed that the EDL effect enhances solute diffusion in Carreau fluid under combined pressure‑driven flow and EOF.~\citet{SGNR24} developed a novel effective dispersion model accounting for Taylor–Aris dispersion, EO-induced dispersion and their coupling. They theoretically proved the positivity of the dispersion coefficient and the existence and uniqueness of strong solutions, and numerically demonstrated the feasibility of charged species separation.
Additionally, researchers have persistently pursued inquiries in this domain to examine the impacts of diverse supplementary facets of electrokinetic phenomena on solute transport and mixing dynamics~\citep{Arcos2018,Mozafari2025,SMSD20}.

Beyond conventional DC-EOF systems, alternating-current electroosmosis flow (AC-EOF) governs dynamic dispersion through electric field and EDL interactions to optimize dispersion and mixing processes in microfluidic systems. \citet{HLYJ17} presented a focused investigation into solute dispersion induced by AC-EOF of viscoelastic fluids, with series expansion and transform techniques employed for analysis. \citet{MRHS19} found that in oscillatory EO-Poiseuille flow through polyelectrolyte-grafted capillaries, PEL enhances solute advection and dispersion. Thicker charged PEL strengthens electroosmotic advection, flattening transverse concentration and increasing radial differences.~\citet{MABM20} theoretically investigated mass transfer in a microcapillary connecting two reservoirs with different concentrations of an electro-neutral solute. Results show fluid elasticity-oscillatory flow interplay can enhance mass transfer rates by orders of magnitude compared to Newtonian oscillatory EOF.~\citet{PAMB20} investigated the mass transfer of Maxwell fluids in concentric annular microchannels driven by AC-EOF. The analysis demonstrates that optimizing the elasticity number, cylindrical gap, and angular Reynolds number rationally can augment the total mass transfer rate and facilitate component separation. Subsequently, extensive research on solute transport driven by AC-EOF has primarily focused on the following key directions: dispersion analysis in microchannels with diverse geometries~\citep{VMDE23}, solute dispersion in nanochannels enclosed by dielectric liquid layer~\citep{CPGP25}, the coupling effects of non-uniform walls~\citep{DRHS21}, regulatory mechanisms of asymmetric wall zeta potentials and slip interfaces~\citep{HKKH22}, as well as the regulatory mechanisms of high zeta potentials and slip interfaces~\citep{SSKB22} on solute transport properties. In this work, an AC electric field is employed due to its superior performance in solute dispersion and microfluidic control. Its periodic polarity reversal suppresses electrode polarization and electrolysis, thereby avoiding the unstable EOF and band broadening associated with DC fields. The oscillating field drives solutes into reciprocating motion, enhancing mixing and dispersion efficiency. Electrokinetic effects can be flexibly tuned by adjusting the frequency and amplitude, while high-frequency operation reduces Joule heating and improves system stability. These merits make AC electric fields highly suitable for microfluidic and biological applications.

In recent years, electric fields, magnetic fields and their synergy are widely used to precisely control microfluidic flows, with broad prospects in chemical and biomedical engineering involving mass and heat transport. In the study of EO and MHD flows, Joule heating and viscous dissipation constitute key thermal and mechanical effects. \citet{Chakraborty2013} reported that under constant wall heat flux, both effects reduce the Nusselt number via distinct mechanisms: Joule heating uniformly elevates bulk fluid temperature and weakens wall temperature gradient, while viscous dissipation intensifies near the wall and raises wall temperature, both deteriorating heat transfer. \citet{ZYYJ17} quantified entropy generation in two-layer MHD and EOF with Joule heating, finding that the entropy generation rate decays from wall to interface and is dominated by Hartmann number and viscous dissipation. \citet{YJXL19} investigated MHD-EOF and heat transfer in rectangular microchannels, demonstrating that heat transfer is jointly affected by Joule heating, viscous dissipation and electromagnetic coupling, where increased viscous dissipation impairs heat transfer while a larger electrokinetic width promotes it. \citet{Rilwan2024} further examined the influences of Joule heating and viscous dissipation on electromagnetohydrodynamic flow and heat transfer in porous microchannels.
Furthermore, the coupled driving of MHD and EOF enables precise regulation of solute dispersion paths, thereby offering a new paradigm for the multifield synergistic control of microscale mass transfer processes. \citet{Poddar24} showed that strong magnetic fields significantly reduce flow velocity and the Taylor dispersion coefficient in two-dimensional MHD Taylor dispersion through parallel channels. \citet{Vargas2017} and \citet{DPKR2024} applied Aris's method of moments to Newtonian solute dispersion in EO-MHD flows through parallel plate microchannels. Specifically,~\citet{Vargas2017} found that slow sinusoidal wall zeta potential variations, combined with EO-MHD forces, significantly alter the axial distribution of the effective dispersion coefficient. \citet{DPKR2024} explored the regulatory effects of symmetric and asymmetric zeta potentials on solute distribution, demonstrating that asymmetric zeta potentials accelerate the dispersion process. Subsequently, \citet{SMRR2025} introduced heterogeneous first-order boundary reactions and transverse electric fields to further investigate the solute dispersion characteristics of non-Newtonian fluids under the combined effects of MHD and EOF in parallel-plate microchannels. All their core conclusions are fundamentally based on Aris's method.
Notably, Joule heating has been consistently neglected in existing studies on solute dispersion in MHD and EOF-driven flows. This indicates that while Joule heating is indispensable in heat transfer-focused analyses of MHD and EOF systems, neglecting it in studies concentrating on solute dispersion is a common and reasonable practice. This simplification is not an oversight, but rather reflects the physical consensus in the field that thermal effects are always secondary to the hydrodynamic mechanisms of interest in such problems. In line with this established research paradigm, the present study focuses on elucidating the fundamental regulatory mechanism of electromagnetic forces on solute dispersion over a wide range of magnetic field intensities. Within the parameter range explored, the temperature rise induced by Joule heating is negligible, and it is therefore justifiably excluded from the analysis.

\subsection{Microchannel interfacial slip and slip-dependent zeta potential}
In microscale fluid mechanics, the extremely small characteristic scales (at the micrometer level) and high surface-area-to-volume ratios make surface properties a core determinant of flow behavior. Notably distinct from macroscopic flows, hydrodynamic slip at microscale liquid-solid interfaces is a prevalent phenomenon, whereby traditional no-slip models often require modification due to their inadequacy~\citep{PASM97,YZSG01,YXGS02}. This attribute is critical in microscale  EOF. First, interfacial slip directly alters the flow field. In hydrophobic microchannels, non-wetting conditions exacerbate the breakdown of the no-slip assumption, making slip length an essential parameter~\citep{MGMW2024,Sujith2025}. Second, wall slip modifies the relationship between surface potential and zeta potential. Specifically, slip leads to an apparent increase in zeta potential, which manifests as a reduced velocity gradient within the EDL~\citep{Joly2006}.

The interplay between interfacial slip and zeta potential has been a focal point of research in recent decades. Owing to the strong coupling between these two factors, the independent measurement of electrokinetic phenomena on hydrophobic surfaces remains challenging. Early investigations by~\citet{CRSS02} established a linear relationship model between slip length and zeta potential through experiments on methylated quartz capillaries. Subsequent studies by~\citet{JYDY02} and~\citet{VTBJ08} further refined this theoretical framework. This mechanistic understanding highlights that traditional surface potential models based on the no-slip assumption are no longer capable of accurately describing microscale transport characteristics, with the interaction between interfacial slip and electrochemical properties emerging as a critical factor in explaining physical processes within microfluidic devices.

Recently, research on surface slip and slip-dependent zeta potential has made significant progress in the field of microscale fluid dynamics, spanning fluid flow, heat transfer mechanisms, energy conversion systems, and solute dispersion processes in EOF, MHD flow, and EO-MHD coupled flow. \citet{BMPB21} showed that traditional surface potential models (neglecting slip) underestimate core flow velocity in combined electroosmotic-pressure-driven microchannels. \citet{BPPB2023} found that slip-induced zeta potential increase strengthens EO force and enhances near-wall velocity in pulsatile EO and shear-driven flows. \citet{SDJS24} confirmed that slip-dependent zeta potential governs rotational flow and volume transport in EO-MHD couple stress flows in rotating microchannels. \citet{DXLC2024} demonstrated that slip-dependent zeta potential significantly enhances the sensitivity of two-layer flow parameters in unsteady cylindrical EOF. Other recent studies on the EO and MHD transport of viscoelastic fluids in microchannels with slip-dependent zeta potential have revealed several critical advances.
The slip-dependent zeta potential significantly optimizes the electrokinetic energy conversion process \citep{Saha2023,LIUWL2024,SSKB2025}; Strengthening the coupling between interfacial charges and fluid slip remarkably modifies the electromagnetic transport properties \citep{DVRK2025} and thermal entropy characteristics \citep{SJBK2023} in microchannel.

 Notably, the influence of interfacial slip on solute dispersion has sparked extensive discussion~\citep{HKKH22,FQYJ19,SSKB22}. The investigation of coupled slip-dependent zeta potential and viscoelastic effects of Carreau fluids in two-fluid EOF demonstrated that slip modifies diffusion kinetics by enhancing the apparent zeta potential, with shear-thinning/thickening behavior further amplifying this effect~\citep{KGTA2025}.
 However, the solute dispersion characteristics in the combined driving regime of  AC-EOF and MHD with slip-dependent zeta potential remain unexplored in existing literature.

\subsection{The position and structure of the current work}
Based on a comprehensive review of recent advancements in solute dispersion studies within microfluidic systems, it has been found that EOF and the hybrid MHD-EO approach for fluid flow and mass transport have emerged as a prominent frontier in multi-physics microscale transport research. This synergistic strategy integrates the high driving efficiency of EO with the superior dispersion-suppression capability of MHD, enabling better controllability than single actuation modes. It exhibits great potential for applications including biochemical separation, targeted drug delivery, and microfluidic reaction systems~\citep{ZZCFFL2025,Sahore2018,Chaudhuri2018}, where stable transport and tunable dispersion are essential. Nevertheless, current research still suffers from several limitations, as detailed below.

(i) A key challenge in this field is that, whereas research has predominantly focused on Newtonian fluid~\citep{MRHS19,SSMS2020,DPKR2024,CCMS25} and select non-Newtonian fluid~\citep{SMRR2025} within microchannels, the governing mechanisms of coupled magneto-electrodynamic fields on solute dispersion in viscoelastic fluids remain elusive. Given the prevalence of such fluids in industrial applications of axial dispersion, elucidating their behavior is critical. This lack of mechanistic insight across broad parameter spaces hinders the development of general principles for dispersion.

(ii) Current research on the influence of micro-scale interfacial characteristics on solute dispersion processes remains insufficient. Most existing studies are based on the no-slip assumption~\citep{MRHS19,SSMS2020,DPKR2024,CCMS25,SMRR2025}, which exhibits significant limitations in micro-scale flows. Moreover, the majority of related work has focused only on the role of single factors, such as heterogeneous or asymmetric zeta potentials at the wall~\citep{Vargas2017,DPKR2024}, without incorporating the combined effects of interfacial slip and zeta potentials.

(iii) Although Gill's generalized dispersion model shows higher reliability in describing the transient processes of solute dispersion, its inherent complexity in mathematical derivation has resulted in limited adoption in practical microfluidic systems.

Against this backdrop, building on an extension of Gill's generalized dispersion model, this study seeks to: (1) systematically investigate the solute dispersion behavior in polymer solutions within microchannels under the drive of MHD-AC-EOF multiphysics coupling; and (2) further elucidate the regulatory role of microscale interfacial effects, specifically interfacial slip characteristics and slip-dependent zeta potential, in this dispersion process. The findings will offer more comprehensive theoretical underpinnings for optimizing the separation and mixing performance of microfluidic devices. 

The remainder of this paper is structured as follows: The subsequent section first systematically describes the flow geometry configuration and explicitly outlines the physical problem formulation along with key assumptions. Section~\ref{VC} outlines the governing equations for MHD-AC-EOF and solute dispersion in polymer solutions; specifically, the development of models for several types of complex polymer solutions and the estimation of their corresponding parameters are detailed in Appendix~\ref{appA}; additionally, the corresponding analytical solution algorithms are proposed. Section~\ref{Results-discussions} presents the core results of this study, including the validation of the proposed model and corresponding simulation results, alongside an in-depth mechanism analysis underlying the effects of major dimensionless parameters on solute dispersion behavior under both AC-EOF and mixed MHD-AC-EOF regimes. Finally, section~\ref{Conclusion} concludes the paper with discussions on the results.
\begin{figure}
  \centering
  \includegraphics[width=0.6\linewidth]{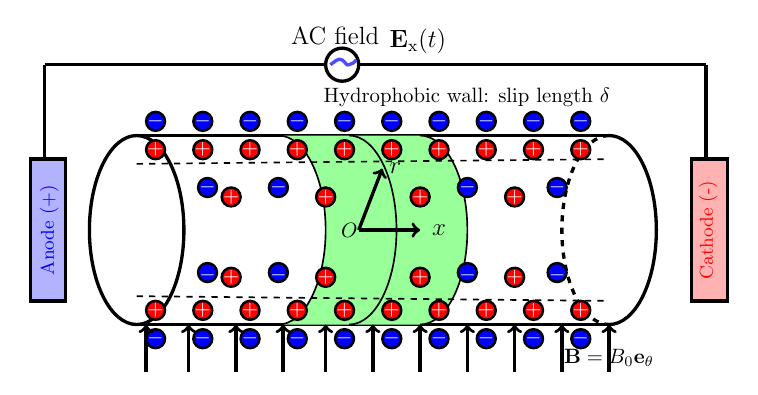}
  \caption{The diagram of the MHD-AC-EOF in a microtube with slip-dependent zeta potential.} 
  \label{fig1}        
\end{figure}

\section{Problem description and assumptions}
As shown in figure~\ref{fig1}, the unsteady flow characteristics and dispersion phenomena in viscoelastic solutions (such as polymer solutions) through a microtube under the coordinated driving of an AC electric field and a magnetic field are considered. A cylindrical coordinate system is adopted, with its origin precisely situated at the center of the tube. In this coordinate system, $r$ represents the radial distance measured from the central axis of the tube, while $x$ denotes the axial coordinate. The microtube has a radius $R$ and is sufficiently long such that the flow is fully developed and all physical quantities are independent of the axial coordinate $x$. The dispersion process commences at $t=0$ with the instantaneous injection of a minute quantity of solute. This solute is uniformly distributed across the cross-section at $x=0$ and subsequently propelled by the MHD-AC-EOF within the flexible microtube. It is postulated that the injected solute band is sufficiently dilute such that the flow of the carrier fluid remains unaffected by the presence of trace-level soluble substances. In this study, the research objectives are pursued in a sequential manner. Initially, an analysis of the dynamic behaviors of MHD-AC-EOF has been performed, which holds significant importance for studies on solute dispersion driven by MHD-AC-EOF. After that, the dispersion characteristics of the solutes injected into the tube will be analyzed under two distinct scenarios: pure AC-EOF and mixed MHD-AC-EOF. The following sections elaborate on these aspects in detail.

\section{Methematic analysis of electric potential, velocity and concentration ﬁelds}\label{VC}
Currently, we are preparing to conduct an analysis of  MHD-AC-EOF and solute dispersion in a narrow tube with slip-dependent zeta potential.  The velocity for the MHD-AC-EOF and concentration for solute dispersion are determined by momentum conservation and mass transport. In the case of considering a symmetric electrolyte solution, the continuity equation, the Navier-Stokes equation, and the convective diffusion equation can be given in the following forms
\begin{equation}
 \bar{ \nabla} \cdot \mathbf{\bar{V}}=0,
\end{equation}
\begin{equation}\label{cauchy}
  \rho\frac{\partial \mathbf{\bar{V}}}{\partial t}+\rho\left(\mathbf{\bar{V}}\cdot \bar{\nabla}\mathbf{\bar{V}}\right)=\nabla \cdot \bm{\bar{T}}+\mathbf{\bar{F}},
\end{equation}
\begin{equation}\label{C}
\frac{\partial \bar{C}}{\partial t}+\bar{\nabla}\cdot  \left(\mathbf{\bar{V}} \bar{C}\right)=D\bar{\nabla }\cdot (\bar{\nabla} \bar{C}).
\end{equation}
In the above equations, $\mathbf{\bar{V}}$ and $\bm{\bar{T}}$ are velocity vector and stress tensor, respectively, $\rho$ is fluid density, $t$ is time, $\bar{C}$ and $D$ are the concentration field and molecular diffusivity, respectively. Moreover, $\mathbf{\bar{F}}$ represents the body force acting on the fluid, originating from the electromagnetic interactions imposed on the system and potentially contributed by
  \begin{equation}\label{F}
    \mathbf{\bar{F}}=\bar{\rho}_{e}\mathbf{\bar{E}}+\mathbf{\bar{J}}\times \mathbf{\bar{B}}.
  \end{equation}
In this formulation, $\bar{\rho}_{e}=e\sum_lz_ln_l$ is the charge density, in which $e$ is the basic electric charge,  $n_l$ and $z_l$  are the molar concentration and charge valence of space $l$, respectively. 
$\mathbf{\bar{E}}$ is the applied electric field with a strength of $E_x(t)$ in the $x$-direction. $\mathbf{\bar{B}}$ represents the applied magnetic field with magnitude $B_0$ in the $\theta$-direction, which is symmetrically distributed around the circumference of the microtube cross-section. At each circumferential position, the magnetic field vector is locally tangential, perpendicular to the plane formed by the central axis and the radial direction, and acts uniformly on all fluid elements across the entire cross-section. $\mathbf{\bar{J}}$ signifies the local ion current density defined by Ohm's law
   \begin{equation}\label{J}
    \mathbf{\bar{J}}=\sigma_e(\mathbf{\bar{E}}+\mathbf{\bar{V}}\times \mathbf{\bar{B}}),
  \end{equation}
where $\sigma_e$ is the electrical conductivity.

\subsection{Electric potential distribution with slip dependent zeta potential}
We first evaluate the magnetic Reynolds number $Re_m$ to assess the validity of neglecting the induced magnetic field. For MHD flow in the microtube, the magnetic Reynolds number is defined as $Re_m=U_mL/\eta_m$, where $U_m$ is the characteristic velocity, $L$ is the characteristic length, and $\eta_m=1/(\mu\sigma_e)$ is the magnetic diffusivity, with $\sigma_e$ the electrical conductivity of the fluid and $\mu$ the magnetic permeability. Using the typical parameter ranges given in Section~\ref{Results-discussions} (e.g., $L=10$--$250\ \mathrm{\mu m}$, $\sigma_e=2.2\times10^{-4}$--$10^6\ \mathrm{S\cdot m^{-1}}$, $U_m\approx 1\ \mathrm{m\cdot s^{-1}}$, and $\mu=\mu_0=4\pi\times10^{-7}\ \mathrm{H\cdot m^{-1}}$), the calculated magnetic Reynolds number ranges from approximately $2.8\times10^{-15}$ to $3.1\times10^{-4}$, which is much less than unity. Under this condition, the induced magnetic field is negligible, and the applied external magnetic field remains unaffected by the flow field.
The potential $\Psi$ in the channel can be considered as the electrostatic potential and is generated by the superposition of external potential $\bar{\varphi}$ and surface charge potential (EDL potential) $\bar{\psi}$ on the microchannel wall. Assuming that the microchannel is very long and any end effects is ignored, then the EDL potential can be considered as independent of the axial position in the microchannel, the total potential can be expressed as follows
\begin{equation}\label{potential1}
    \bar{\Psi}=\bar{\varphi}+\bar{\psi}.
\end{equation}
Following Gauss's law, the electric potential equation can be decomposed into two equations
\begin{equation}\label{potential2}
   \varepsilon\bar{\nabla}^2\bar{\varphi}=0,
\end{equation}
\begin{equation}\label{potential3}
    \varepsilon\bar{\nabla}^2\bar{\psi}=-\bar{\rho}_{e},
\end{equation}
where $\varepsilon$ is the permittivity of the fluid.  According to the electrostatic theory of aqueous electrolyte solutions adjacent to charged surfaces, the ionic concentration $n_l$ display a spatial gradient extending from the capillary wall to its central axis. Under the assumption of equilibrium conditions for electrolyte solution across the entire capillary cross-section, the ionic concentration conform to a Boltzmann distribution
\begin{equation}\label{potential4}
    n_l=n_l^\infty\exp\left(-\frac{ez_l\bar{\psi}}{k_bT}\right).
\end{equation}
Here,  $n_l^\infty$ is defined as the bulk concentration of ion j under the neutral condition where the electric potential $\bar{\psi}$ equals zero.
By combining equations (\ref{potential3}) and (\ref{potential4}), the Poisson-Boltzmann equation governing the electric potential distribution in the microchannel is mathematically derived as follows
 \begin{equation}\label{P-B}
   \varepsilon \bar{\nabla}^{2}\bar{\psi}=-e\sum_lz_ln_l^\infty\exp\left(-\frac{ez_l\bar{\psi}}{k_bT}\right).
\end{equation}
This model provides a fundamental framework for describing ion transport phenomena in microfluidic systems~\citep{MWQK10}.  Given the challenges associated with obtaining analytical solutions to the complete set of electrokinetic equations, three primary approximation frameworks have been established over the years. These methodologies are specifically tailored to scenarios characterized by thin EDLs, weak field/flow conditions, and low surface electric potential, respectively. (i) If the thickness of the EDL (approximated by the Debye length $\lambda_D$) is significantly smaller than the system's characteristic length scale ($\lambda_D /L\ll 1$), specifically, fluid flow of a Newtonian electrolyte (Debye length $\lambda_D\sim \mathcal{O}(10^{-9} m)$) in a microchannel with complex geometry ($L\sim  \mathcal{O}(10^{-5} m)$), and under the assumption of quasi-laminar flow within the EDL, then the EDL effect can be approximated by a surface slip velocity. This approach circumvents the need to resolve the flow field within the EDL. (ii) The Helmholtz-Smoluchowski theory, $\mathbf{u}_{HS}=\mu\mathbf{E}$, is canonically employed to model this slip velocity under such scenarios~\citep{CJYC23,ZHHT23}. Importantly, slip boundary conditions inherently fail to resolve microscale phenomena occurring within EDL. (iii) The low-surface-potential assumption (i.e., $|ez_l\psi/k_bT| < 1$) can be adopted. This enables the application of the Debye-H\"{u}ckel linearization to equation (\ref{P-B})~\citep{LID2004,MJBS2006}. In this paper, for symmetric electrolyte solutions ($z^+=-z^-=z, D^+=D^-=D$) confined within micro/nanotubular geometries under low surface potential conditions, the Poisson-Boltzmann equation can be expressed in cylindrical coordinates and linearized via the Debye-H\"{u}ckel approximation as the following dimensionless form
\begin{equation}\label{slip-zeta1}
    \frac{\partial^{2} \psi}{\partial r^{2}}+\frac{1}{r}\frac{\partial \psi}{\partial r}=\kappa^2\psi.
\end{equation}
Here, the subsequent collection of dimensionless quantities is utilized: $\psi=ez\bar{\psi}/k_bT$,  $r=\bar{r}/R$, and $\kappa=R/\lambda_D$ is known as the Debye parameter, where $\lambda_D=(2n^\infty e^2z^2/\varepsilon k_bT)^{-1/2}$ is the Debye length.

The quantitative relationship between the surface zeta potential and the slip length was first derived by~\citet{CRSS02} as an analytical model, which can be expressed in the dimensionless form: $\zeta_\delta=\zeta(1+\delta \kappa)$. Here, $\zeta_\delta=\bar{\zeta}_\delta ez/k_bT$ and $\zeta=\bar{\zeta}ez/k_bT$ denote the dimensionless form of zeta potentials under slip and no-slip boundary conditions, respectively, with $\delta=\bar{\delta}/R$ defined as the dimensionless slip length parameter governing interfacial hydrodynamic effects. Subsequently,~\citet{JYDY02} introduced a modification factor $f_m$ to establish the adjusted zeta potential formulation $\zeta_\delta=\zeta(1+\delta \kappa)/f_m$. Most recently,~\citet{VTBJ08} developed a nonlinear slip-dependent zeta potential model for a symmetric $z:z$ electrolyte expressed as:
\begin{equation}\label{zeta-beta}
 \zeta_\delta=\zeta\left(1+\delta \kappa\frac{\sinh\zeta}{\zeta}\right).
\end{equation}
The core physical significance of this model is that it quantitatively reveals the strong coupling between solid-liquid interfacial slip and EDL electrokinetic properties in micro/nano-fluidic systems, providing a unified, precise theoretical framework for interpreting complex electrokinetic transport at hydrophobic interfaces. This analytical framework has been extensively adopted in contemporary investigations of electroosmotic transport within microfluidic systems~\citep{VMPS21,KSSP22,BMPB21,BPPB2023,SDJS24,DXLC2024}. Particularly, we utilizes this theoretical relationship to quantitatively determine the interfacial electric potential distribution along the microchannel surfaces. The suitable boundary conditions  required to solve the equation (\ref{slip-zeta1}) are enumerated as follows:
\begin{equation}\label{slip-zeta3}
   \left. \psi\right|_{r=0}=\mathrm{finite},\;\;   \left.\frac{\mathrm{d} \psi}{\mathrm{d} r}\right|_{r=0}=0,\;\;\;\left.\psi\right|_{r=1}=\zeta_\delta.
\end{equation}
Under the condition of slip-dependent zeta potential, the Debye-H{\"u}ckel approximation is valid when $|\zeta_\delta|<1$, and this condition can be guaranteed by imposing the constraint $\delta\kappa<(1-|\zeta|)/\sinh|\zeta|$ (with $|\zeta|<1$ for the non-slip zeta potential).
Consequently, by applying the stated boundary conditions, the dimensionless analytical solutions to equation (\ref{slip-zeta1}) and the expression for charge density $\rho_e$ are obtained below:
\begin{equation}\label{slip-zeta4}
      \psi=\zeta\left(1+\delta \kappa\frac{\sinh\zeta}{\zeta}\right)\frac{I_0(\kappa r)}{I_0(\kappa )},\;\;\rho_e=\kappa^2\psi,
\end{equation}
where $I_0(r)$ is the modified Bessel function of the first kind of order zero~\citep{GNWA1995}. Analogously, with respect to the slip-independent zeta potential, the mathematical representation of $\psi$ is given by
\begin{equation}\label{slip-zeta5}
      \psi=\zeta\frac{I_0(\kappa r)}{I_0(\kappa )}.
\end{equation}

\subsection{Velocity distribution}
With the electric potential distribution within the microtube now determined, we can compute the electrokinetic body force for incorporation into the Navier-Stokes equation. This enables us to analytically solve for the transient velocity profile in the microtube. Given the axisymmetric geometry and the elongated, narrow configuration characteristic of microfluidic channels, the axial velocity component exhibits independence on $\theta$ and $x$. The velocity field therefore reduces to $\mathbf{V}=(0, 0, u(r,t))$ in cylindrical coordinates, inherently satisfying the continuity equation for this constrained flow. Through simplification of the viscoelastic constitutive relation (equation (\ref{FMaxwell})) for the rectilinear flow geometry and subsequent substitution into the momentum equation, governing equations for the velocity distribution can be reformulated as
\begin{align}\label{FMaxwell-ACEOF}
  & \left(1+\lambda^{\alpha-\beta}\frac{\partial^{\alpha-\beta}}{\partial \bar{t}^{\alpha-\beta}}\right)\left (\rho\frac{\partial \bar{u}}{\partial\bar{t}}+\sigma_e B_0^2\bar{u}\right)\\ \nonumber
   &=\eta\lambda^{\alpha-1}\frac{\partial^{\alpha-1}}{\partial \bar{t}^{\alpha-1}}\frac{1}{\bar{r}}\frac{\partial }{\partial \bar{ r}}\left(\bar{r}\frac{\partial \bar{u}}{\partial \bar{r}}\right)+\bar{\rho}_{e}(\bar{r})\left(1+\lambda^{\alpha-\beta}\frac{\partial^{\alpha-\beta}}{\partial \bar{t}^{\alpha-\beta}}\right)\bar{E}_x(\bar{t}).
\end{align}

As previously discussed, the applied electric field, and resulting flow velocity each contain harmonic oscillating components with identical angular frequency ($\omega$). Consequently, these quantities can be represented in complex form as follows:
\begin{equation}\label{u-E-complex}
   \left \langle\bar{u}(\bar{r},\bar{t}), \;\bar{E}_x(\bar{t})\right\rangle= \operatorname{Im}\left\{\left \langle\bar{u}_0(\bar{r}), \;\bar{E}_0\right\rangle \exp(\mathrm{i}\bar{\omega} \bar{t})\right\}.
\end{equation}
In this mathematical formulation, $\mathrm{i}=\sqrt{-1}$ denotes the imaginary unit, while $\operatorname{Im}$ designates the imaginary component of a complex number. Substituting equation (\ref{u-E-complex}) into equations (\ref{FMaxwell-ACEOF}) and equating the coefficients of the exponential term
 in the resulting equation on both sides, the following expression can be derived
\begin{align}\label{u-complex1}
    &\eta \lambda^{\alpha-1}(\mathrm{i} \bar{\omega})^{\alpha-1}\left(\frac{\mathrm{d}^{2} \bar{u}_0(\bar{r})}{\mathrm{d} \bar{r}^{2}}+\frac{1}{\bar{r}}\frac{\mathrm{d} \bar{u}_0(\bar{r})}{\mathrm{d} \bar{r}}\right)\\ \nonumber
    &-\left[1+\lambda^{\alpha-\beta}(\mathrm{i} \bar{\omega})^{\alpha-\beta}\right]\left(\rho \mathrm{i}\bar{\omega} \bar{u}_0(\bar{r})+\sigma_e B_0^2\bar{u}_0(\bar{r})\right)
    =-\bar{\rho}_e(r)\left[1+\lambda^{\alpha-\beta}(\mathrm{i} \bar{\omega})^{\alpha-\beta}\right].
\end{align}
For analytical tractability, substituting the charge density expression via equation (\ref{slip-zeta4}) and performing proper scaling, the above equation can be further cast into the dimensionless form below.
\begin{align}\label{dimensionless1}
   & \frac{\mathrm{d}^2 u_0(r)}{\mathrm{d}r^2}+\frac{1}{r}\frac{\mathrm{d}u_0(r)}{\mathrm{d}r}-\left(\mathrm{i}^{1-\alpha}De^{1-\alpha}+\mathrm{i}^{1-\beta}De^{1-\beta}\right)
    (Re \mathrm{i}+Ha^2)u_0(r)\\ \nonumber
    &=-\kappa^2\zeta\left(\mathrm{i}^{1-\alpha}De^{1-\alpha}+\mathrm{i}^{1-\beta}De^{1-\beta}\right)\left(1+\kappa \delta\frac{\sinh\zeta}{\zeta}\right)\frac{I_0(\kappa r)}{I_0(\kappa)}.
\end{align}
Within the preceding equation, the dimensionless variables are hereinafter denoted as follows: $r=\bar{r}/R,\;\psi =\bar{\psi}ez/k_bT,\;\kappa= \bar{\kappa}R,\; \;u=\bar{u}/U_{EO},\;\omega=\bar{\omega}/\omega_{ref},\;t=\bar{t}\omega_{ref},\;De=\lambda \bar{\omega},\; Re=\rho \bar{\omega} R^{2}/\eta,$ \;$Ha=B_0R\sqrt{\frac{\sigma_e}{\eta}}$, where $U_{EO}=-\varepsilon k_bTE_0/\eta ze$ represent the Helmholtz-Smoluchowski velocity, $\omega_{ref}=D/R^2$ is the reference frequency, $De$, $Re$ and $Ha$ denote the Deborah number, oscillation Reynolds number and Hartmann number, respectively. To solve equation (\ref{dimensionless1}), the following conditions are defined: The value of velocity should be finite at the center of the capillary; velocity needs to meet the axial-symmetry requirement; a slip condition holds at the capillary wall. Each of these conditions can be expressed in the following dimensionless forms, respectively
\begin{equation}
\left. \begin{array}{ll}
\displaystyle  \left. u_0(r)\right|_{r=0}=\mbox{finite},\;\;\;\;\left.\frac{\mathrm{d} u_0(r)}{\mathrm{d} t}\right|_{r=0}=0,\\[8pt]
\displaystyle \left[u_0(r)+\delta\left.\frac{\mathrm{d} u_0(r)}{\mathrm{d}r} \right]\right|_{r=1}=0.
 \end{array}\right\}
 \label{dimensionless2}
\end{equation}
The solution of the equations (\ref{dimensionless1}) and (\ref{dimensionless2}) can be derived as
\begin{align}\label{solution1}
    u_0(r)=C_1&\left[C_2\mathrm{J}_0(i\sqrt{A}r)+C_3r\mathrm{I}_0(\kappa r)\mathrm{I}_1(\sqrt{A}r)\mathrm{Y}_0(-\mathrm{i}\sqrt{A}r) \right.\\ \nonumber
    &-\mathrm{i}C_3r\mathrm{I}_0(\kappa r)\mathrm{J}_0(i\sqrt{A}r)\mathrm{Y}_1(-i\sqrt{A}r)\\ \nonumber
  &\left.+C_4r\mathrm{I}_1(\kappa r)\mathrm{J}_0(i\sqrt{A}r)\mathrm{Y}_0(-i\sqrt{A}r)-C_4r\mathrm{I}_0(\sqrt{A}r)\mathrm{I}_1(\kappa r)\mathrm{Y}_0(-i\sqrt{A}r)\right].
\end{align}
The explicit expressions for coefficients $C_1-C_4$ and the coupled parameter $A$ are provided in Appendix~\ref{appB}. Since the analytical velocity solution is mathematically involved, we have performed an equivalent mathematical transformation to obtain a more compact form to facilitate the subsequent derivation and evaluation of the concentration distribution, with detailed derivations presented in Appendix~\ref{appB}.

\subsection{Analysis of solute dispersion}
The generalized dispersion model developed by \cite{Gill1970} employs time-varying dispersion coefficients to ensure applicability at all times after solute injection and is readily solvable for fully-developed steady tube flow. In contrast, this study deals with a multi-field driven oscillatory flow with a complex velocity distribution. Due to this complexity, quantities such as $K_2(t)$ and $g_1(r,t)$ are solved using a method different from the reference (see equations \eqref{eq:F}-\eqref{K2-tau} and Appendix \ref{appC}). The model is expressed in cylindrical coordinates as follows:
\begin{equation}\label{C-dispersion}
   \frac{\partial \bar{C}}{\partial \bar{t}}+\bar{u}(\bar{r},\bar{t}) \frac{\partial \bar{C}}{\partial \bar{x}}=D\left[  \frac{\partial^{2} \bar{C}}{\partial \bar{x}^{2}}+\frac{1}{\bar{r}}\frac{\partial}{\partial \bar{r}}\left(\bar{r}\frac{\partial \bar{C}}{\partial \bar{r}}\right)\right],
\end{equation}
where $\bar{u}(\bar{r},\bar{t}) $  is the velocity of the MHD-AC-EOF. We assume that the injected solute is introduced at the initial time $\bar{t}=0$ within a bandwidth of $x_s$, and its concentration is set to $C_0$. The concentration approaches zero as the spatial position extends to infinity. No mass transfer occurs at the wall surface, and the concentration distribution exhibits perfect symmetry about the central axis. Then, the associated conditions can be rewritten as

\begin{subequations}\label{eq:group}
    \begin{equation}
       \bar{C}(\bar{x}, \bar{r}, 0)=\bar{C}_0,    \left(|x|\leq \frac{1}{2}x_s\right),\;\; \bar{C}(\bar{x}, \bar{r}, 0)=0,\;\;\left (|x|> \frac{1}{2}x_s\right),
        \label{eq:C-condition1}
    \end{equation}
    \begin{equation}
       \bar{C}(\infty, \bar{r}, \bar{t})=0,\;\; \;\;\frac{\partial \bar{C}}{\partial \bar{r}}(\bar{x},0,\bar{t})=\frac{\partial \bar{C}}{\partial \bar{r}}(\bar{x},R,\bar{t})=0.
        \label{eq:C-condition2}
    \end{equation}
\end{subequations}
The subsequent dimensionless variables are introduced for further analysis
\begin{equation}\label{dimensionless-varibles}
  C=\frac{\bar{C}}{C_0},\;x=\frac{D\bar{x}}{R^2U_{EO}},\;\;x_s=\frac{D\bar{x}_s}{R^2U_{EO}},\;\;r=\frac{\bar{r}}{R},\;\;
  t=\frac{D\bar{t}}{R^2},\;\;Pe=\frac{RU_{EO}}{D}.
\end{equation}
Then, the aforesaid equation is capable of being expressed in the following dimensionless form:
\begin{subequations}\label{eq:group}
    \begin{equation}
        \frac{\partial C}{\partial t}+u(r,t) \frac{\partial C}{\partial x}=\frac{1}{Pe^2} \frac{\partial^{2} C}{\partial x^{2}}+\frac{1}{r}\frac{\partial}{\partial r}\left(r\frac{\partial C}{\partial r}\right),
        \label{eq:dimensionless-C}
    \end{equation}
    \begin{equation}
        C(x, r, 0)=1,    \left(|x|\leq \frac{1}{2}x_s\right),\;\;C(x, r, 0)=0,\;\; \left(|x|> \frac{1}{2}x_s\right),
        \label{eq:dimensionless-condition1}
    \end{equation}
    \begin{equation}
      C(\infty, r, t)=0,\;\; \frac{\partial C}{\partial r}(x,0,t)=\frac{\partial C}{\partial r}(x,R,t)=0,
        \label{eq:dimensionless-condition2}
    \end{equation}
\end{subequations}
where $Pe$ is the Peclet number.

For analyzing the solute dispersion process, a new coordinate system is set up, in which the $x$-axis moves a distance determined by calculating the mean velocity of the flow
\begin{equation}\label{x1}
    x_1=x-\int_0^tU(t)dt, \;\;\mathrm{ where}  \;\; U(t)=2\int_0^1ru\mathrm{d}r=\operatorname{Im}\{U_1 e^{i\omega t}\},
\end{equation}
where $U(t)$ denotes the instantaneous cross-sectionally averaged velocity (with the radius normalized to unity), and $U_1 = 2\int_0^1 r u_0 \mathrm{d}r$ represents its complex amplitude. Given the new coordinate system, equation \eqref{eq:dimensionless-C} can be reformulated as
\begin{equation}\label{C-dimensionless3}
    \frac{\partial C}{\partial t}+(u-U)\frac{\partial C}{\partial x_1}=\frac{1}{Pe^2} \frac{\partial^{2} C}{\partial x_1^{2}}+\frac{1}{r}\frac{\partial}{\partial r}\left(r\frac{\partial C}{\partial r}\right).
\end{equation}

Adopting the method proposed by~\citet{Gill1970}, the local concentration $C$ is currently defined by a series expansion based on axis gradients of different orders of the area-average concentration $C_m$
\begin{equation}\label{C-Cm}
  C=\sum_{p=0}^{\infty}g_p(r,t)\frac{\partial^p C_m}{\partial x_1^p},
\end{equation}
where
\begin{equation}\label{Cm}
  C_m=2\int_{0}^{1} rC\mathrm{d}r,
\end{equation}
the function $g_p(r,t)$ bears resemblance to the Taylor expansion coefficient. By default, $g_0=1$. After substituting equation (\ref{C-Cm}) into (\ref{C-dimensionless3}) , we obtain
\begin{align}\label{Cm-2}
  &\frac{\partial C_m}{\partial t}+(u-U)\frac{\partial C_m}{\partial x_1}-\frac{1}{Pe^2}\frac{\partial^2 C_m}{\partial x_1^2}
  +\sum_{p=1}^{\infty}\left\{\left[\frac{\partial g_p}{\partial t}-\frac{1}{r}\frac{\partial}{\partial r}\left(r\frac{\partial g_p}{\partial r}\right)\right]\frac{\partial^p C_m}{\partial x_1^p}\right. \\ \nonumber
  &\left.+(u-U)g_p\frac{\partial^{p+1} C_m}{\partial x_1^{p+1}} -\frac{1}{Pe^2}g_p\frac{\partial^{p+2} C_m}{\partial x_1^{p+2}}+g_p\frac{\partial^{p+1} C_m}{\partial t\partial x_1^{p}}\right\}=0.
\end{align}

Assuming the evolution of $C_m$ is diffusive from time zero, we can formulate the generalized dispersion model in dimensionless form with time-dependent dispersion coefficients as follows
 \begin{equation}\label{Cm-K}
  \frac{\partial C_m}{\partial t}=\sum_{j=0}^{\infty}K_j(t)\frac{\partial^{j} C_m}{\partial x_1^{j}}.
\end{equation}
The coefficients $K_0(t)$, $K_1(t)$, and $K_2(t)$ denote the exchange, convection, and dispersion coefficients, respectively, while $K_3(t)$ and $K_4(t)$ in equation (\ref{Cm-K}) characterize the skewness and kurtosis of the solute concentration distribution during the dispersion process. Since the terms involving $K_0-K_2$ are several orders of magnitude larger than the higher-order terms, the cross-sectional mean concentration profile can be accurately approximated using only the first three terms~\citep{Gill1970}. This convention is widely adopted in the relevant literature~\citep{SSMS2020,MRHS19,RanaM16}. In the present study, because the wall is impermeable and there is no mass exchange at the boundary, we have $K_0(t)=0$. The coefficients $K_1(t)$ and $K_2(t)$ are determined next.
Upon the substitution of equation (\ref{Cm-K}) into equation (\ref{Cm-2}), followed by the rearrangement of terms, and taking into account the condition
\begin{equation}\label{Cm-condition}
   \frac{\partial^{p+1} C_m}{\partial t\partial x_1^{p}}=\sum_{j=1}^{\infty}K_j(t)\frac{\partial^{j+p} C_m}{\partial x_1^{j+p}},
\end{equation}
an infinite set of differential equations is generated by equating the coefficients of  $\partial^{p} C_m/\partial x_1^{p}$  to zero
\begin{subequations}\label{g-K}
\begin{align}
\frac{\partial g_1}{\partial t}=&\frac{1}{r}\frac{\partial}{\partial r}\left(r\frac{\partial g_1}{\partial r}\right)-\left(u-U+K_1(t)\right),\label{g-K:g1}\\
	 \frac{\partial g_2}{\partial t}=&\frac{1}{r}\frac{\partial}{\partial r}\left(r\frac{\partial g_2}{\partial r}\right) -\left(u-U+K_1(t)\right)g_1-\left(K_2(t)-\frac{1}{Pe^2}\right),\label{g-K:g2}\\ 
 \frac{\partial g_{p+2}}{\partial t}=&\frac{1}{r}\frac{\partial}{\partial r}\left(r\frac{\partial g_{p+2}}{\partial r}\right) -\left(u-U+K_1(t)\right)g_{p+1}\label{g-K:gp}\\  \nonumber
&-\left(K_2(t)-\frac{1}{Pe^2}\right)g_p-\sum_{j=3}^{p+2}K_j(t)g_{p+2-j},\;\;\;p=1,2,3,\cdots.
\end{align}
\end{subequations}
Equations \eqref{Cm-K} and \eqref{g-K} formulate the transient dispersion problem associated with time-varying flow. In view of the initial and boundary conditions for $C$ as presented in  equation \eqref{eq:group}, and the definition of $C_m$ in (\ref{Cm}), the corresponding conditions for $g_p$ can be derived as
\begin{subequations}\label{g-conditions}
\begin{align}
   g_p(0,\;r)&=0,\label{g-conditions:a}\\
   \frac{\partial g_p}{\partial r} (t,\;0)&=\frac{\partial g_p}{\partial r} (t,\;1)=0, \label{g-conditions:b}\\
   \int_{0}^{1}rg_p\mathrm{d}r&=0. \label{g-conditions:c}
\end{align}
\end{subequations}

Multiplying equation \eqref{g-K:g1} by $r$ and subsequently integrating $r$ from $0$ to $1$ gives the corresponding result
\begin{equation}\label{K1-t}
    K_1(t)=-2\int_{0}^{1}(u-U)r\mathrm{d}r=0.
\end{equation}
It is worth reminding here that $U$ is the average velocity, which has been defined previously, as shown in equation (\ref{x1}). Moving forward, we compute $K_2(t)$, an essential coefficient in equation (\ref{Cm-K}). Its relationship to $g_1$ and $g_2$ makes determination non-trivial. Hence, we will now employ an alternative method to calculate it. Initially, we intend to utilize the homogenization principle presented by Duhamel's theorem~\citep{CFRV42}. Actually, since $g_1$ is known (as derived from equation \eqref{g-K:g1}, with the full derivation provided in Appendix~\ref{appC}), combining Duhamel's theorem with the superposition principle represents an optimal approach for obtaining $K_2(t)$ without needing to determine the function $g_2$. Denote $F(r,\;t,\;\tau)$ as the solutions of the equation below, along with its initial and boundary conditions
\begin{subequations}\label{eq:F}
\begin{equation}\label{eq:F:Frttao}
    \frac{\partial F}{\partial t}(r,t,\tau)+\left[u(r,\tau)-U(\tau)\right]g_1(r,\tau)+K_2(\tau)-\frac{1}{Pe^2}=\frac{1}{r}\frac{\partial}{\partial r}\left(r\frac{\partial F(r,t,\tau)}{\partial r}\right),
\end{equation}
\begin{equation}\label{eq:F:Frttao-condi}
    F(r,0,\tau)=0,\;\;\frac{\partial F}{\partial r}(0,t,\tau)=\frac{\partial F}{\partial r}(1,t,\tau)=0.
\end{equation}
\end{subequations}
In this context, $\tau$ acts as a character temporarily treated as a variable for the purpose of substituting $\tau$ in $u(r,t)$, $U(t)$, $g_1(r,t)$ and $K(t)$. By decomposing equation \eqref{eq:F} into two separate equations and determining the solutions $F_1(r,\tau)$ and $F_2(r,t,\tau)$ for each of equation respectively, one can readily acquire the solution $F(r,t,\tau)=F_1(r,\tau)+F_2(r,t,\tau)$ to equations \eqref{eq:F:Frttao} and \eqref{eq:F:Frttao-condi}.
\begin{subequations}\label{eq:F12}
 \begin{equation}\label{eq:F12:F1}
    \frac{1}{r}\frac{\partial}{\partial r}\left(r\frac{\partial F_1(r,\tau)}{\partial r}\right)=\left[u(r,\tau)-U(\tau)\right]g_1(r,\tau)+K_2(\tau)-\frac{1}{Pe^2},
\end{equation}
\begin{equation}\label{eq:F12:F2}
     \frac{\partial F_2(r,t,\tau)}{\partial t}=\frac{1}{r}\frac{\partial}{\partial r}\left(r\frac{\partial F_2(r,t,\tau)}{\partial r}\right).
\end{equation}
\end{subequations}
Furthermore, if we multiply both sides of equation \eqref{eq:F12:F1} by $r$ and then perform integration from $0$ to $1$, the expression for $K_2(\tau)$ can be obtained as follows
\begin{equation}\label{K2-tau}
    K_2(\tau)=\frac{1}{Pe^2}-2\int^1_0r[u(r,\tau)-U(\tau)]g_1(r,\tau)\mathrm{d}r+\left[r\frac{\partial F_1(r,\tau)}{\partial r}\right]^{r=1}_{r=0}.
\end{equation}
To obtain $ K_2(\tau)$, we need to solve the last term on the right side of the equation (\ref{K2-tau}), which involves the boundary conditions of function $F_1$. The boundary condition $\partial F(1,t,\tau)/\partial r=0$ necessitates $\partial F_2(1,t,\tau)/\partial r=0$. Otherwise,
if $\partial F_2(1,t,\tau)/\partial r\neq 0$, by the superposition $\partial F/\partial r=\partial F_1/\partial r+\partial F_2/\partial r$, this would force
$\partial F(1,t,\tau)/\partial r$ to become a non-zero function of $t$, directly contradicting the original boundary condition. Therefore, we can deduce $\partial F_1(1,\tau)/\partial r=0$. Then, the equation (\ref{K2-tau}) can be rewritten in the following form by replacing $\tau$ with $t$
\begin{equation}\label{K2-t}
    K_2(t)=\frac{1}{Pe^2}-2\int^1_0r[u(r,t)-U(t)]g_1(r,t)\mathrm{d}r.
\end{equation}
Under the imposed flow conditions, the convection coefficient vanishes, i.e., $K_1(t)=0$, see equation (\ref{K1-t}). Consequently, the generalized dispersion model reduces to a simplified form
 \begin{equation}\label{Cm-K-2}
  \frac{\partial C_m}{\partial t}=K_2(t)\frac{\partial^{2} C_m}{\partial x_1^{2}}.
\end{equation}

Next, to derive the average concentration distribution $C_m$, we firstly define a new variable $\xi=\int_{0}^{t}K_2(t')\mathrm{d}t'$ such that the governing equation (\ref{Cm-K-2}) and corresponding initial-boundary conditions transform into
\begin{subequations}\label{eq:Cm}
 \begin{equation}\label{eq:Cm-xi}
  \frac{\partial C_m}{\partial \xi}=\frac{\partial^{2} C_m}{\partial x_1^{2}},
\end{equation}
 \begin{equation}\label{eq:Cm-xi-conditions}
   C_m(x_1,0)=x_s\cdot \delta_{x_s},\;\;\;
   \delta_{x_s}=
     \begin{cases}
          \frac{1}{x_s}& \text{ $ x_1\in \left[-\frac{x_s}{2},\frac{x_s}{2}\right] $ } \\
         0,& \text{ others }
     \end{cases}.
\end{equation}
\end{subequations}
By applying the Fourier transform to equations (\ref{eq:Cm-xi}) and (\ref{eq:Cm-xi-conditions}), we have
 \begin{equation}\label{eq:42}
  \tilde{C}_m(\mu, \xi)=\int_{-\infty}^{\infty}x_s\cdot \delta_{x_s}\cdot \mathrm{e}^{-\mathrm{i}\mu \nu}\mathrm{d}\nu\cdot \mathrm{e}^{-\mu^2 \xi}.
\end{equation}
Ultimately, the solutions to the system (\ref{eq:Cm}) are given by the following expression via the inverse Fourier transform.
\begin{equation}\label{eq:43}
   C_m(x_1, \xi)=\frac{1}{2}\left[ \erf\left(\frac{\frac{1}{2}x_s+x_1}{2\xi^\frac{1}{2}}\right)
   +\erf\left(\frac{\frac{1}{2}x_s-x_1}{2\xi^\frac{1}{2}}\right)\right],
\end{equation}
where $\erf(x)$ is the error function.
\begin{table}
    \begin{center}
    \def~{\hphantom{0}}
    \begin{tabular}{lcccccccc}
    \hline
 & Parameter & &&&&Value &\\

  &Order of the fractional derivative $\alpha$  & &&&& 0--1 & \\
 & Order of the fractional derivative $\beta$ & &&&& 0--1 ($<\alpha$)&\\
 & Deborah number $De$ & &&&&0--$1$&\\
 & Debye parameter $\kappa$& &&&&10--30& \\
 & Oscillation Reynolds number $Re$ & &&&& $0-50$&  \\
 & Peclet number $Pe$ & &&&& 50&  \\
 & Hartmann number $Ha$ & &&&&0--3 &  \\
 & Interface slip coefficient $\delta$ & &&&&0--0.05&\\
 & Surface slip-independent zeta  potential  $\zeta$&& &&&0.5&\\
 & Value of band wide $x_s$ & &&&&0.019& \\
  \hline
  \end{tabular}
  \caption{Normalized ranges of major dimensionless parameters.}\label{mytab3}
  \end{center}
\end{table}

\begin{figure}
  \centering
  \subfigure
  {\includegraphics[width=0.42\textwidth]{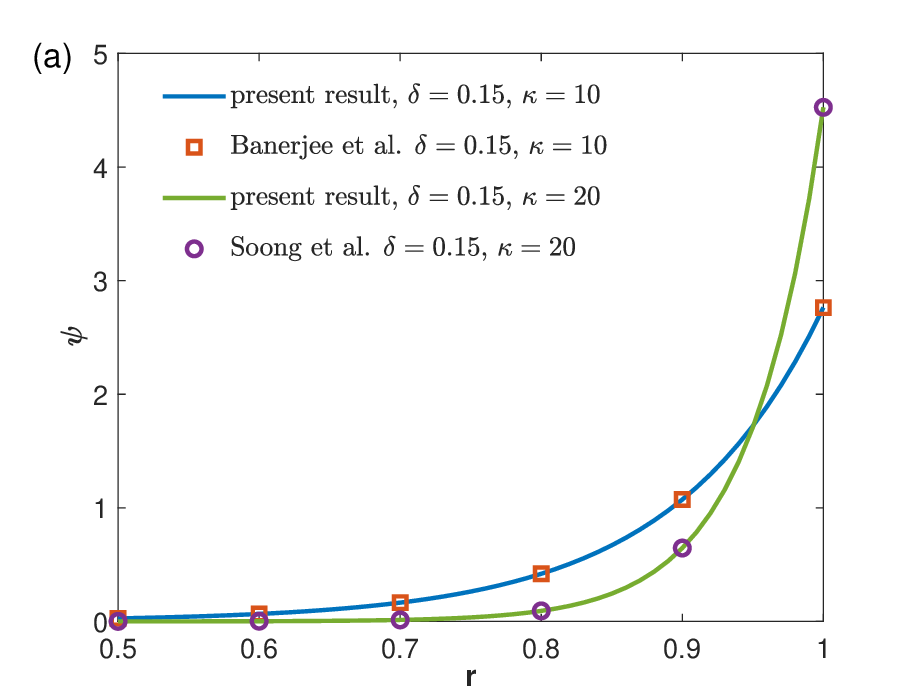}}
  \subfigure
  {\includegraphics[width=0.42\textwidth]{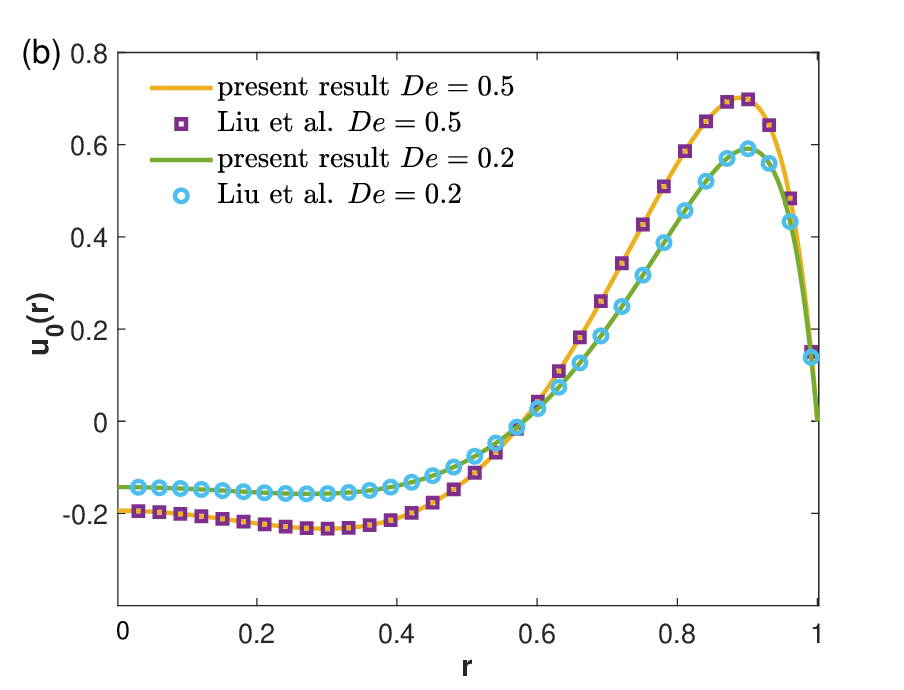}}
  \subfigure
  {\includegraphics[width=0.42\textwidth]{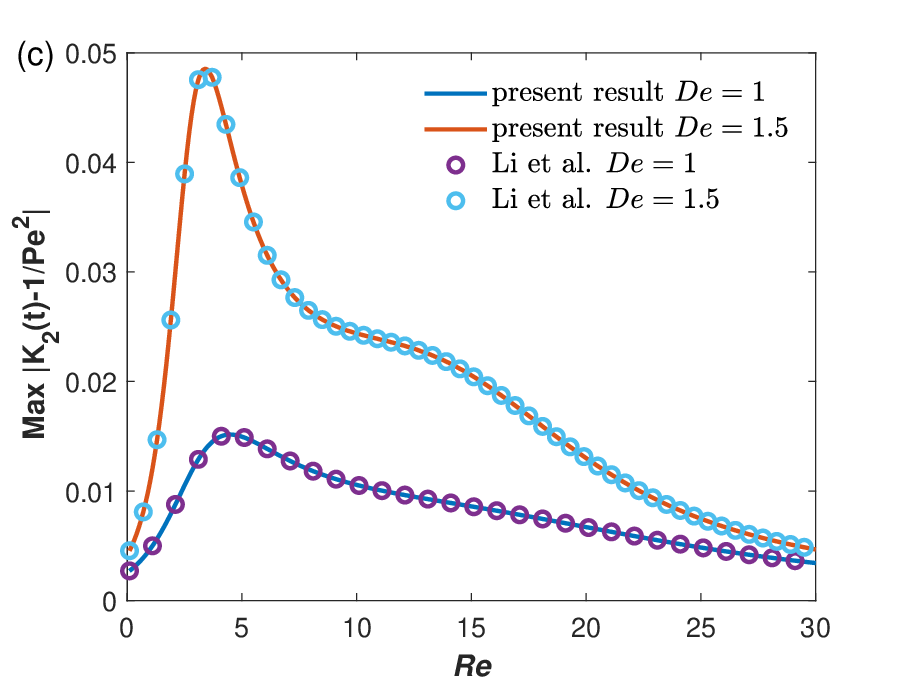}}
    \subfigure
  {\includegraphics[width=0.42\textwidth]{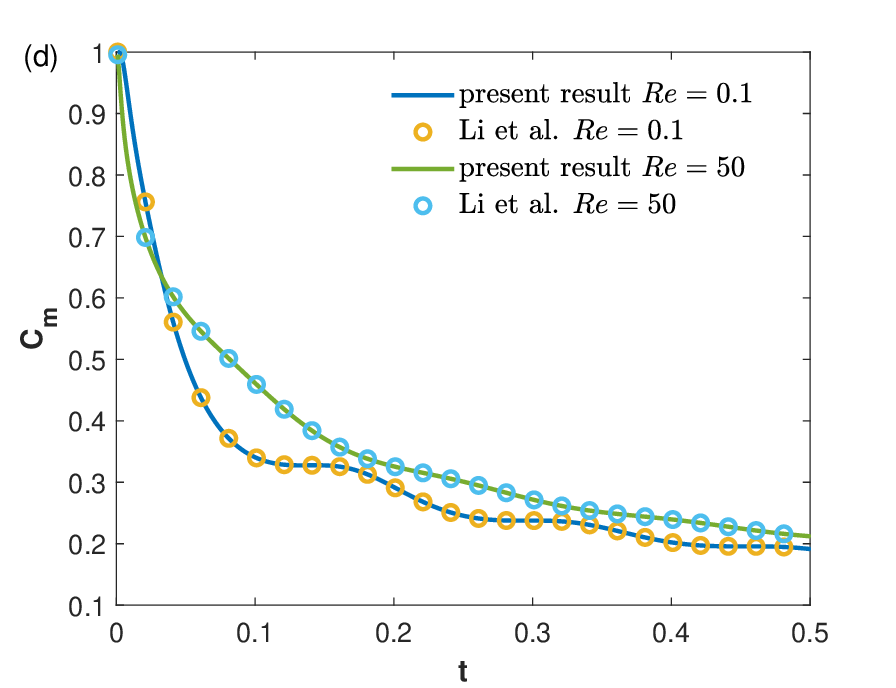}}
  \caption{(a) Comparison of the present potential distribution with~\citet{BMPB21,Soong10} for the case of a slip-dependent zeta potential at $\zeta=1$;  (b) Comparison of present velocity profiles with~\citet{LJCY12} at $\alpha=1$, $\beta=0$, $\kappa=20$, $\zeta=1$, $\delta=0$, $Re=10\pi$; (c)  Comparison of mean dispersion coefficient between the present study and~\citet{HLYJ17} at $\alpha=1$, $\beta=0$, $\kappa=10$, $\omega=20$, $\zeta=1$, $\delta=0$; (d)  Comparison of mean concentration between the present study and~\citet{HLYJ17} at $\alpha=1$, $\beta=0$, $\kappa=10$, $\omega=20$, $De=1$, $Pe=50$, $\zeta=1$, $\delta=0$, $x_s=0.019$, $x_1=0.005$.\label{fig2}}
\end{figure}

\section{Results and discussions}\label{Results-discussions}
The primary objective of this investigation is to comprehensively characterize the temporal evolution of solute dispersion in microtube driven by the MHD-AC-EOF. Consequently, critical parameters governing fluid flow and solute transport must be precisely determined. Based on experiments in microfluidic systems, the radius of the tube $R$ is selected within the range of $10-250$ $\mathrm{\mu m}$~\citep{Sparks2003}, the fluid density $\rho$ is set to $10^{3}$  $\mathrm{kg\cdot m^{-3}}$~\citep{WHYC2011,Nghe2011} and fluid viscosity $\eta$ is in the range of $10^{-3}-2\times10^{-3}$  $\mathrm{kg\cdot m^{-1}s^{-1}}$~\citep{WHYC2011,Nghe2011}. Based on the works of EOF and MHD~\citep{LJCY12,HLYJ17}, Helmholtz-Smoluchowski velocity $U_{EO}$ is set to $10^{-5}-2.5\times 10^{-4}$ $\mathrm{m\cdot s^{-1}}$. Referring to the analysis of solute dispersion in microchannel by~\cite{RanaM16,PVSN16}, the constant diffusion coefficient  $D$  is taken as $10^{-10}-10^{-9} \;\mathrm{m^2\cdot s^{-1}}$, the local concentration $C_0$ as $10^{-3} \;\mathrm{mol \cdot L^{-1}}$, and  the injection band width $\bar{x}_s$ as $10^{-4}$  $\mathrm{m}$. Additionally, the interface slip coefficient $\delta$ is considered to be $0-0.5$ $\mathrm{\mu m}$~\citep{YZSG01,VMPS21}, the imposed magnetic field $B_0$ as $0-0.44$  $\mathrm{T} $~\citep{Jang2000,YJXL19,ZYYJ17}. 
The electrical conductivity $\sigma_e$ is $2.2 \times10^{-4}-10^{6}$ $\mathrm{S\cdot m^{-1}}$~\citep{ZYYJ17,YJXL19,SJBK2023}, where the high conductivity values are set to explore the evolution laws of fluid flow and solute transport across a broad electromagnetic effect parameter range. The relaxation time $\lambda$ is simulated within a range of $4 \times 10^{-4}$ $\mathrm{s}$ to $5 \times 10^{-3}$ $\mathrm{s}$. This interval safely aligns with the typical relaxation time domain $10^{-4}-10^3$ $\mathrm{s}$ documented in \citet{BSEN01}. Crucially, the relaxation time must satisfy the condition $\lambda\leq2\pi/\bar{\omega} $ (or $De<2\pi$), ensuring it remains smaller than the oscillating period of the electric field, as stipulated in \citet{BSEN01,LJLL15}, thus the oscillating angular frequency $\bar{\omega}$ is selected within $ 0.2\pi-5\pi\times 10^3 \;\mathrm{rad \cdot s^{-1}}$. Therefore, the dimensionless parameter values derived from real-world microfluidic system data and used in our simulation are listed in the table~\ref{mytab3}.

Prior to investigating the solute dispersion under the combined MHD-AC-EOF drive, the accuracy of the analytical solutions from the previous sections must be validated. Figure \ref{fig2}(a–d) provide a comparison between the present analytical results of electric potential, velocity, dispersion coefficient, and mean concentration distribution with results from existing literature~\citep{BMPB21,Soong10,LJCY12,HLYJ17}, across varying critical governing parameter values. As evident from figure \ref{fig2}(a–d), all sets of results exhibit excellent consistency, which confirms both the reliability of the current model and the precision of the analytical algorithm utilized in this study. Subsequent sections will present a comparative analysis of the effects of major dimensionless parameters on the dispersion coefficient and mean concentration for two flow configurations: AC electroosmotic-driven flow ($Ha=0$) and mixed MHD and AC electroosmotic-driven flow ($Ha\neq0$).

\subsection{Velocity distribution}\label{velocity}
\begin{figure}
  \center
  \subfigure
  {\includegraphics[width=0.3\textwidth]{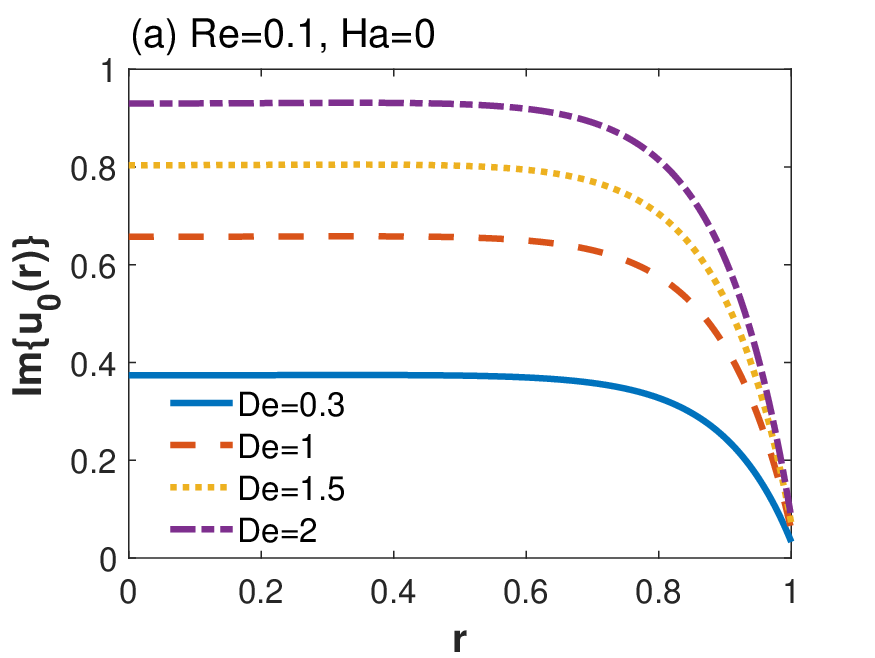}}
  \subfigure
  {\includegraphics[width=0.3\textwidth]{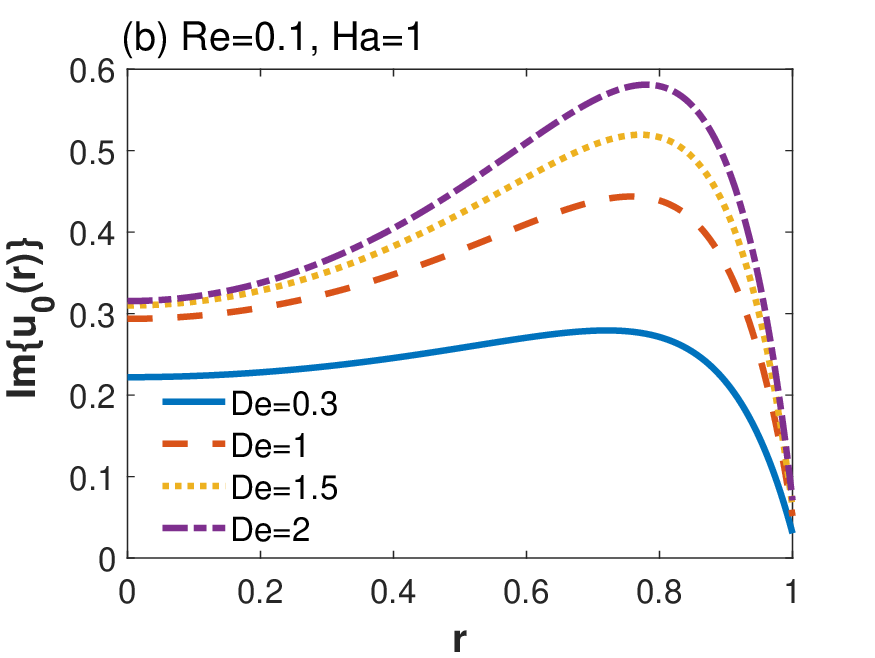}}
   \subfigure
  {\includegraphics[width=0.3\textwidth]{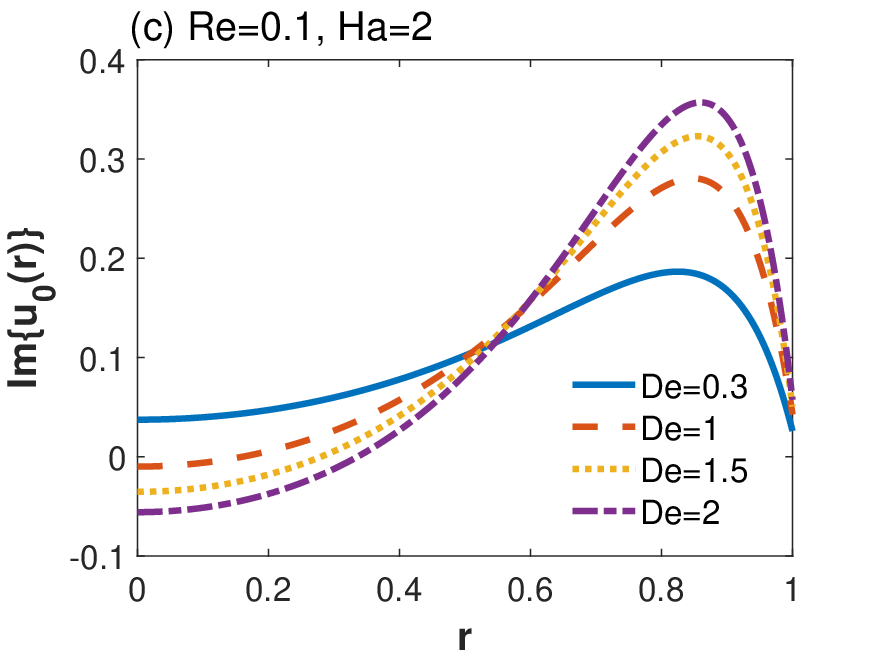}}
    \subfigure
  {\includegraphics[width=0.3\textwidth]{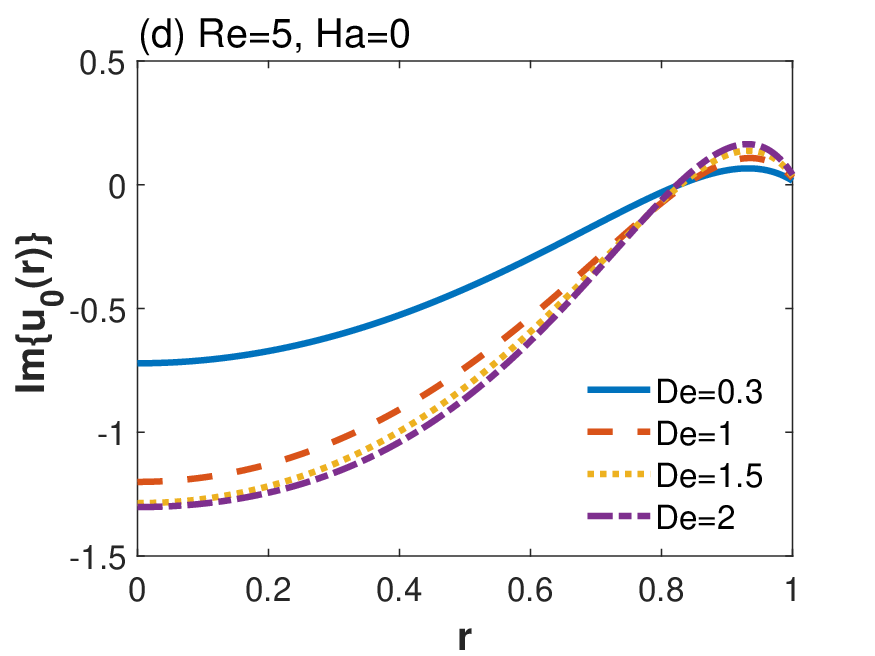}}
    \subfigure
  {\includegraphics[width=0.3\textwidth]{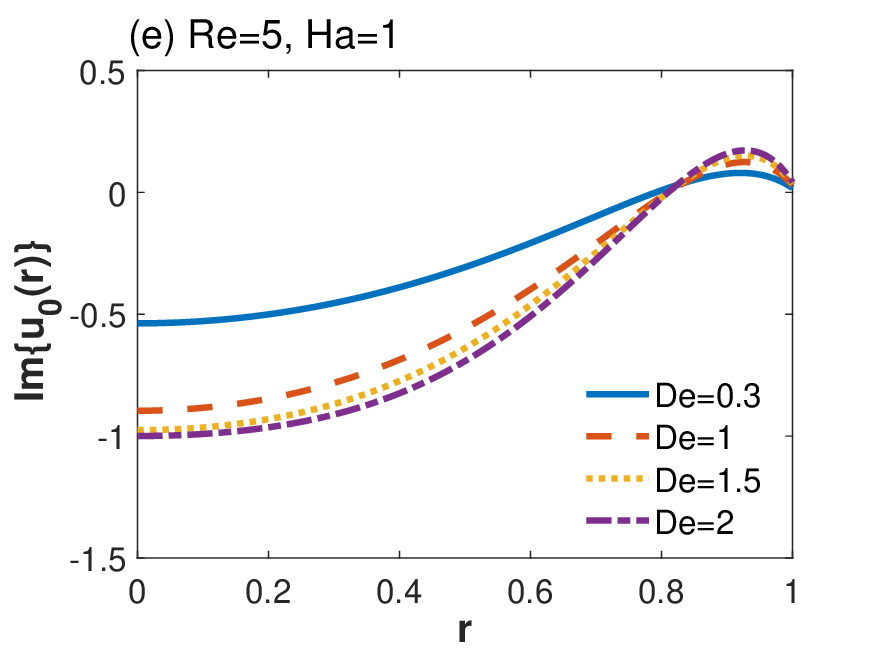}}
   \subfigure
  {\includegraphics[width=0.3\textwidth]{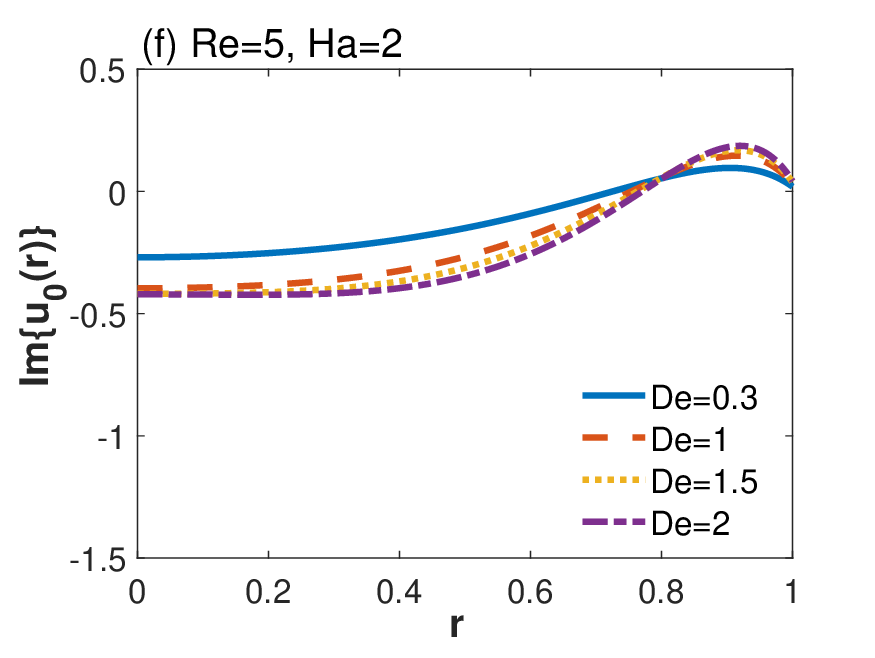}}
  \subfigure
  {\includegraphics[width=0.3\textwidth]{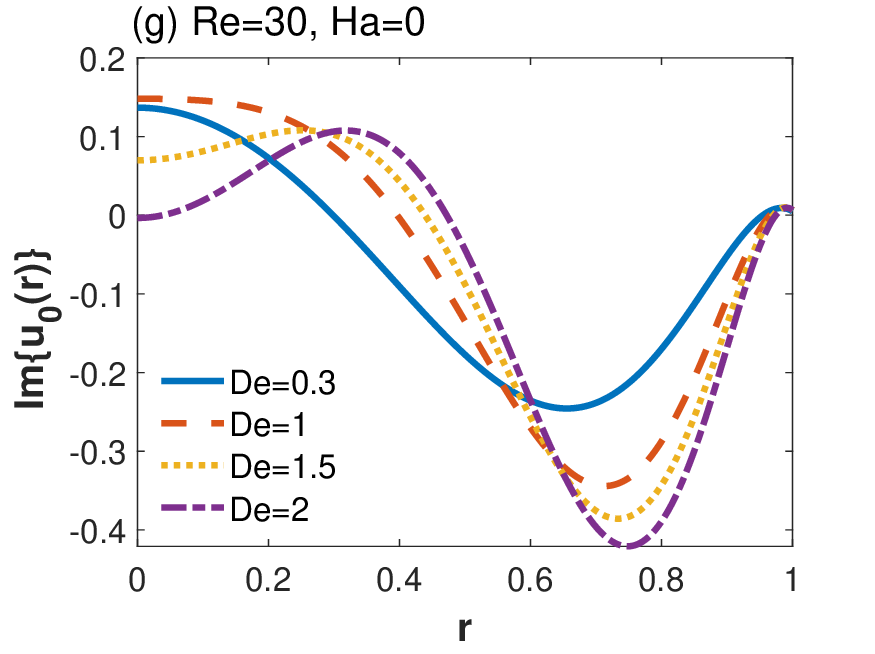}}
    \subfigure
  {\includegraphics[width=0.3\textwidth]{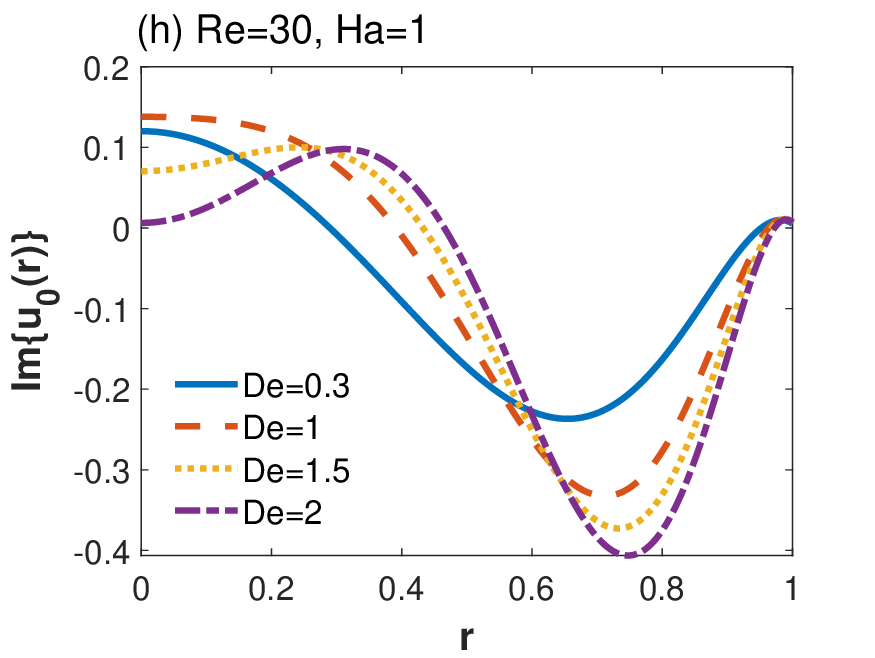}}
   \subfigure
  {\includegraphics[width=0.3\textwidth]{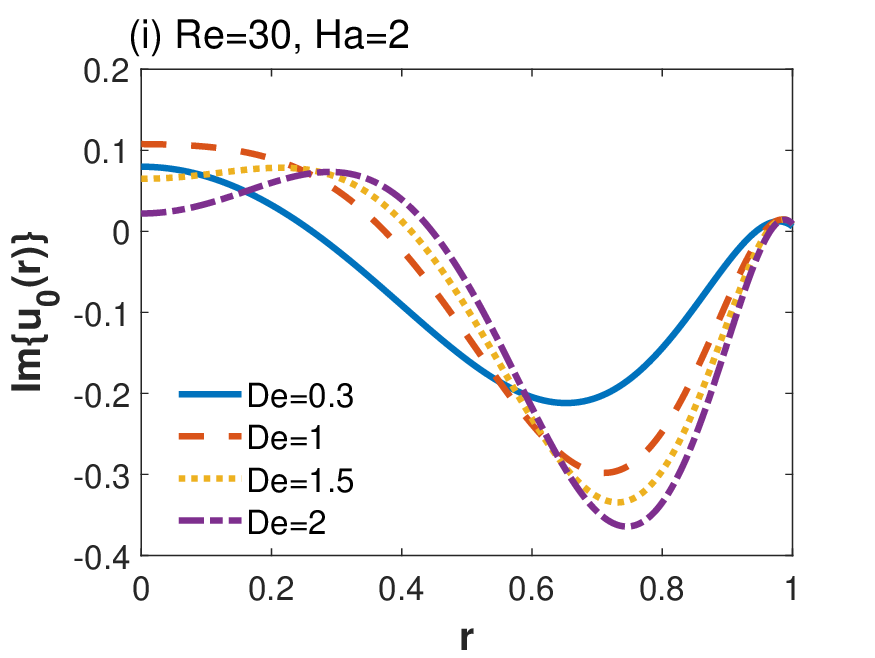}}
  \caption{Spatial variation of the imaginary part of the velocity complex amplitude $\operatorname{Im}\{u_0(r)\}$ with $r$ for Deborah number $De$ and Hartmann number $Ha$ at different values of Reynolds number $Re$ ($\kappa=10$, $\zeta=0.5$, $\delta=0.01$, $\alpha=0.8$, $\beta=0.4$).\label{fig3}}
\end{figure}

This subsection focuses on investigating the rules governing how key parameters affect the velocity distribution, and thus provides support for the in-depth analysis of the solute dispersion mechanism.
First, we aim to analyze the velocity distribution under both pure AC-EOF actuation ($Ha=0$) and mixed MHD-AC-EOF actuation ($Ha\neq 0$), see figure~\ref{fig3}(a-i). A broad range of Hartmann numbers ($0\leq Ha\leq 3$) is employed to explore the underlying mechanism of magnetic field-fluid interaction and reveal the evolution laws under both weak and strong electromagnetic effects. As anticipated, when the Hartmann number ($Ha$) is set to zero and the oscillation Reynolds number ($Re$) is small, the velocity profile exhibits a plug-like morphology of classical EOF, see figure~\ref{fig3}(a). In the presence of a magnetic field ($Ha=1, 2$), the imaginary part of the velocity complex amplitude begins to decrease, and the alteration in the velocity profile becomes more substantial, see figure~\ref{fig3}(b,c).  The underlying cause is that the Lorentz force in the axial direction, with a magnitude of $\sigma_e B_0^2u_0$ in equation~(\ref{FMaxwell-ACEOF}), acts as an opposing force, thus leading to the reduction of velocity. Then, as the oscillation Reynolds number $Re$ increases, the influence of the magnetic field on the flow velocity gradually weakens. Although the trend of flow velocity change does not show obvious alteration, it is still observable that the presence of the magnetic field leads to a decrease in the velocity amplitude, as specifically shown in  figure~\ref{fig3}(d-f). Moreover, it can be seen that the continuous increase of the $Re$ will result in the magnetic field having a negligible impact on both the trend of velocity change and the amplitude, see figure~\ref{fig3}(g-i). That is because the dimensionless ratio $Ha^2/Re$ governs the capacity of electromagnetic forces to modulate flow behavior~\citep{SVGG10}. Additionally, we can perceive that an upward trend in $Re$ leads to the emergence of wave-like velocity profiles. For low Reynolds numbers ($Re=0.1, 5$), the velocity amplitude increases monotonically with $De$. However, at high $Re=30$, the velocity field undergoes significant oscillations with large peak-to-valley differences. In this regime, the magnitude of the oscillatory velocity becomes large, but the variation with $De$ is non-monotonic due to the complex coupling effects. This observation can be theoretically elucidated through the following physical mechanisms. In viscoelastic fluids, as the relaxation time increases proportionally with $De$, the energy-storage capacity of macromolecules is significantly enhanced. Meanwhile, these macromolecules fail to dissipate energy effectively during the oscillation time period~\citep{LJCY12,VMPS21}. Consequently, the polymer chains are elongated along the flow direction, which causes a decrease in the cross-sectional area of the polymer molecules. This elongation restricts the molecular flow to the direction of stretching, thereby promoting enhanced shear-thinning behavior~\citep{CBKD07}.
\begin{figure}
  \centering
  \subfigure
  {\includegraphics[width=0.49\textwidth]{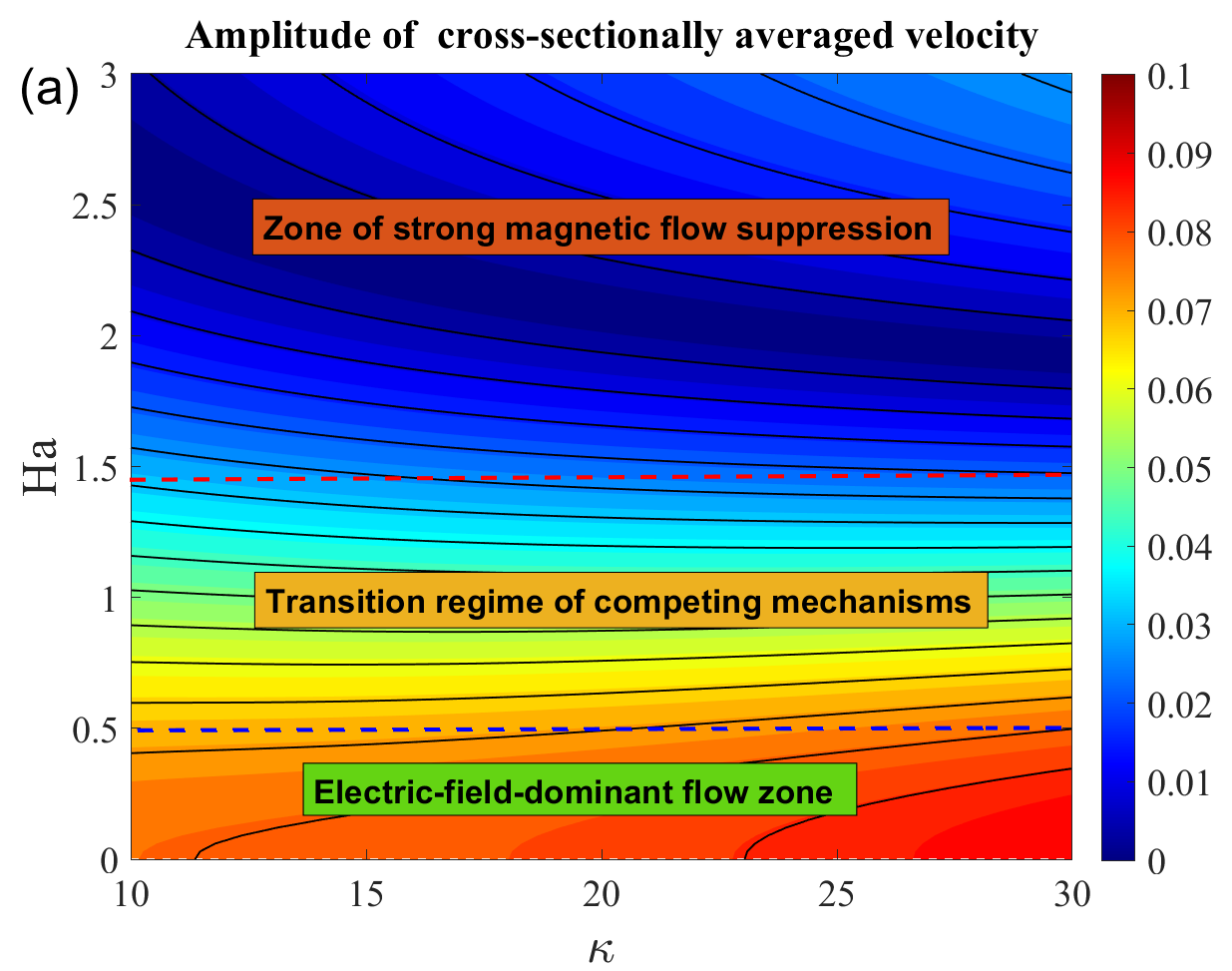}}
  \subfigure
  {\includegraphics[width=0.49\textwidth]{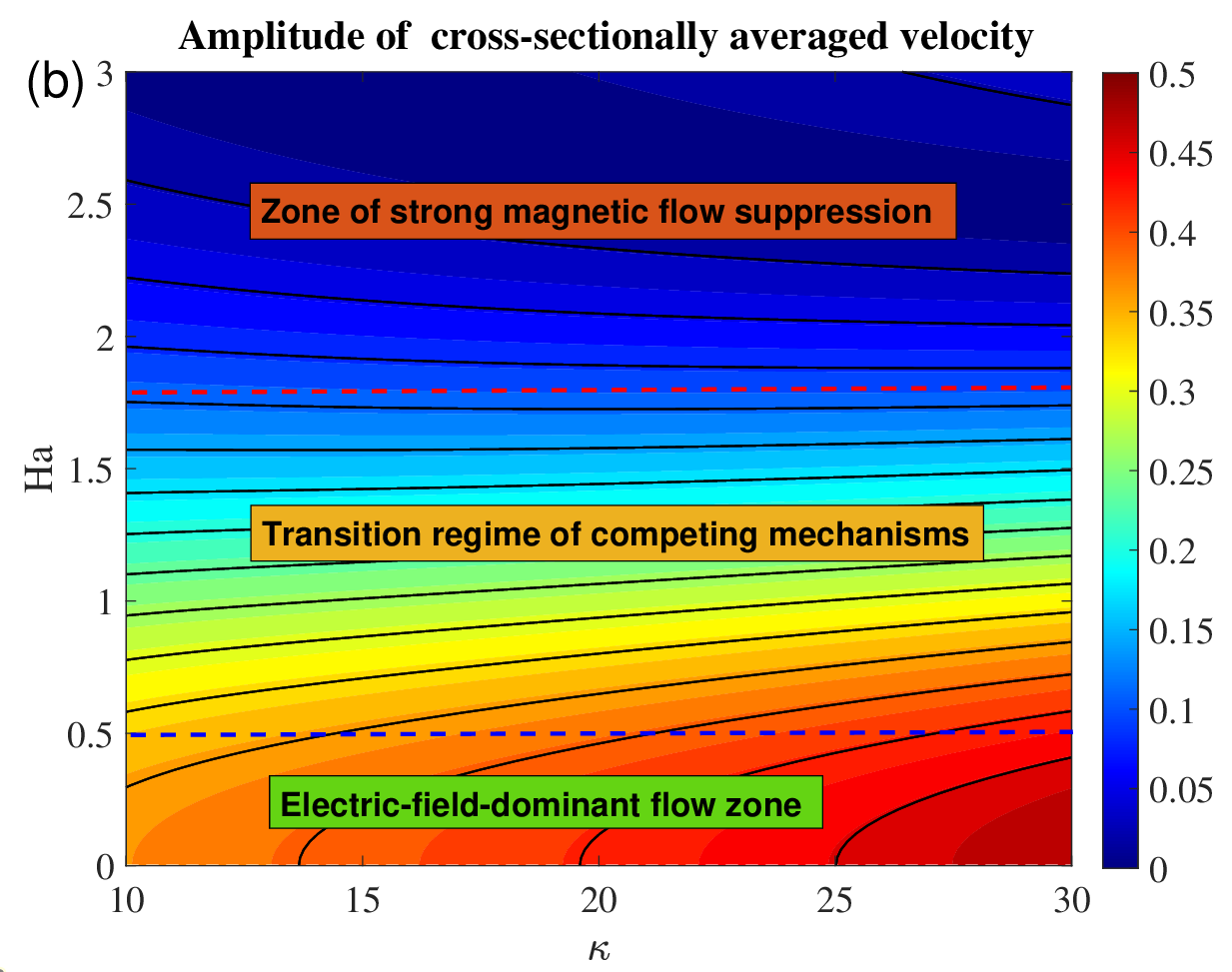}}
   \subfigure
  {\includegraphics[width=0.55\textwidth]{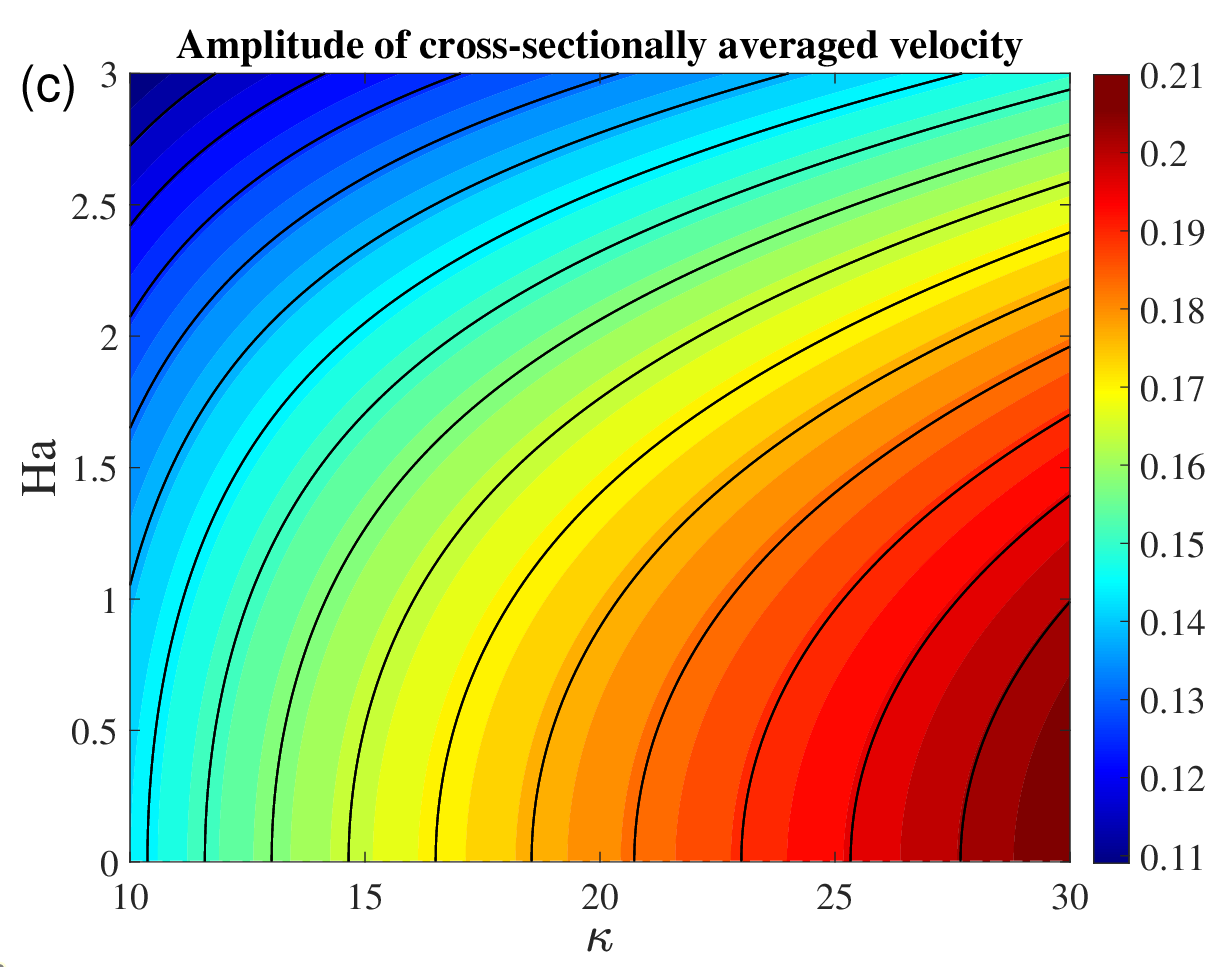}}
  \caption{ Contour plot of the cross-sectionally averaged velocity for (a) $De=0.1$, $Re=5$; (b) $De=1$, $Re=5$; (c) $De=1$, $Re=30$ under $\alpha=0.8$, $\beta=0.4$, $\zeta=0.5$, $\delta=0.01$.\label{fig4}}
\end{figure}

Next, to analyze the velocity flow patterns under the combined action of MHD and AC-EOF, we show the absolute amplitude of the cross-sectionally averaged velocity ($|U_1|$) on the plane of the electrokinetic parameter $\kappa$ and magnetohydrodynamic parameter $Ha$, in the contour plot of figure~\ref{fig4}. The $U_1$ is defined below equation (\ref{x1}). It can be observed from figure~\ref{fig4} (a,b) that the $\kappa$-$Ha$ plane can be divided into three regions: at low $Ha$ ($<0.5$), the region is electric field-dominated, where the flow velocity increases with $\kappa$; at moderate $Ha$, the electric force and magnetic forces compete with each other, leading to the flow velocity being sensitive to both $\kappa$ and $Ha$ while decreasing overall; at high $Ha$ ($> 1.5$ or $2$), the region exhibits magnetic field-induced flow restriction, where the flow velocity is significantly suppressed. 
Furthermore, the flow response of the polymer solution, governed by its viscoelasticity through the Deborah number, differs markedly from Newtonian fluids. At $De=0.1$, polymer chains respond incompletely to the rapid electric field, limiting velocity at low-to-moderate $Ha$. However, even at high $Ha$, increasing the electrokinetic parameter $\kappa$ still enhances flow, indicating that the field partially overcomes polymeric relaxation lag. In contrast, at $De=1$, the matched relaxation and field periods enable efficient elastic energy conversion, yielding higher velocities at low-to-moderate $Ha$ before saturation at high 
$Ha$. Consequently, a higher $De$ enhances the electric-field-driven flow and elevates the critical 
$Ha$ for magnetic suppression.

A comparison of figure~\ref{fig4}(b) and (c) shows that increasing the Reynolds number ($Re$) from $5$ to $30$ markedly reduces the amplitude of cross-sectionally averaged velocity and weakens the magnetic confinement effect. In figure~\ref{fig4}(b), the flow is highly sensitive to both $\kappa$ and $Ha$, as shown by the dense, steep contours and the appearance of a strong confinement zone for $Ha>2$, indicating distinct flow regimes. In contrast, at $Re=30$, see figure~\ref{fig4}(c), the electrokinetic-dominated region extends to higher $Ha$ (velocity remains above 0.3 even at $Ha=3$), the confinement zone shrinks, and contours become nearly parallel to the $Ha$-axis. This transition occurs because higher inertial forces at elevated $Re$ enhance momentum transport, partially offsetting the Lorentz force damping and stabilizing the tripartite coupling among electric, magnetic, and inertial forces, thereby reducing the flow's sensitivity to $\kappa$ and $Ha$.

\subsection{Dispersion coefficient $K_2(t)$} 

\begin{figure}
  \centering
  \subfigure
  {\includegraphics[width=0.49\textwidth]{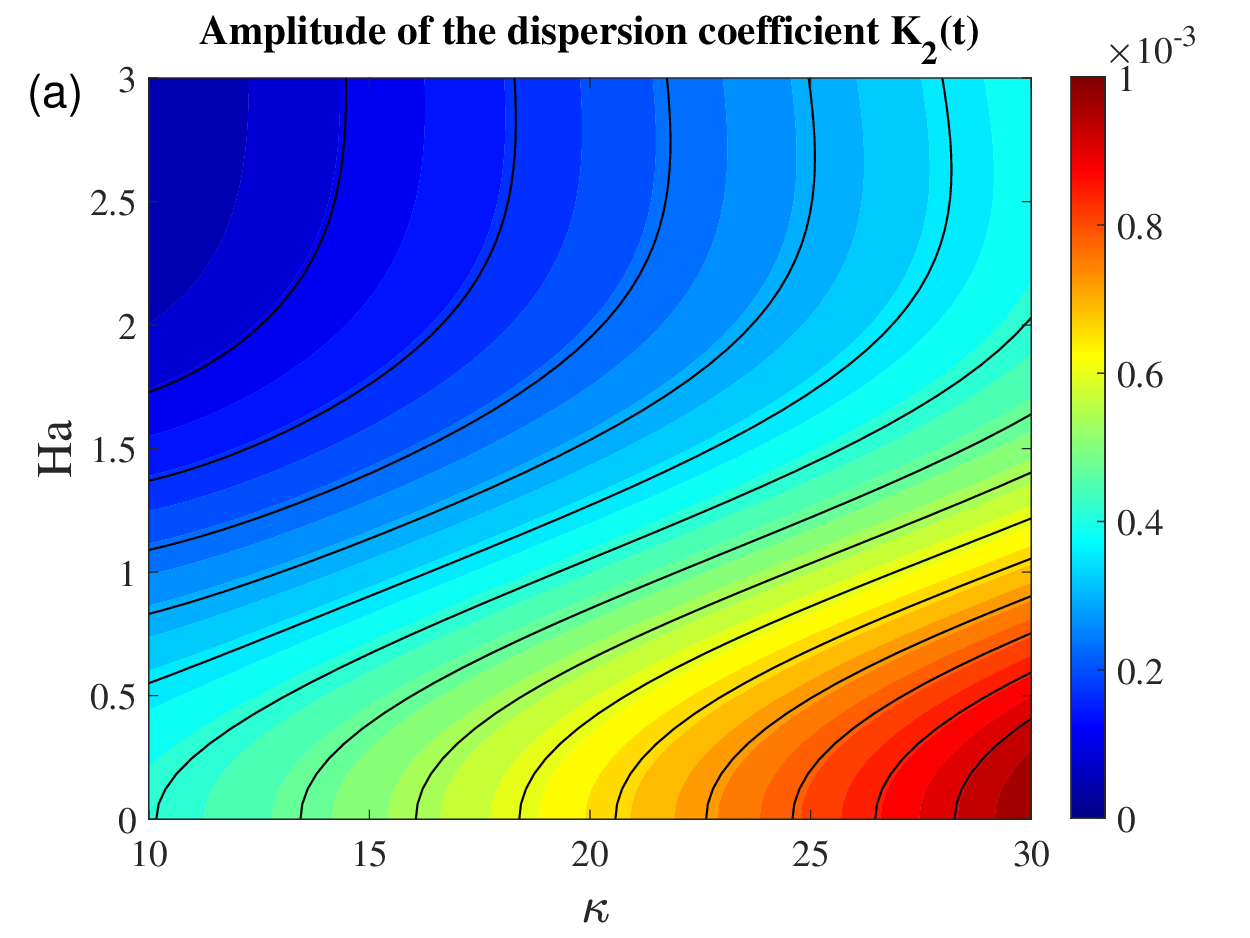}}
  \subfigure
  {\includegraphics[width=0.49\textwidth]{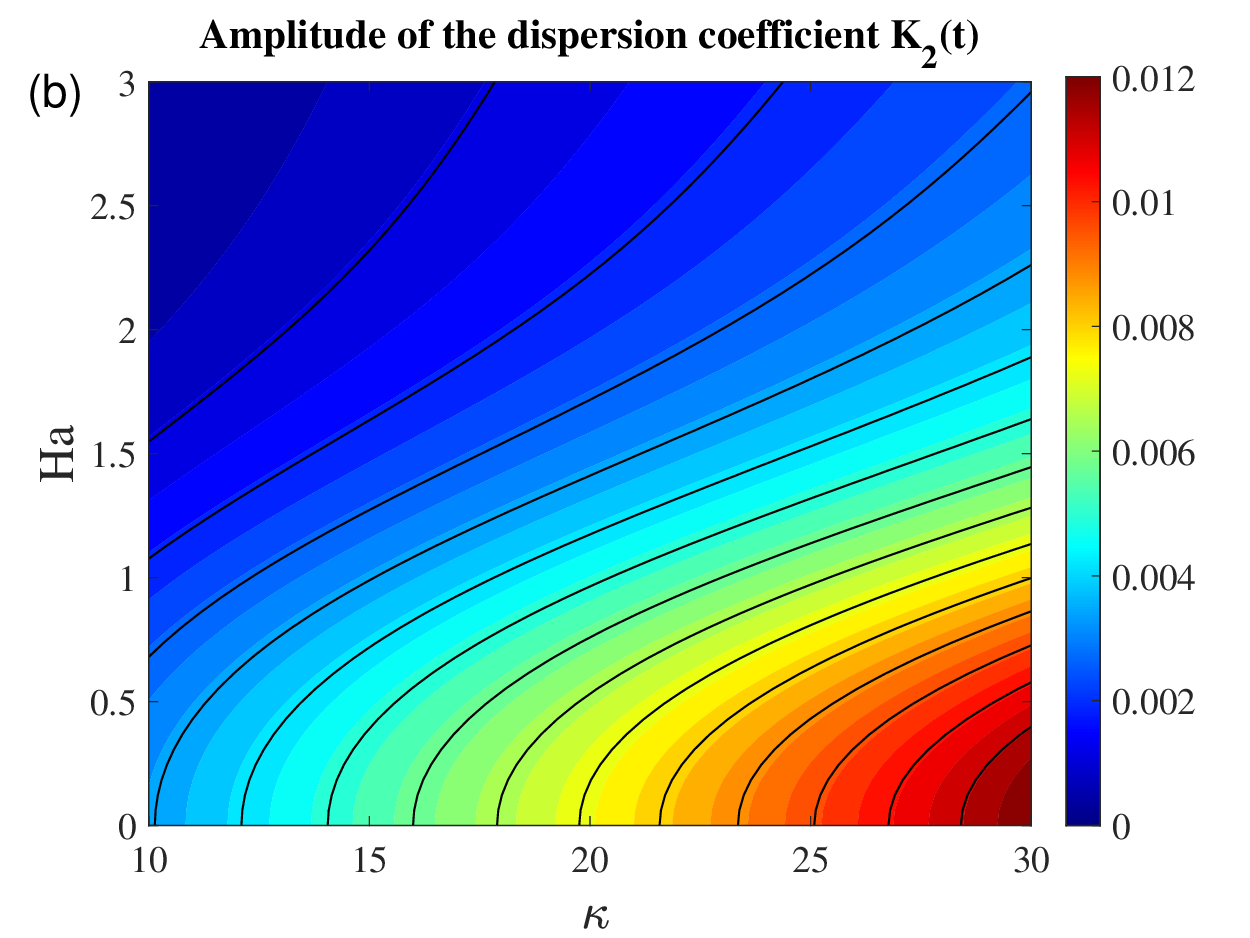}}
   \subfigure
  {\includegraphics[width=0.55\textwidth]{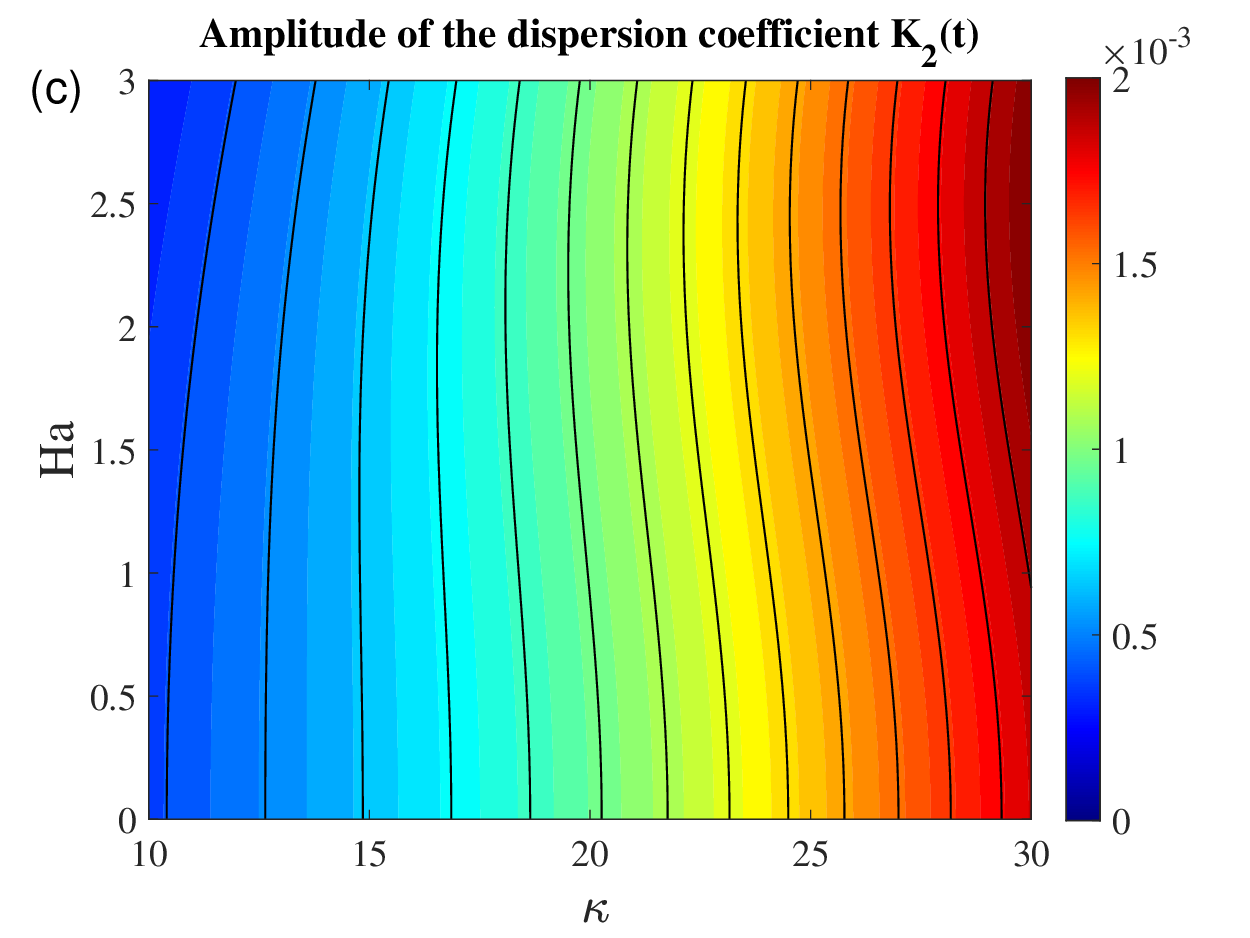}}
  \caption{Contour plot of the amplitude of dispersion coefficient for (a) $De=0.1$, $Re=5$; (b) $De=1$, $Re=5$; (c) $De=1$, $Re=30$ under $\alpha=0.8$, $\beta=0.4$, $\zeta=0.5$, $\delta=0.01$.\label{fig5}}
\end{figure}

This subsection commences by examining the effects of electroosmotic and magnetohydrodynamic drivings on solute dispersion. To this end, figure~\ref{fig5} presents contour maps of the dispersion coefficient amplitude across the parameter space defined by the electrokinetic parameter $\kappa$, and the magnetic field parameter $Ha$. The amplitude of the dispersion coefficient is defined as $(K_2(t))_{\text{amp}} = \langle \max(K_2(t)) - \min(K_2(t)) \rangle / 2$, where $\langle \cdot \rangle$ denotes the periodic average. 
As shown in the figure, the dispersion coefficient amplitude responds distinctly to $\kappa$ across $Ha$ regimes. In the low-$Ha$ region, the amplitude increases substantially with $\kappa$ (color transitions from blue to red) due to weakened Lorentz force damping, which enables effective electrokinetic driving and enhances convective dispersion. This trend reverses in the high-$Ha$ region, where behavior becomes $Re$-dependent. Under low-$Re$ (figure~\ref{fig5}(a,b)), the dominant Lorentz force suppresses flow motion and diminishes the enhancement effect of $\kappa$, resulting in limited amplitude growth (the color shifts to blue). Conversely, at high $Re$ (figure~\ref{fig5}(c)), enhanced inertial forces partially compensate for magnetic confinement, maintaining considerable dispersion levels even at high $\kappa$.

A comparative analysis of figure~\ref{fig5}(b) and (c) further elucidates the distinct dispersion characteristics at different $Re$. Figure~\ref{fig5}(b), corresponding to a low $Re$ regime, displays strongly curved contours and a large dispersion coefficient amplitude. It also exhibits a marked magnetic suppression effect at high $Ha$, as indicated by the abrupt color transition toward blue with increasing $Ha$. In contrast, figure~\ref{fig5}(c), obtained at high $Re$, features flattened contours and a significantly lower dispersion amplitude, along with attenuated magnetic suppression under high $Ha$ (manifested as a gradual color shift). These morphological and amplitude differences stem from enhanced fluid inertia at elevated $Re$, which mitigates magnetic confinement while simultaneously intensifying the $\kappa$-driven flow. As a result, electric field effects persist even under high $Ha$ conditions, leading to clearly divergent dispersion regimes.
Additionally, the sensitivity of the dispersion coefficient to the Deborah number ($De$) is primarily governed by the ``temporal synchronization" between the electric field and the viscoelastic response of the polymer solution. This is clearly evidenced by comparing figure~\ref{fig5}(a) and (b). At 
$De=1$, the polymer relaxation time matches the field period, allowing the elastic polymer chains to fully respond and effectively transfer energy. This resonant interaction generates a highly complex oscillatory flow, which dramatically enhances axial dispersion through strong advection-diffusion coupling (figure~\ref{fig5}(b)). In contrast, at $De=0.1$, the rapid field oscillations prevent the polymer chains from responding fully, resulting in weak dispersion effects and generally lower amplitude (figure~\ref{fig5}(a)).
 Therefore, in microfluidic applications such as separation and mixing processes, precise control of these key parameters is essential for optimizing the outcome and efficiency.

\begin{figure}
  \centering
  \subfigure
  {\includegraphics[width=0.43\textwidth]{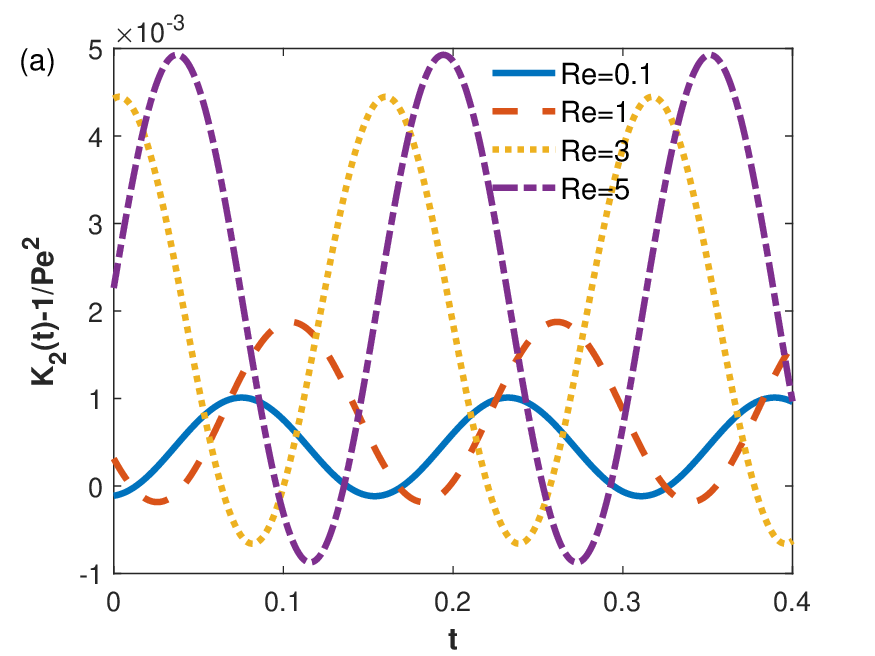}}
  \subfigure
  {\includegraphics[width=0.43\textwidth]{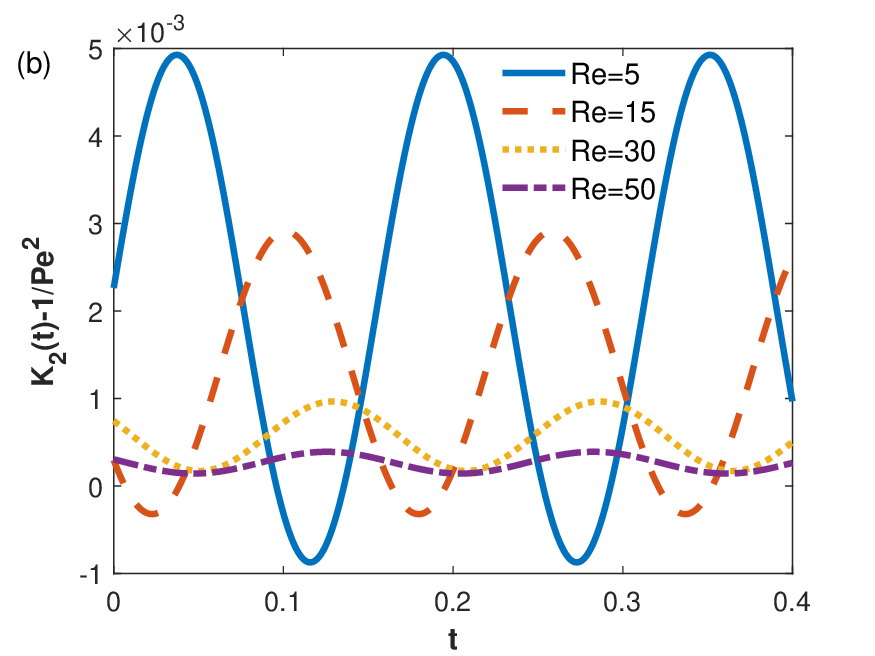}}
  \subfigure
  {\includegraphics[width=0.43\textwidth]{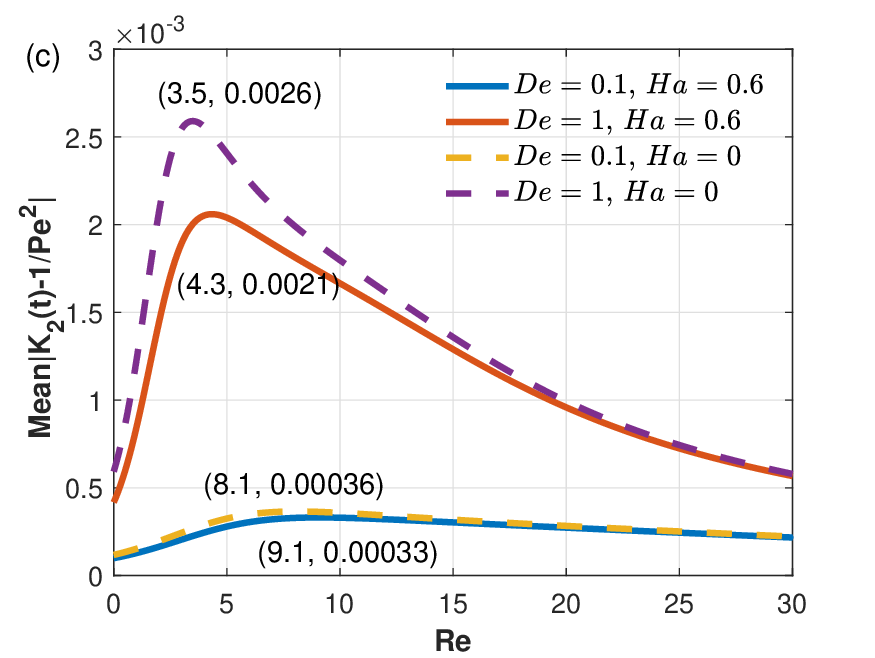}}
    \subfigure
  {\includegraphics[width=0.43\textwidth]{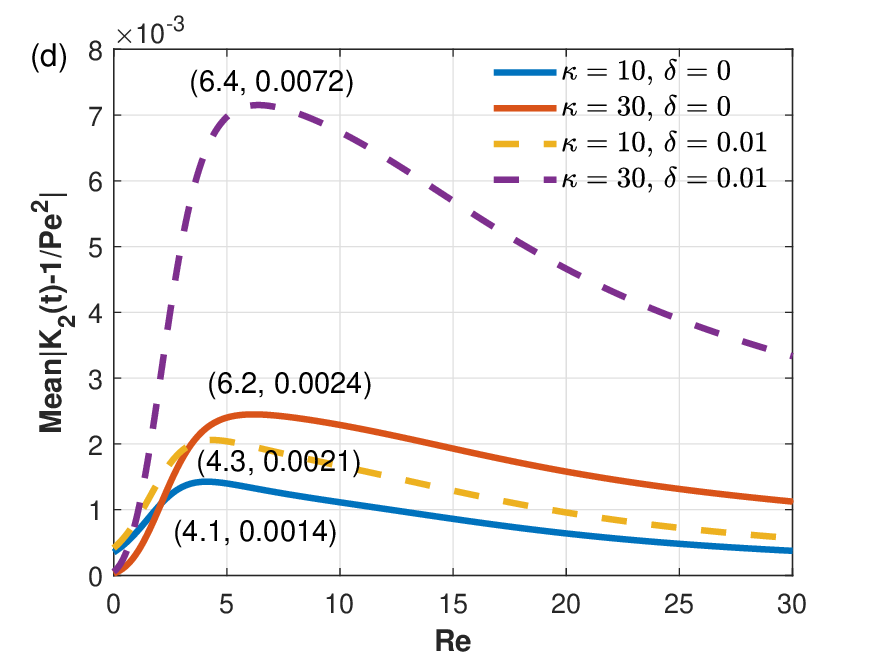}}
    \subfigure
  {\includegraphics[width=0.43\textwidth]{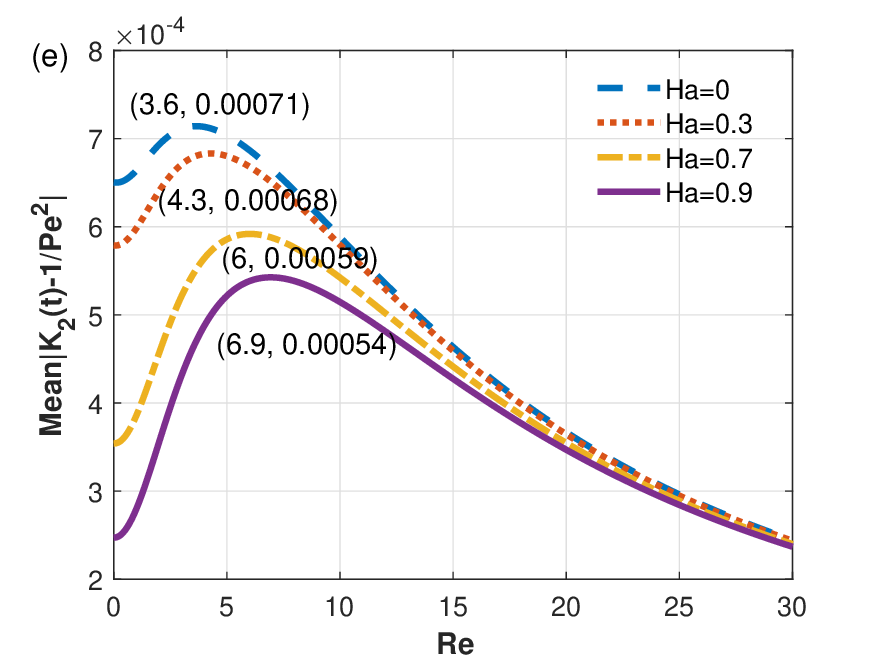}}
    \subfigure
  {\includegraphics[width=0.43\textwidth]{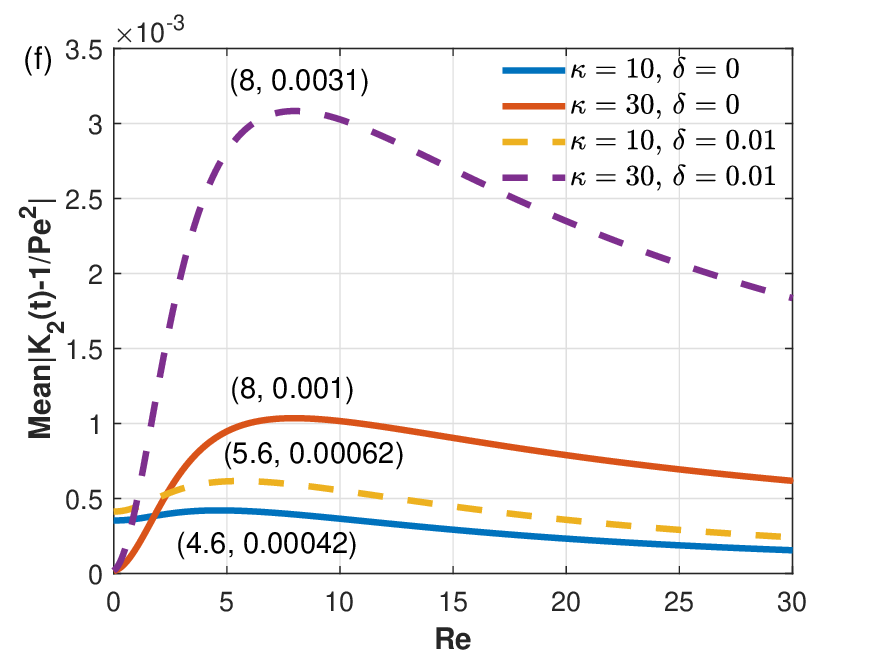}}
  \caption{(a,b) Variations of dispersion coefficient with $t$ for different values of oscillation Reynolds number $Re$ under $\kappa=10$, $Ha=0.6$, $De=1$,  $\omega=20$, $\zeta=0.5$, $\delta=0.01$, $\alpha=0.8$, $\beta=0.4$. (c,d) Mean dispersion coefficient of a fractional Maxwell fluid as a function of oscillation Reynolds number $Re$ for different parameters under $\omega=20$, $\zeta=0.5$, $\alpha=0.8$, $\beta=0.4$. (e,f) Mean dispersion coefficient of a Newtonian fluid as a function of oscillation Reynolds number $Re$ for different parameters under $\omega=20$, $\zeta=0.5$, $\alpha=1$, $\beta=1$.\label{fig6}}
\end{figure}

As mentioned previously, the Reynolds number significantly influences the dispersion coefficient. Therefore, we will next examine the variation of the dispersion coefficient at different Reynolds number values. Figure~\ref{fig6}(a,b) displays the dispersion coefficient under low ($Re < 5$) and high ($Re > 5$) Reynolds number conditions, respectively. The additive contribution of axial diffusion ($1/Pe^2$) has been subtracted from these values, following the data processing approach of \citet{RanaM16} (their figures 11–32) and \citet{SSMS2020} (their figures 11–16). In the low oscillation Reynolds number regime, viscous and inertial forces operate at comparable magnitudes. As $Re$ increases, the amplitude of the oscillatory inertial force grows significantly and dominates fluid flow, thereby nonlinearly modulating the dispersion coefficient and amplifying the amplitude of $K_2(t)-1/Pe^2$. Conversely, in the high-$Re$ regime, inertial forces overwhelm viscous forces, and their rapid oscillation compresses the growth for the $K_2(t)-1/Pe^2$ amplitude, manifesting an opposing evolutionary trend. Notably, these findings indicate the existence of a critical oscillation Reynolds number, $Re_{c}$, which optimally modulates the amplitude of $K_2(t)-1/Pe^2$. This observation aligns with the research outcomes presented by~\cite{PVSN16,HLYJ17}. This critical  value serves as a pivotal threshold, demarcating the transition between the two disparate response patterns of $K_2(t)-1/Pe^2$ to changes in $Re$. Consequently, we need to analyze the influence of relevant parameters on the critical oscillation Reynolds number. 
Figures~\ref{fig6}(c,d) and (e,f) illustrate the variation of the mean dispersion coefficient with $Re$ for the fractional Maxwell fluid and the Newtonian fluid over the time interval $0-5$, respectively. For both fluid models, the mean dispersion coefficient exhibits a unimodal trend of initial increase followed by decrease with $Re$, corresponding to a well-defined critical Reynolds number $Re_{c}$: when $Re < Re_{c}$, inertial effects enhance transverse mixing of fluid elements, thereby increasing the dispersion coefficient; when $Re>Re_{c}$, inertia dominates and regularizes the flow field, causing the dispersion coefficient to decay with further increase in $Re$. Notably, $Re_{c}$ decreases with increasing Deborah number $De$, while it increases with larger Hartmann number $Ha$, Debye parameter $\kappa$, and slip length $\delta$.
For the fractional Maxwell fluid, the Deborah number $De$ characterizes its viscoelasticity: a higher $De$ suppresses inertial mixing at lower $Re$, thereby reducing $Re_{c}$. In contrast, the Newtonian fluid, lacking viscoelasticity, exhibits a generally higher $Re_{c}$ under the same $Ha$, $\kappa$, and $\delta$ conditions. Furthermore, viscoelasticity amplifies the deformation of fluid elements and enhances mixing, leading to a considerably higher peak dispersion coefficient for the fractional Maxwell fluid.

\begin{figure}
  \centering
  \subfigure
  {\includegraphics[width=4.4cm, height=3.8cm]{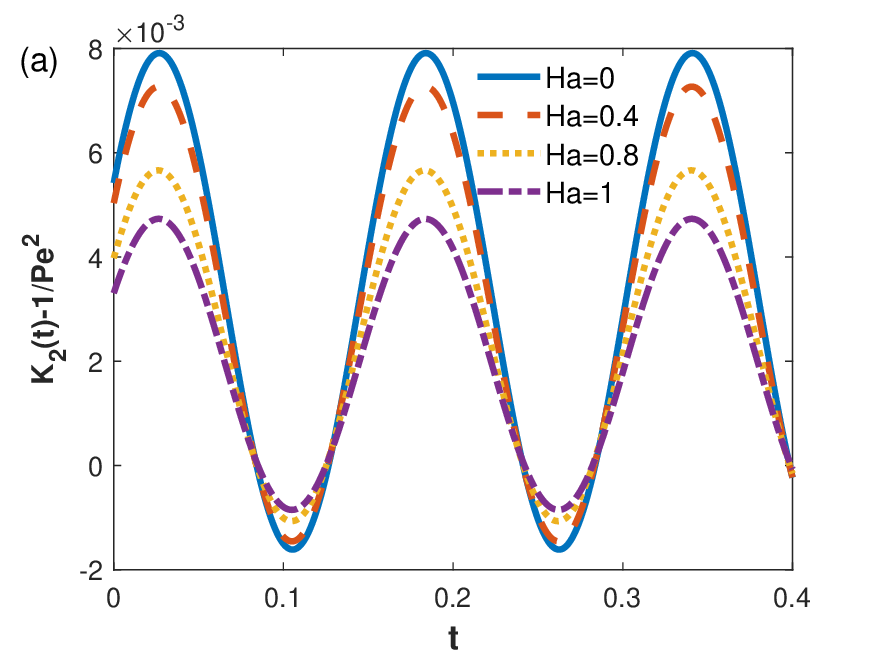}}
  \subfigure
  {\includegraphics[width=4.4cm, height=3.8cm]{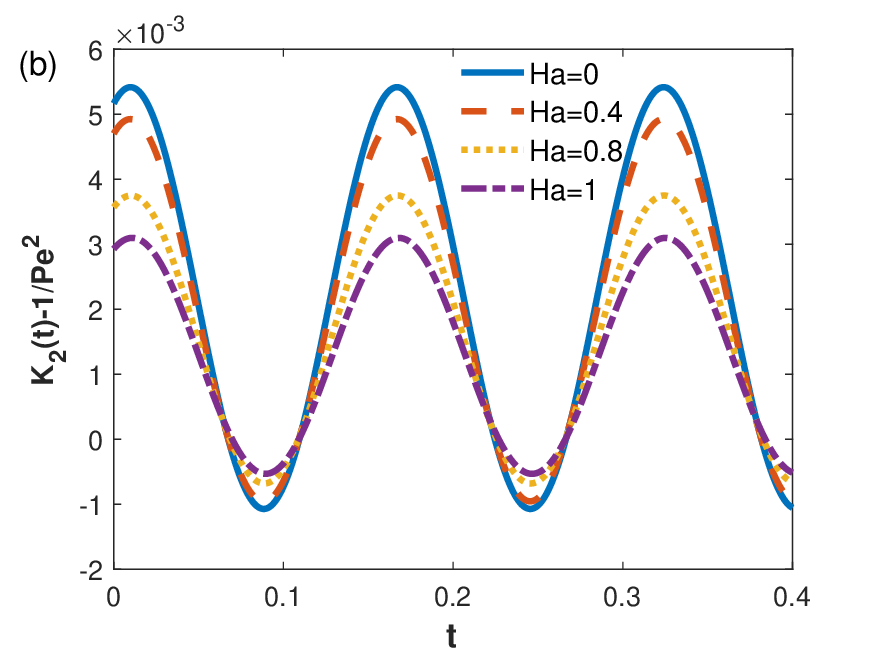}}
  \subfigure
  {\includegraphics[width=4.4cm, height=3.8cm]{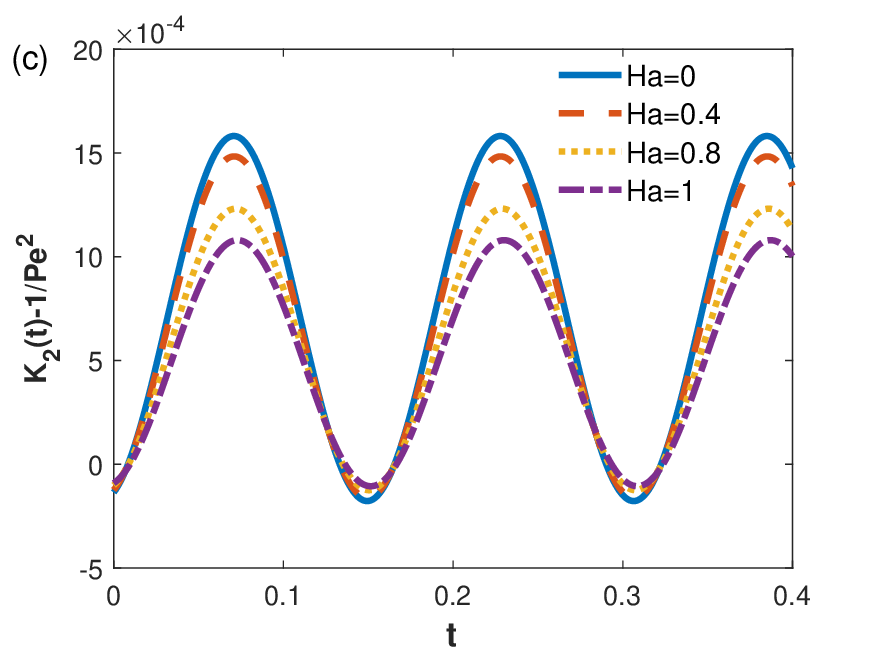}}
  \caption{Variations of dispersion coefficient with different Hartmann number $Ha$ for (a) fractional Maxwell fluid ($\alpha=0.8$, $\beta=0.2$); (b) classical Maxwell fluid ($\alpha=1$, $\beta=0$); (c) Newtonian Fluid ($\alpha=1$, $\beta=1$) under $\kappa=10$, $Re=5$, $De=1$, $\omega=20$, $\zeta=0.5$, $\delta=0.01$.\label{fig7}}
\end{figure}

\begin{figure}
  \centering
  \subfigure
  {\includegraphics[width=0.43\textwidth]{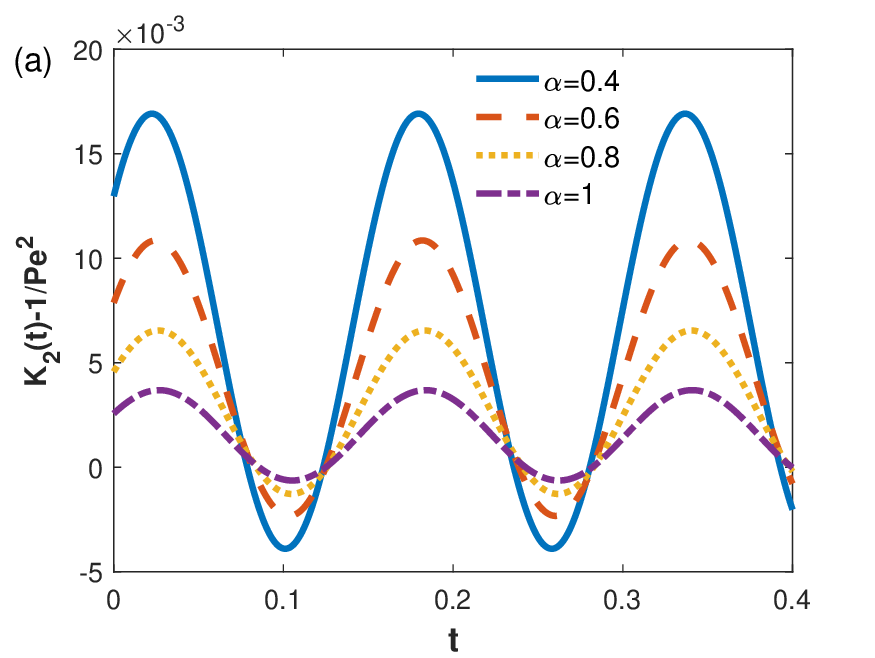}}
  \subfigure
  {\includegraphics[width=0.43\textwidth]{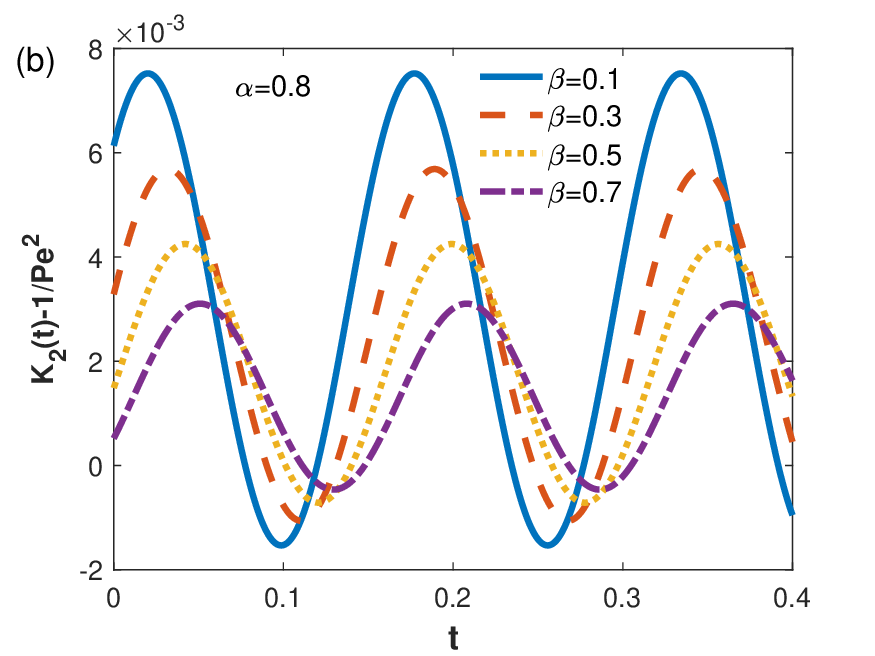}}
  \caption{Variations of dispersion coefficient with $t$ for (a) different fractional parameter $\beta$ and different fractional parameter $\alpha$ under $\kappa=10$, $Re=5$, $De=1$, $\omega=20$, $Ha=0.6$, $\zeta=0.5$, $\delta=0.01$.\label{fig8}}
\end{figure}

Figure~\ref{fig7} illustrates the influence of the Hartmann number $Ha$ on $K_2(t)-1/Pe^2$ under three fluid models, characterized by different parameter combinations: the fractional Maxwell model ($\alpha=0.8$, $\beta=0.2$), the classical integer-order Maxwell model ($\alpha=1$, $\beta=0$), and the Newtonian fluid model ($\alpha=1$, $\beta=1$). We can find that the amplitude of the dispersion coefficient for fractional Maxwell model is the largest, compared with that of the classical integer Maxwell model and the Newtonian fluid model. This phenomenon can be attributed to the distinct values of fractional parameters, which modulate the relative contributions of viscous and elastic behaviors within the fractional viscoelastic model. Specifically, under the current parameter regime ($\alpha=0.8$, $\beta=0.2$), the elastic component exerts a dominant influence, significantly intensifying solution convection and subsequently facilitating the dispersion process. The dispersion coefficients for other values of the fractional parameters are presented in figure~\ref{fig8}. It can be observed that as the fractional parameters $\alpha$ and $\beta$ increase, the amplitude of the dispersion coefficient gradually decreases. The reason is that increasing $\alpha$ and $\beta$ leads to reductions in both velocity magnitude and gradients, rendering the velocity field more uniform. Such uniformity weakens dispersion and decreases the magnitude of the dispersion coefficient.


 \begin{figure}
  \centering
  \subfigure
  {\includegraphics[width=0.43\textwidth]{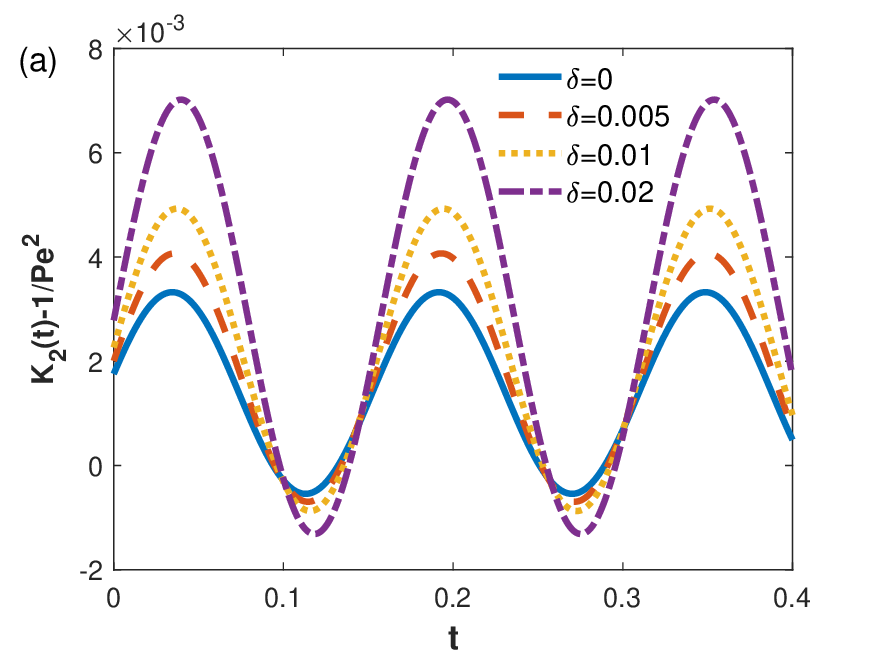}}
    \subfigure
  {\includegraphics[width=0.43\textwidth]{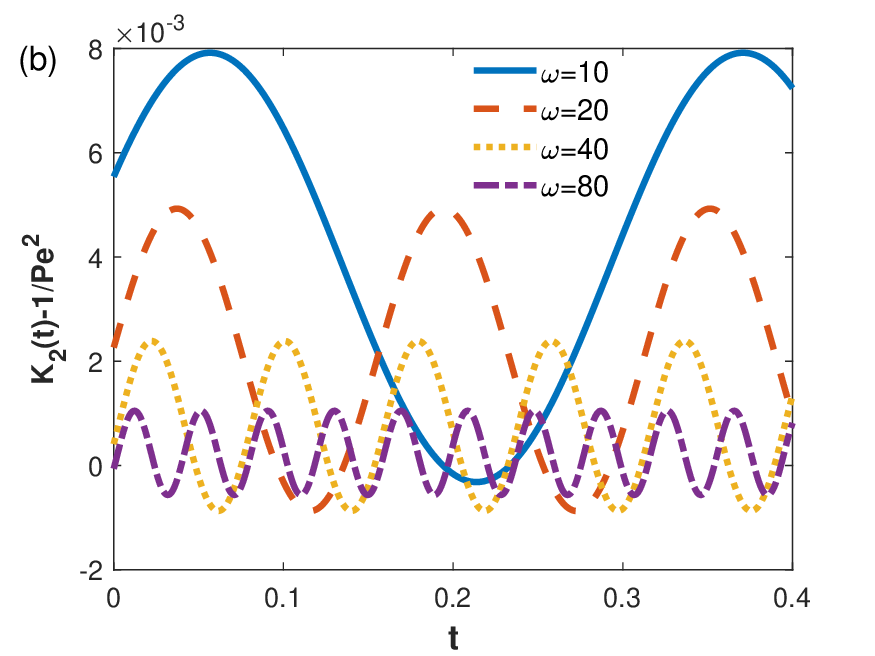}}
  \caption{Variations of dispersion coefficient with $t$ for different parameters (a) $\omega=20$; (b)  $\delta=0.01$ at $\kappa=10$, $De=1$, $Re=5$, $Ha=0.6$, $\zeta=0.5$, $\alpha=0.8$, $\beta=0.4$.\label{fig9}}
\end{figure}

The impacts of the remaining parameters (slip length $\delta$ and oscillation frequency $\omega$) on the dispersion coefficient are comprehensively illustrated in figure~\ref{fig9}. It is crucial to emphasize that the degree of dispersion is predominantly governed by the relative significance of the flow velocity magnitude and the configuration of the flow velocity profile. As observed in figure~\ref{fig9}(a), an increase in the slip coefficient leads to a higher amplitude of the dispersion coefficient. The underlying mechanisms are as follows: i) An increased slip coefficient accelerates the flow near the channel wall, bringing its velocity closer to that in the central region and thus disturbing the uniformity of the velocity distribution; ii) Owing to the interdependence between the slip coefficient and the flow potential, a larger slip coefficient elevates the potential, resulting in greater flow velocity and intensified convective effects.
Figure~\ref{fig9}(b) illustrates the evolution of dispersion coefficient for specific values of $\omega$. Evidently, as $\omega$ increases, the amplitude of the dispersion coefficient decreases accordingly. This phenomenon can be attributed to the fact that a higher $\omega$ indicates the period of the applied electric field is significantly shorter than the diffusion time scale. Consequently, the more rapid oscillation of the applied electric field severely restricts the effective time available for solute transport.
\begin{figure}
  \centering
 \includegraphics[width=0.6\textwidth]{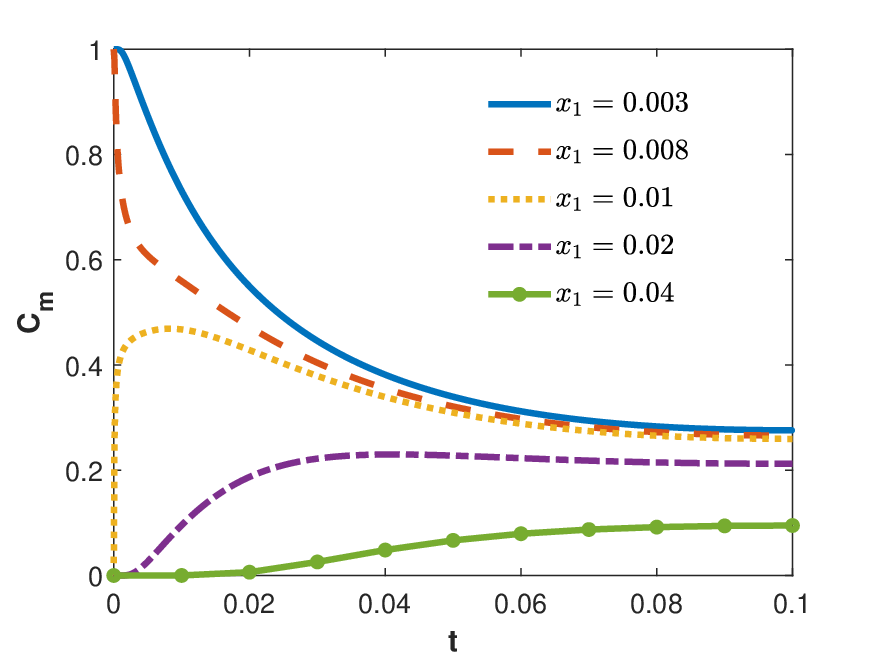}
  \caption{The evolution of the concentration distribution at different positions over time with $Pe=50$, $\kappa=10$, $Re=5$, $De=1$, $Ha=0.6$, $\zeta=0.5$, $\omega=20$, $\delta=0.01$, $\alpha=0.8$, $\beta=0.4$, $x_s=0.019$.\label{fig10}}
\end{figure}

\subsection{Mean concentration $C_m(x_1,t)$}
In this subsection, we will analyze the evolution of the mean concentration distribution $C_m$ with different parameters. As shown in equation (\ref{eq:43}), the mean concentration distribution is related to the dispersion coefficient $K_2(t)$ through the parameter $\xi$. Thus, parameters that influence $K_2(t)$ should also be considered for their impacts on the mean concentration. First, the evolution of the mean concentration at different spatial positions $x_1$ (the formal definition of this quantity can be found in equation (\ref{x1})) over time with the band wide of injection $x_s=0.019$ is illustrated in figure~\ref{fig10}. Analysis indicates that for $x_1=0.003$ and $0.008$, $C_m$ undergoes monotonic decay over time. In contrast, when $x_1=0.01$, $0.02$ and $0.04$, $C_m$ first increases and then decreases with time $t$, and the inflection point between these two trends occurs near the boundary of the solute injection region ($|x_1|\approx x_s/2$). A closer examination reveals that as $x_1$ approaches $x_s/2$, $C_m$ decreases monotonically. Conversely, when  $x_1>x_s/2$,  $C_m$ follows a rise-decay pattern, with the peak concentration decreasing as the magnitude of $x_1$ increases.  For cases where $|x_1|$ is substantially distant from $x_s/2$, $C_m$  exhibits a delay in increasing from zero, which indicates that the solute requires a certain period of time to reach the target location from the injection point through transport processes including diffusion, advection, and dispersion.

\begin{figure}
  \centering
 \includegraphics[width=0.9\textwidth]{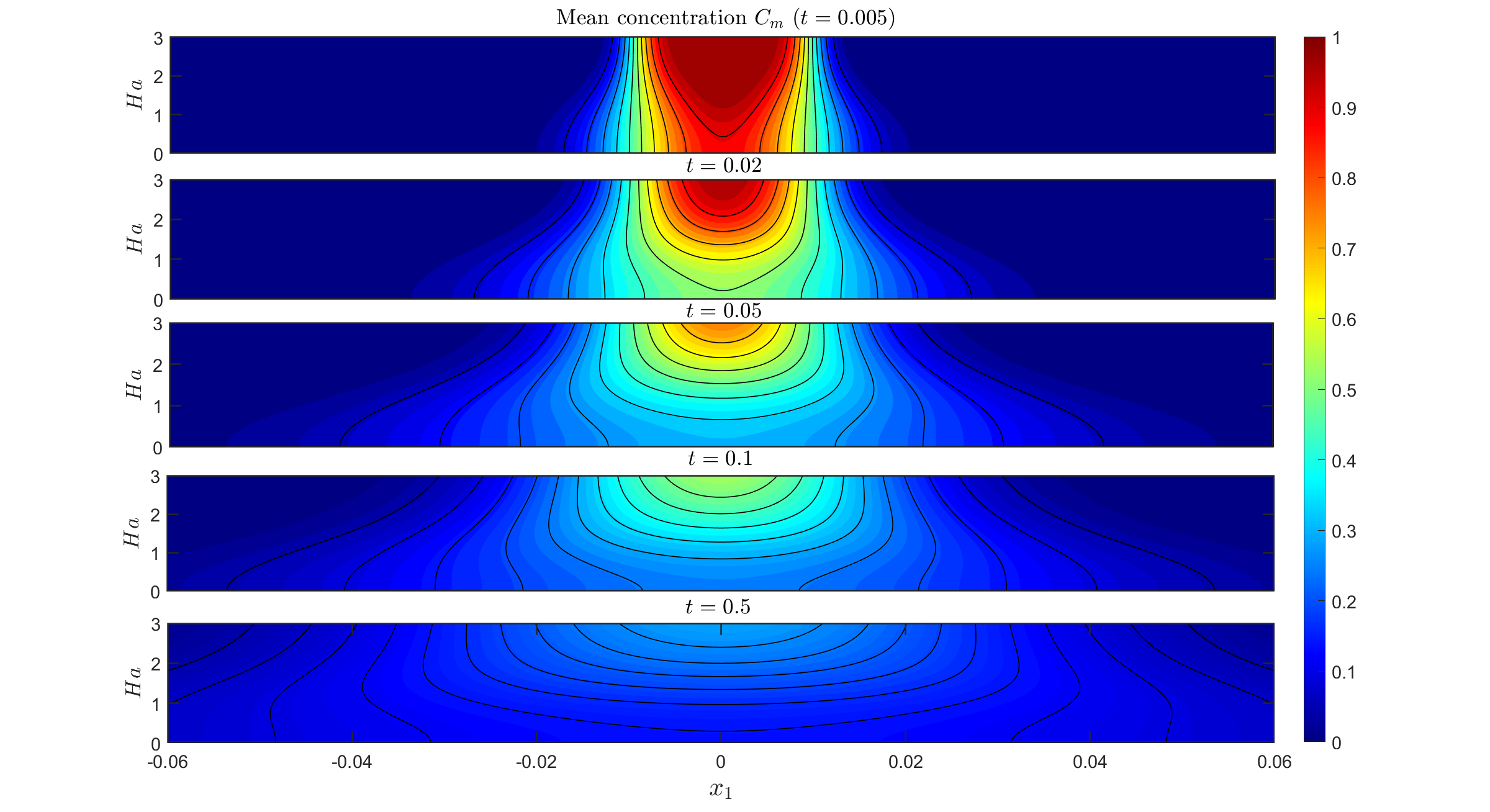}
  \caption{Mean concentration contours at various times for different Hartmann numbers with $Pe=50$, $\kappa=10$, $Re=5$, $De=1$, $\omega=20$,  $\zeta=0.5$, $\delta=0.01$, $\alpha=0.8$, $\beta=0.4$, $x_s=0.019$.\label{fig11}}
\end{figure}

\begin{figure}
  \centering
  \subfigure
  {\includegraphics[width=0.9\textwidth]{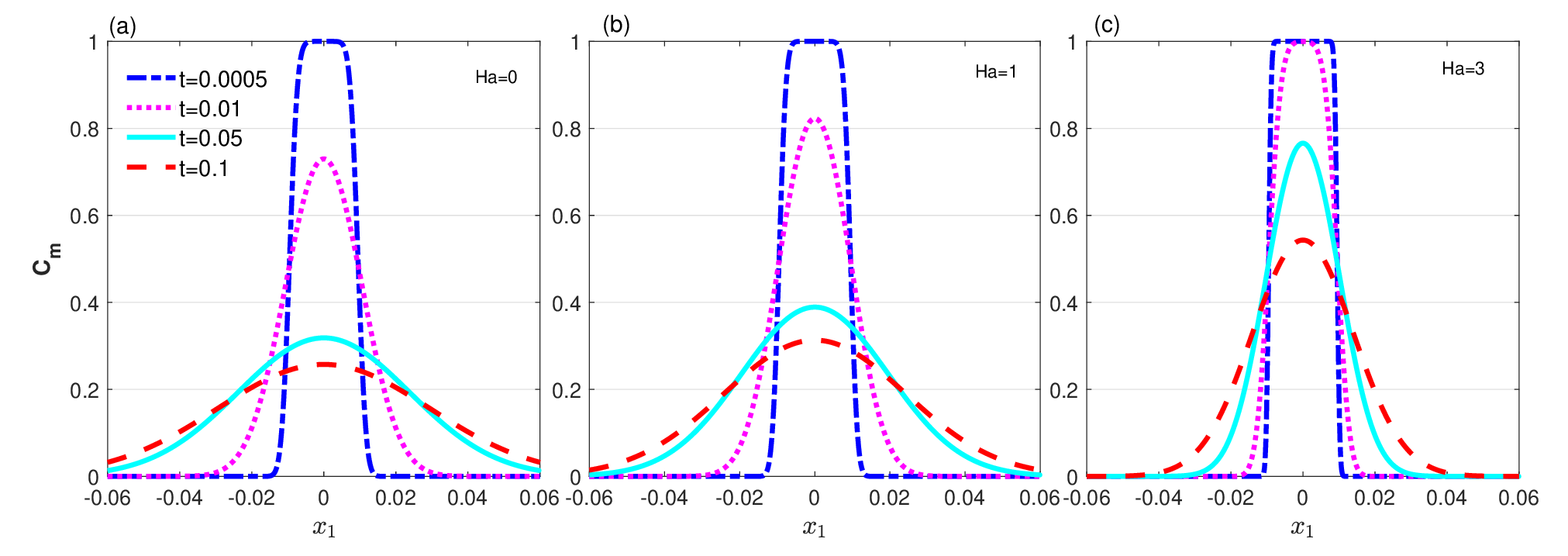}}
    \subfigure
  {\includegraphics[width=\textwidth]{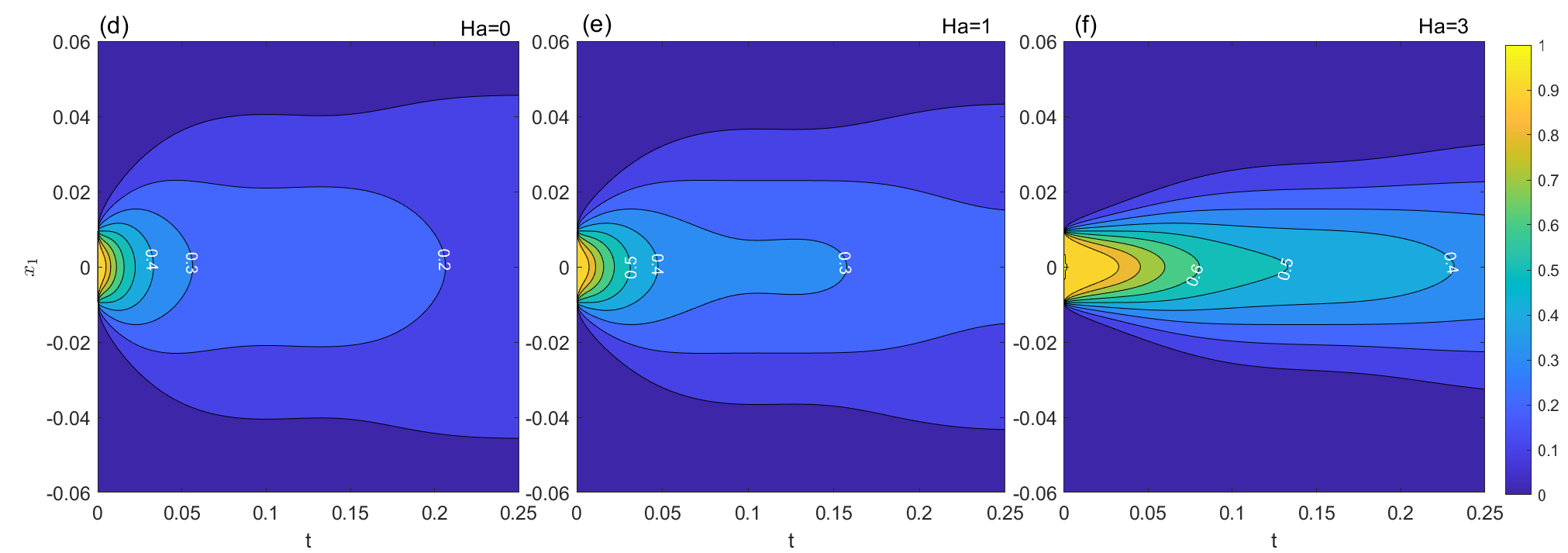}}
  \caption{(a-c) Variation of the mean concentration along the axial distribution with different Hartmann number; (d-f) Mean concentration contours of solute with different Hartmann number at $Pe=50$, $\kappa=10$, $Re=5$, $De=1$, $\omega=20$, $\zeta=0.5$, $\delta=0.01$, $\alpha=0.8$, $\beta=0.4$, $x_s=0.019$.\label{fig12}}
\end{figure}

We next analyze the effect of a magnetic field on solute dispersion in AC-EOF driven systems. Recall that $Ha=0$ corresponds to the absence of a magnetic field (pure AC-EOF), whereas a finite $Ha$ introduces an additional MHD mechanism via the Lorentz force.
Figure~\ref{fig11} presents the mean concentration contours at fixed times. It directly illustrates the magnetic field regulation on solute spatial distribution. At early times ($t=0.005$), solute is highly concentrated near the center $x_1=0$. A higher $Ha$ indicates a stronger magnetic field. It confines the solute to a narrower central region and yields a sharper concentration peak. As time evolves from $t=0.02$ to $0.5$, the concentration gradually broadens and homogenizes. This process is significantly delayed at larger $Ha$. Strong Lorentz forces suppress radial fluid motion. They sustain steeper concentration gradients and prolong the transient dispersion phase.
To further complement the spatial perspective, figure~\ref{fig12} displays the temporal evolution of mean concentration at fixed Hartmann numbers. Figure~\ref{fig12}(a-c) shows the spatial distribution of mean concentration at various times. Under pure AC electroosmotic driving with $Ha=0$, the concentration peak flattens and spreads more rapidly. This behavior indicates enhanced dispersion. In contrast, increasing $Ha$ attenuates dispersion. It notably prolongs the initial transient phase and delays homogenization. Figure~\ref{fig12}(d-f) offers a clearer comparison through concentration profiles and contours. In the absence of a magnetic field with $Ha=0$, AC-EOF induces strong convective transport. This leads to fast dispersion and early homogenization. Applying a magnetic field suppresses EOF through Lorentz forces. It weakens convection and retards homogenization. Under strong magnetic fields, flow velocity is significantly reduced. Concentration gradients can be sustained for extended periods without full homogenization.
Together, figures~\ref{fig11} and~\ref{fig12} provide a comprehensive picture of magnetic field effects: With the electroosmotic effect held constant, higher $Ha$ constrains solute near the center and slows down dispersion, while lower $Ha$ promotes uniform radial distribution and faster homogenization.

\begin{figure}
  \centering
  \subfigure
  {\includegraphics[width=0.9\textwidth]{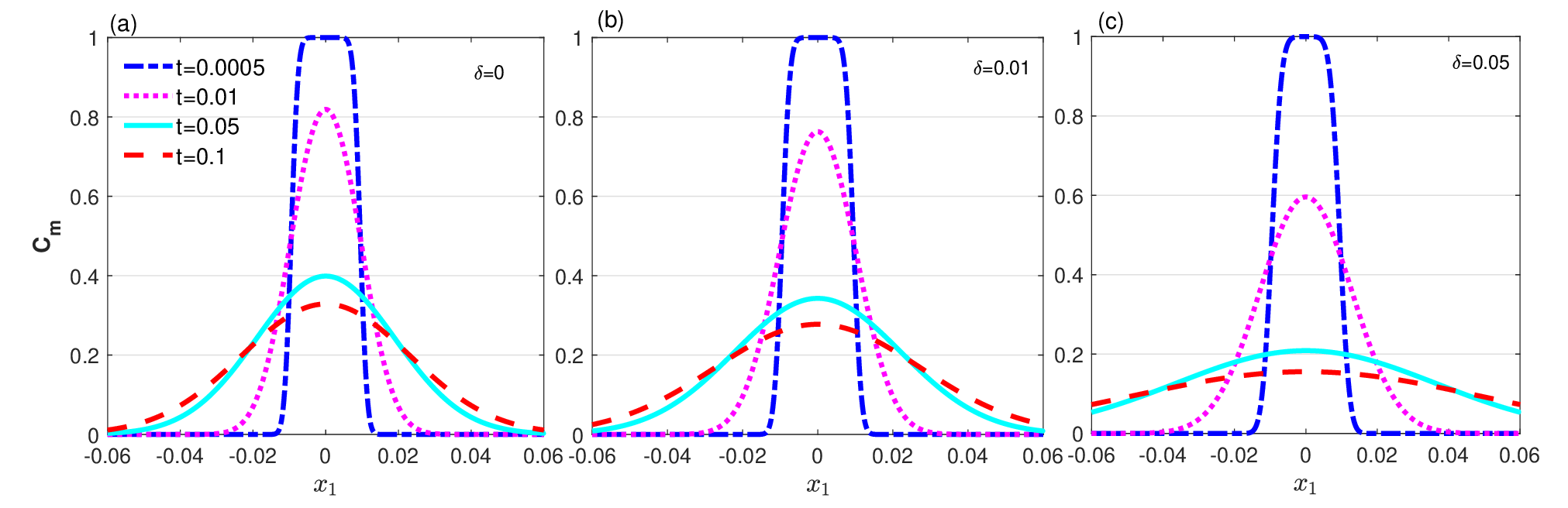}}
  \subfigure
  {\includegraphics[width=\textwidth]{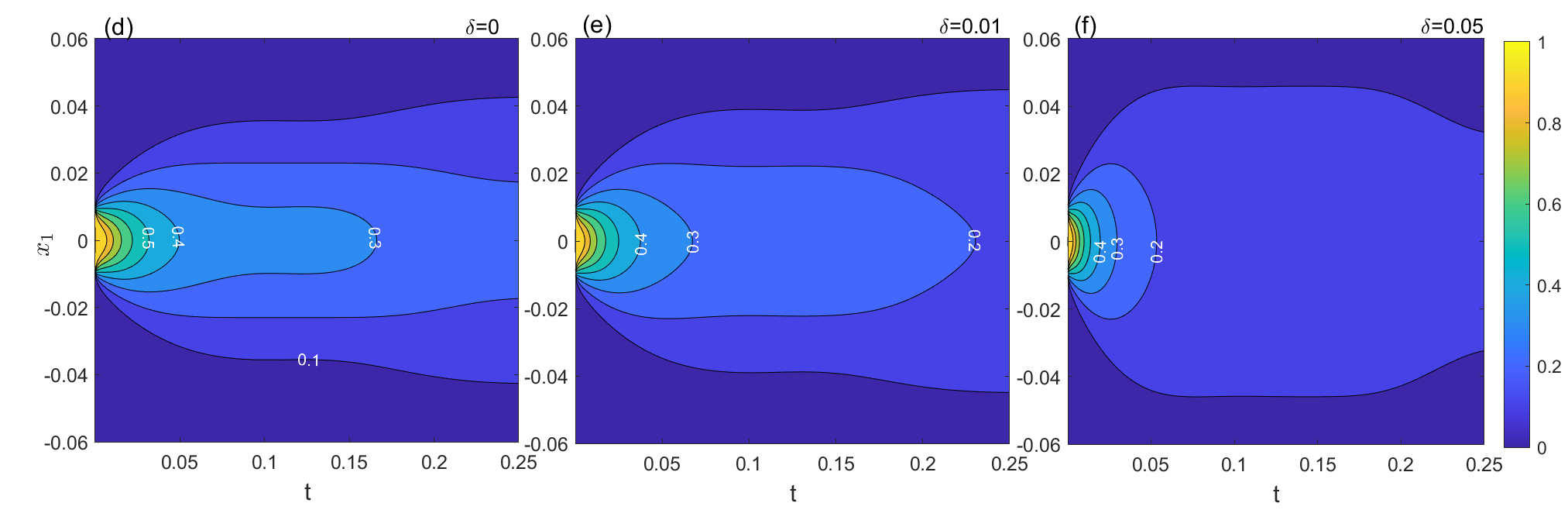}}
  \caption{(a-c) Variation of the mean concentration along the axial distribution with different slip lengths; (d-f) Mean concentration contours of solute with  different slip lengths at $Pe=50$, $\kappa=10$, $Re=5$, $De=1$, $\omega=20$, $Ha=0.6$, $\zeta=0.5$, $\alpha=0.8$, $\beta=0.4$, $x_s=0.019$.\label{fig13}}
\end{figure}

Figure~\ref{fig13}(a–f) illustrates the influence of the interfacial slip coefficient on the mean concentration distribution. The presence of slip accelerates initial solute dispersion and reduces the time required for full homogenization. This behavior stems from slip-enhanced boundary velocity profiles. Through slip-amplified zeta potential, convective transport is intensified and velocity non-uniformity increased, thereby expediting solute homogenization.

\begin{figure}
  \centering
  \subfigure
  {\includegraphics[width=0.9\textwidth]{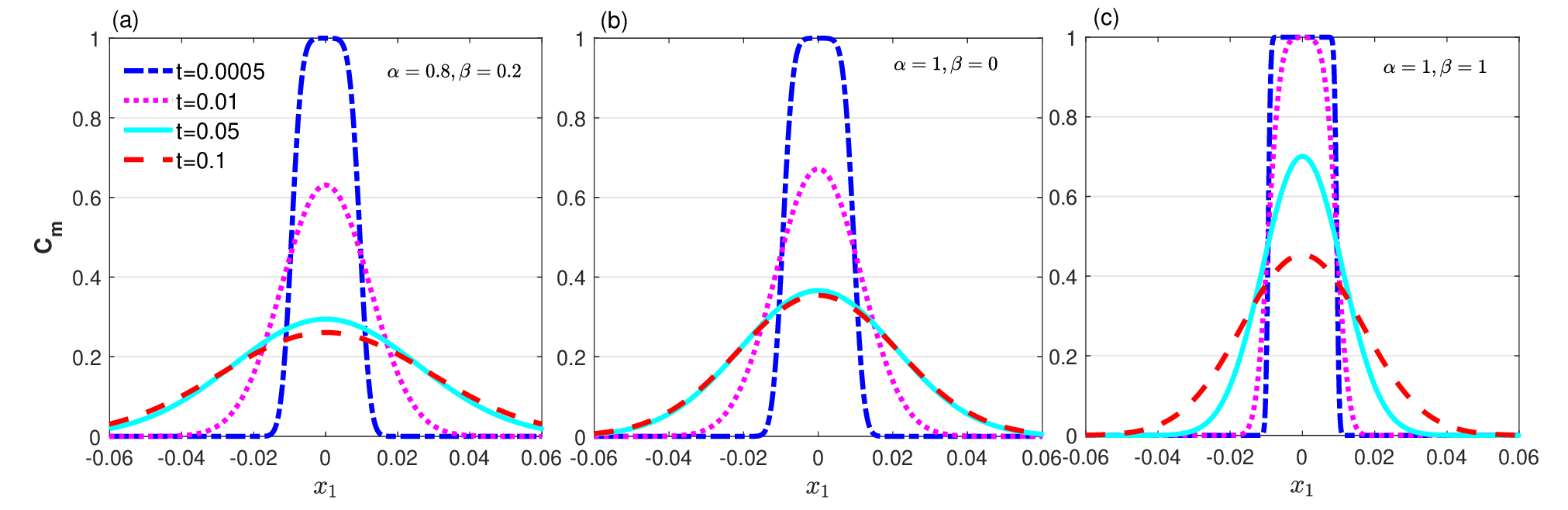}}
  \subfigure
  {\includegraphics[width=\textwidth]{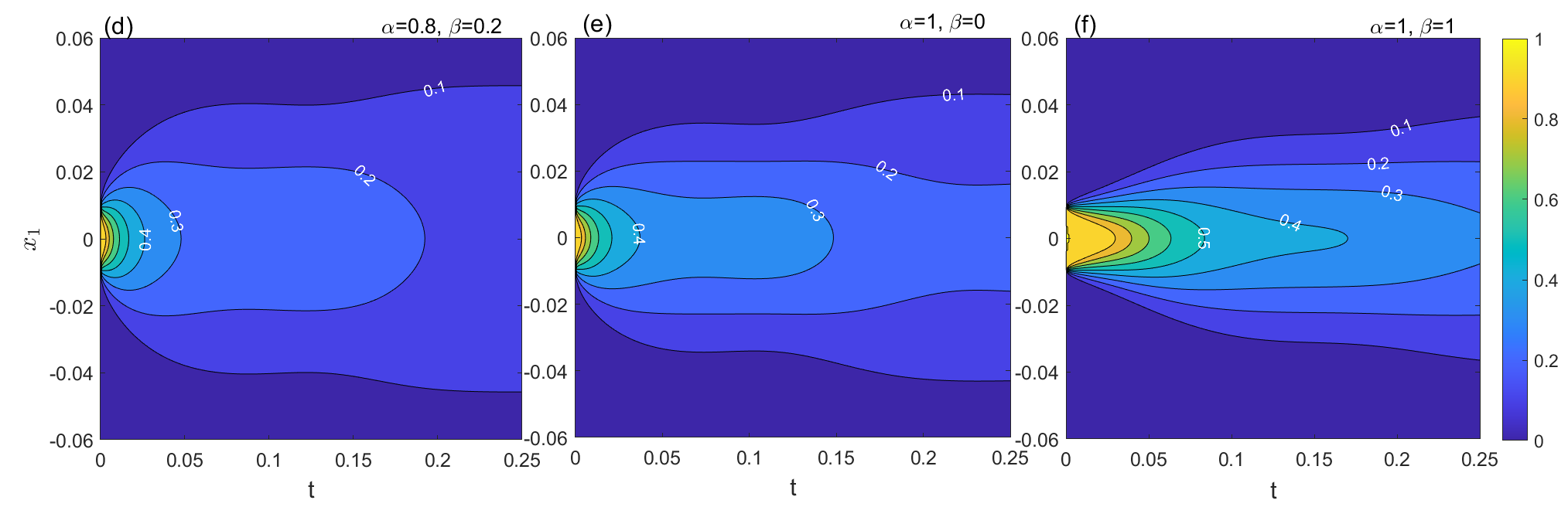}}
  \caption{(a-c) Variation of the mean concentration along the axial distribution with different fractional parameters; (d-f) Mean concentration contours of solute with different fractional parameters at $Pe=50$, $Re=5$, $De=1$, $\kappa=10$, $\omega=20$, $Ha=0.6$, $\zeta=0.5$, $\delta=0.01$, $x_s=0.019$.\label{fig14}}
\end{figure}

The fractional Maxwell model can reduce to either the classical Maxwell model or the Newtonian fluid model depending on the fractional parameters, making a comparative analysis of their dispersion behaviors necessary. As shown in figure~\ref{fig14}, compared with the integer-order Maxwell and Newtonian models, the fractional Maxwell model exhibits rapid initial response and enhanced dispersion depth due to its long-range memory and multi-scale characteristics, leading to the fastest peak decay and the lowest equilibrium concentration. However, its long-tail relaxation prolongs the time to reach steady state. The classical Maxwell model achieves quick elastic-viscous balance via exponential relaxation, resulting in efficient early-stage dispersion, moderate final mixing, and the shortest stabilization time. In contrast, the Newtonian fluid, lacking elasticity and memory, shows the slowest decay and least dispersion. 
This mechanism reflects the crucial influence of the relaxation behavior of the microscopic structure on the macroscopic solute transport in multi-field coupling scenarios, ranging from simple viscous fluids to complex viscoelastic fluids.

\begin{figure}
  \centering
  \subfigure
  {\includegraphics[width=0.35\textwidth]{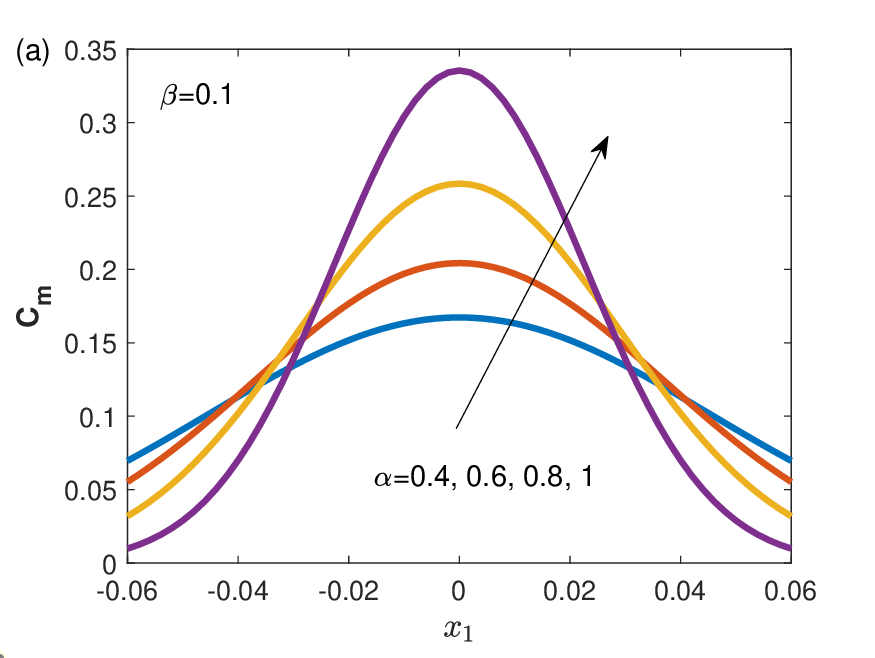}}
  \subfigure
  {\includegraphics[width=0.35\textwidth]{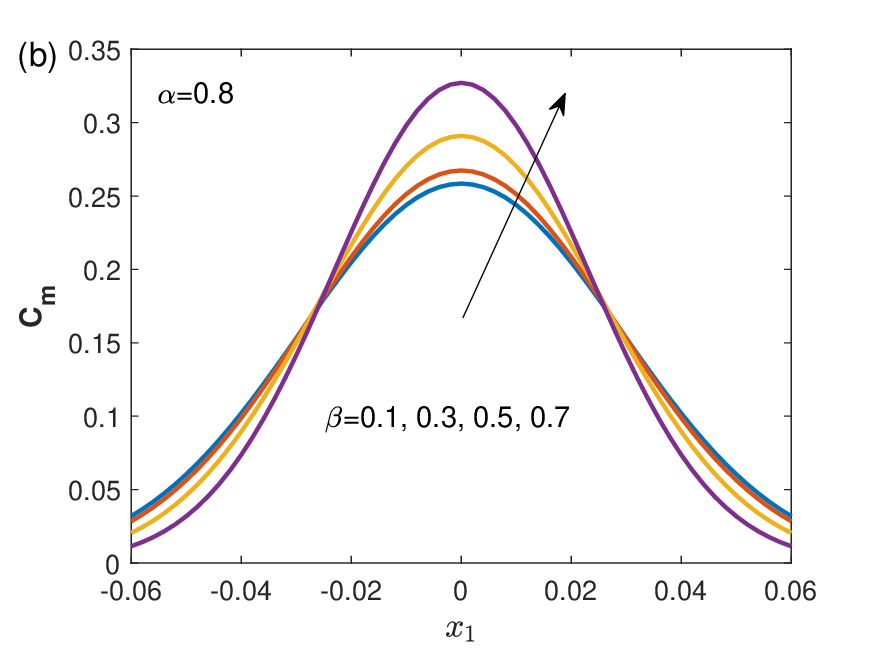}}
   \subfigure
  {\includegraphics[width=0.35\textwidth]{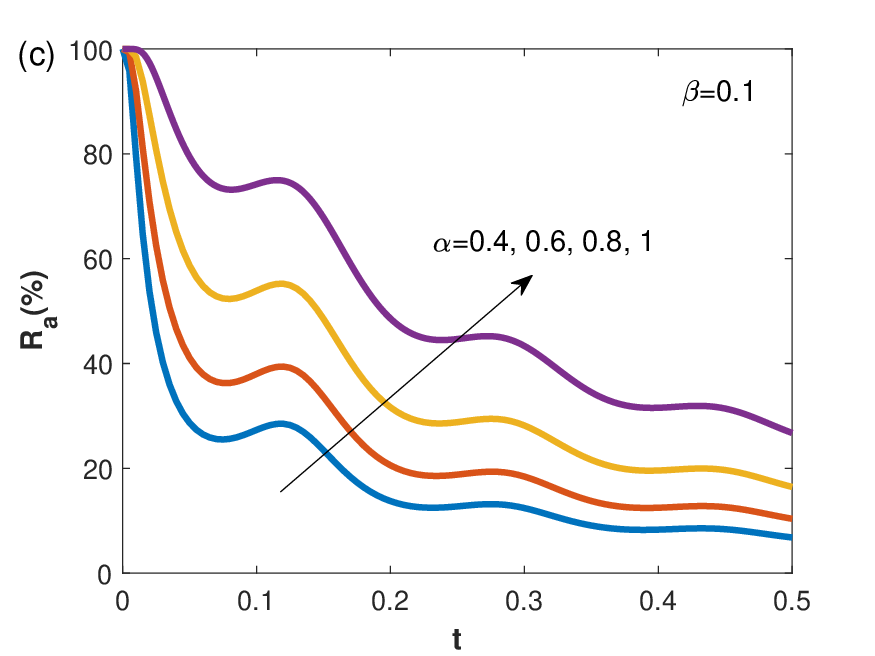}}
  \subfigure
  {\includegraphics[width=0.35\textwidth]{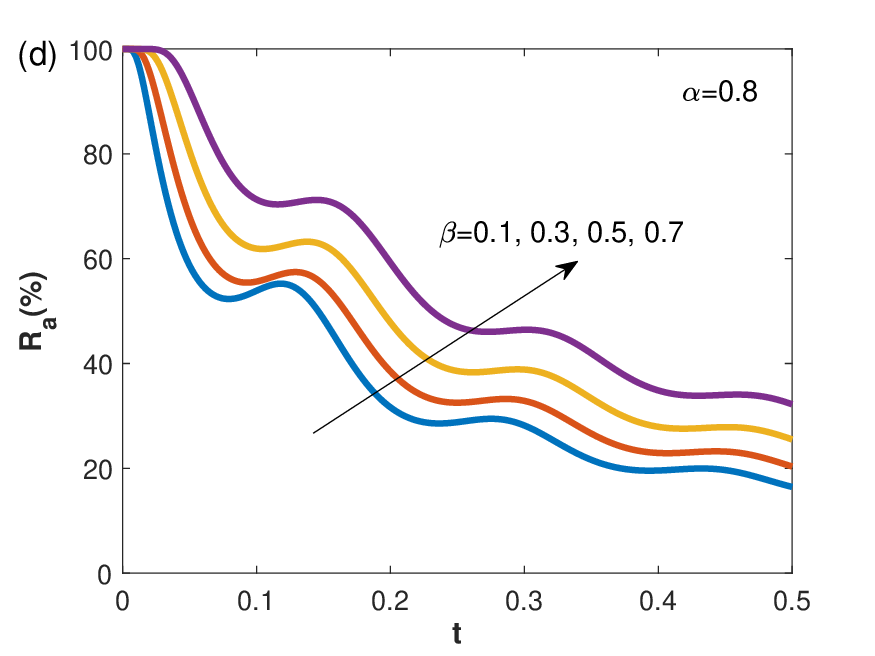}}
  \caption{(a,b) Variation of the mean concentration along the axial distribution with different fractional parameters $\alpha$ and $\beta$ at $t=0.1$; (c,d) the percentage of  axial concentration difference varies as a function of the fractional parameters $\alpha$ and $\beta$, under $Pe=50$, $\kappa=10$, $Re=5$, $De=1$, $\omega=20$, $\delta=0.01$, $Ha=0.6$, $\zeta=0.5$, $x_s=0.019$.\label{fig15}}
\end{figure}

Then, the effect of fractional parameters on the mean concentration distribution is given in figure~\ref{fig15}(a,b). It can be observed that when either fractional parameter $\beta$ (or $\alpha$) is held constant, the peak mean concentration increases as $\alpha$ (or $\beta$) increases. A higher $\alpha$ enhances the fluid's sensitivity to high-frequency oscillations in the AC electric field, reducing phase lag and improving particle tracking. This promotes directed migration and aggregation under periodic forcing. Meanwhile, an increase in $\beta$ intensifies the stress memory effect, effectively suppressing random diffusion. Thus, raising both $\alpha$ and $\beta$ amplifies the concentration peak and attenuates dispersion. To quantitatively evaluate the axial homogenization of solutes under hybrid MHD-AC-EOF over time, we introduce a modified version of the concentration difference percentage index used in previous studies~\citep{ZWGQ14, MRHS19}. The detailed definition is provided below.
\begin{equation}
       \centering
        R_a(t)=\frac{\mathrm{max}\left[C_m(x_1,t)\right]-\mathrm{min}\left[C_m(x_1,t)\right]}{C_m(x_1,t)|_{x_1=0, \;t=t_0}}\times 100\%.
\end{equation}
Here, $R_a$ represents the ratio of the maximum variation in the axially mean concentration over a predefined distance ($x_1\in [-0.06,0.06]$) to the concentration at the centroid of the solute band at a given time, thus serving as a crucial metric for evaluating the degree of axial uniformity of the solute. As demonstrated in figure~\ref{fig15} (c,d), $R_a$ for different fractional parameters exhibits an oscillatory decline over time, indicating that the axial distribution of the solute gradually achieves uniformity. Moreover, an increase in the fractional parameters $\alpha$ and $\beta$ leads to an elevation in $R_a$, signifying greater axial non-uniformity of the solute concentration. 

\begin{figure}
  \centering
  \includegraphics[width=0.95\textwidth]{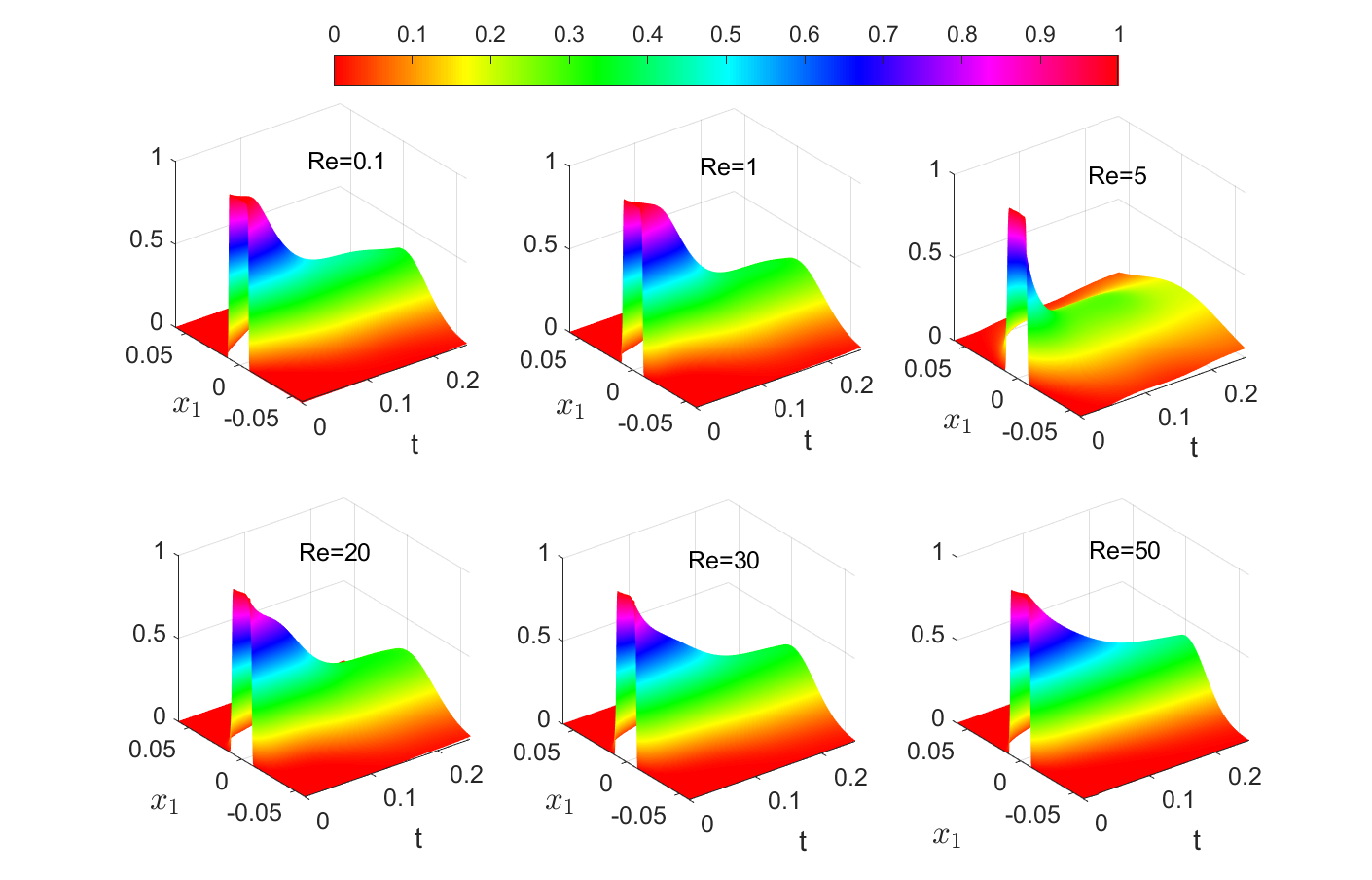}
  \caption{Mean concentration contours of solute with different oscillation Reynolds number $Re$ at $\alpha=0.8$, $\beta=0.4$, $Pe=50$, $\kappa=10$, $De=1$, $\omega=20$, $Ha=0.6$, $\zeta=0.5$, $\delta=0.01$,  $x_s=0.019$.\label{fig16}}
\end{figure}

Figure~\ref{fig16} displays contour plots of the mean concentration distribution at various oscillation Reynolds numbers $Re$. Below the critical value, the concentration gradient decreases gradually, which leads to a more uniform solute distribution and reduces the time required to reach steady state. At $Re = 5$, the concentration profile is nearly homogeneous throughout the entire domain, indicating highly effective dispersion. Above $Re_c$ ($20 \leq Re \leq 50$), increases in $Re$ sharpen concentration gradients and amplify solute non-uniformity, reflecting a reduction in dispersion efficacy. This behavior occurs because, at low $Re$, electromagnetic and viscous forces act together to enhance dispersion. At high $Re$, inertial forces dominate, and nonlinear interactions between electromagnetic and inertial effects reverse the trend in concentration gradient, thereby increasing the spatial heterogeneity of the solute distribution.

\begin{figure}
  \centering
  \includegraphics[width=0.85\textwidth]{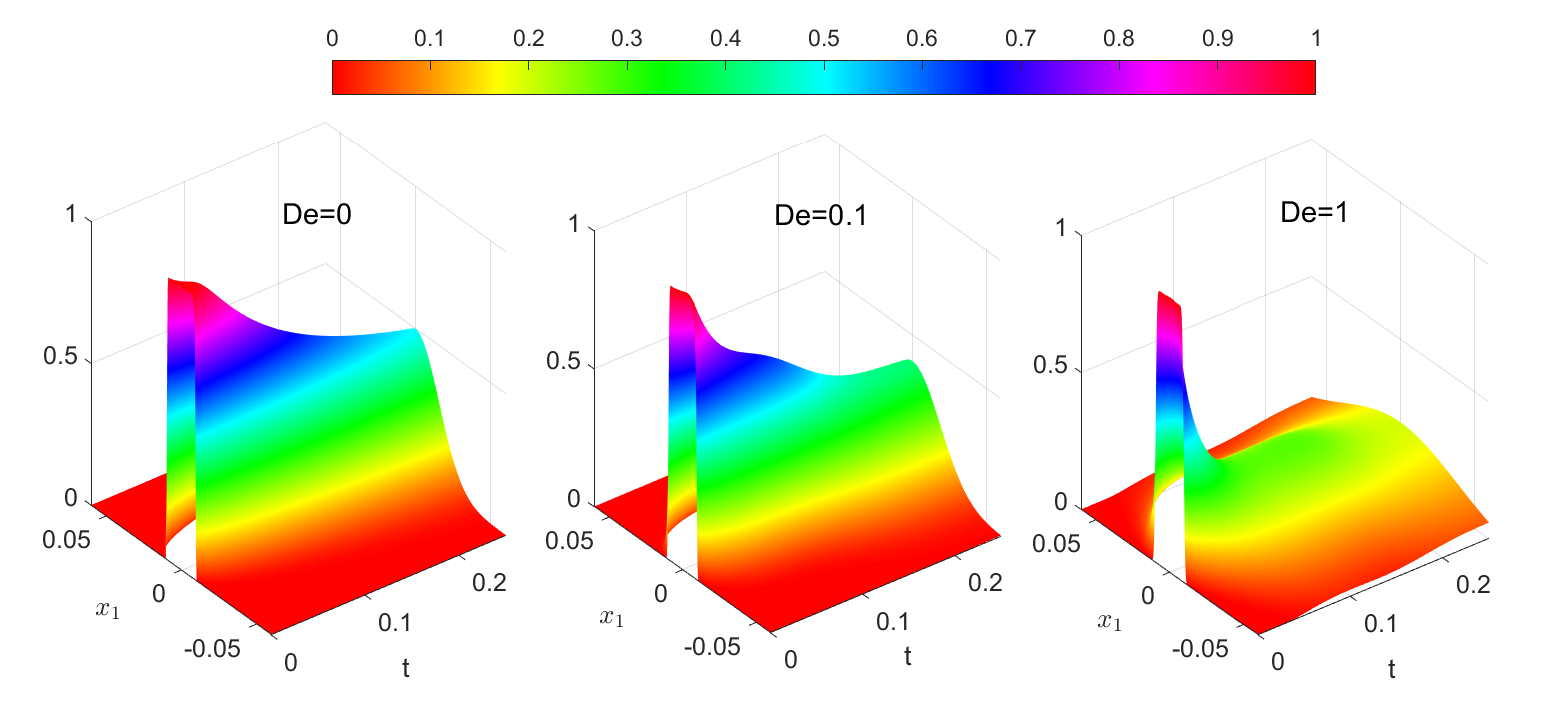}
  \caption{Mean concentration contours of solute with different Deborah number $De$ at $\alpha=0.8$, $\beta=0.4$, $Pe=50$, $\kappa=10$, $Re=5$, $\omega=20$, $Ha=0.6$, $\zeta=0.5$, $\delta=0.01$, $x_s=0.019$.\label{fig17}}
\end{figure}

\begin{figure}
  \centering
  \includegraphics[width=0.85\textwidth]{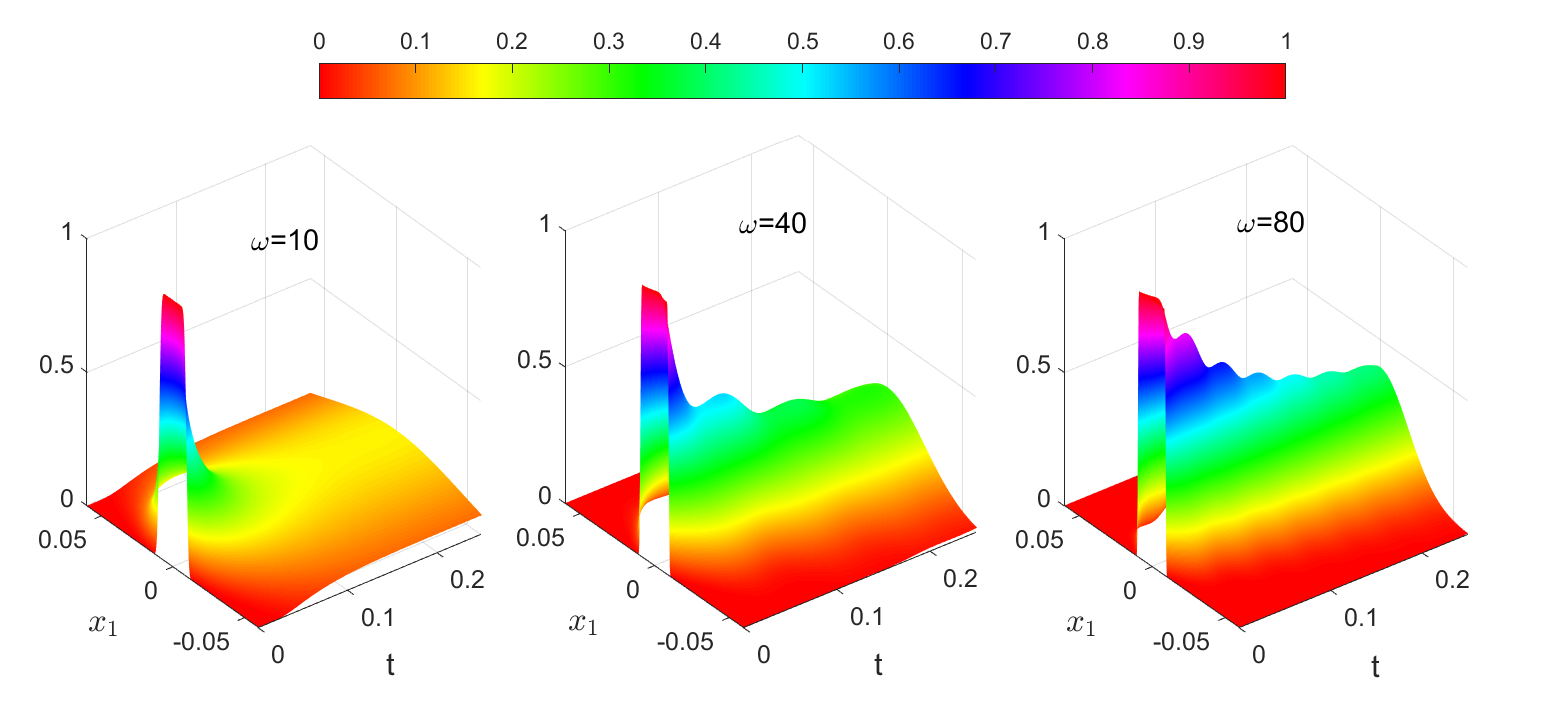}
  \caption{Mean concentration contours for different electric field oscillation frequencies $\omega$ at $Re=5$,  $\alpha=0.8$, $\beta=0.4$, $Pe=50$, $\kappa=10$, $De=1$, $Ha=0.6$, $\zeta=0.5$, $\delta=0.01$, $x_s=0.019$.\label{fig18}}
\end{figure}

Figure~\ref{fig17} illustrates the mean concentration contours of solute at different Deborah number $De$. Clearly, an increase in $De$ results in a reduction in $C_m$, which can be attributed to the enhanced convective effects that render the dispersion process more pronounced. By comparing the concentration distribution characteristics of the fractional Maxwell fluid ($De\neq 0$) and the Newtonian fluid ($De=0$), it is found that the solute dispersion rate in the fractional Maxwell fluid is faster. This difference is directly related to the inherent elastic effect of viscoelastic fluids. The contours of mean concentration with different oscillation frequency $\omega$ is given in figure~\ref{fig18}. When the frequency of the AC electric field is low, solute has more sufficient time to respond to the electric field actuation, accelerating the dispersion process and causing the concentration distribution to rapidly approach uniformity. Conversely, under high-frequency conditions, solute migration fails to keep pace with the rapid oscillations of the electric field, giving rise to periodic concentration fluctuations and a notable enhancement of the concentration gradient.

\section{Conclusion}\label{Conclusion}
An outstanding challenge in microfluidic transport is the accurate modeling of solute dispersion in viscoelastic fluids under multiphysics fields, where key interfacial phenomena such as hydrodynamic slip and zeta potential coupling are critical. The omission of these effects has hindered the development of high-performance mixers and separators. To address this, we introduce a unified theoretical framework that integrates these previously neglected phenomena. Our model provides analytical solutions capable of capturing the full spatiotemporal transport dynamics, thereby establishing a foundational tool for the predictive design of complex microfluidic systems. We demonstrate that the strategic application of a reverse Lorentz force via a magnetic field effectively suppresses solute dispersion in AC-EOF by flattening the underlying velocity profile. This approach proves most effective when the EDL is thick (i.e., for small values of $\kappa$). Our results confirm the cross-system universality of a fundamental principle: as concurrently revealed by~\cite{CCMS25} for EOF driven flows, employing an opposing pressure gradient to modulate flow morphology constitutes a robust and universal strategy for minimizing solute dispersion.

The key findings of this study show that while the magnetohydrodynamic suppression of solute dispersion in polymer solutions aligns with observations in both Newtonian~\citep{DPKR2024} and non-Newtonian systems~\citep{SMRR2025}, the viscoelasticity of these solutions induces a distinct coupled-field response. Specifically, at elevated Deborah numbers ($De$), electric-field-induced polymer chain stretching and alignment enhance electrokinetic flow, intensify axial dispersion, and increase the critical Hartmann number ($Ha$) needed for MHD suppression. Moreover, the dispersion response to the electrokinetic parameter ($\kappa$) exhibits significant sensitivity to $Ha$. Under low-$Ha$ conditions, the dispersion amplitude increases with $\kappa$, indicating electrokinetic dominance. In contrast, at high $Ha$, the behavior transitions to a Reynolds number ($Re$)-dependent regime: Lorentz forces dampen the flow and diminish the effect of $\kappa$ at low $Re$, whereas at high $Re$, inertial forces partially overcome the MHD suppression, thereby sustaining a higher dispersion level.

Furthermore, this study broadens the scope of investigation by exploring the effect of flow inertia on dispersion behavior across an extended $Re$ range of $1$–$30$, which goes beyond the limited low-$Re$ framework ($0.1$–$2$) typically covered in existing literature~\citep{MRHS19,DPKR2024,SMRR2025}. Due to their constrained $Re$ intervals, prior studies could only report a monotonic trend of decreasing dispersion with increasing $Re$. In contrast, our work, spanning a wider $Re$ range, reveals a key nonlinear feature: there exists a critical $Re_c$ value below which the dispersion coefficient increases with $Re$, and above which it decreases. This critical $Re_c$ is collectively influenced by multiple parameters, including $Ha$, $De$, $\kappa$, and slip coefficient ($\delta$). 

Our study also advances microfluidic transport analysis by integrating interfacial slip and its electrokinetic coupling with the zeta potential, extending beyond the conventional no-slip paradigm. Results indicate that the slip-dependent zeta potential model significantly enhances near-wall electrokinetic fields and flow velocities, exacerbating velocity profile non-uniformity and thereby intensifying global solute dispersion. Combined with prior research showing that ``asymmetric channel-wall zeta potential"~\citep{DPKR2024} and ``interfacially grafted polyelectrolyte layers"~\citep{MRHS19,SSMS2020} also enhance dispersion, our findings position the active regulation of microchannel wall properties as a novel approach for controlling axial solute dispersion in microfluidic systems.

Although this study systematically analyzes solute dispersion in MHD and AC-EOF, certain limitations remain, such as the idealized circular tube geometry and the focus on dilute single-solute systems. Despite these, this work advances the fundamental understanding of solute transport in microfluidic systems by elucidating the coupled effects of MHD and EOF. It reveals how key dimensionless parameters, such as the Hartmann number, electrokinetic parameter, and Reynolds number, govern flow patterns and dispersion characteristics, providing theoretical insights and practical guidance for electromagnetically driven microfluidic devices, including lab-on-a-chip systems, targeted drug delivery, and micromixers. Future work will aim to overcome the above limitations and establish a quantitative model correlating the Reynolds number, Hartmann number, and electrokinetic parameter to develop a generalized stability criterion, ultimately supporting precise flow control in such devices.
\begin{figure}
  \centering
  \includegraphics[width=0.55\textwidth]{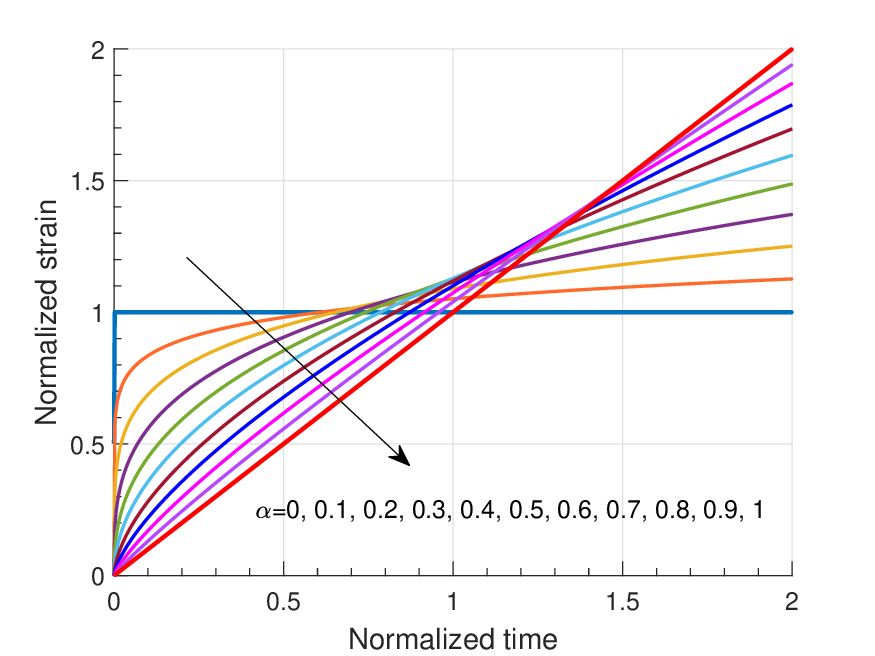}
  \caption{The normalized strain is presented as a function of the normalized time when a constant stress is applied at time $t=0$ to springpots with various exponents $\alpha$.\label{fig19}}
\end{figure}
\section*{Acknowledgments}
 This work was supported by the National Natural Science Foundation of China (NSFC, Grant Nos. 12402300, 12172197, and 12171284) and the Natural Science Foundation of Shandong Province (Grant Nos. ZR2024QA160, and ZR2025MS35).

\section*{Declaration of interests}
 The authors report no conflict of interest.

\appendix
\renewcommand{\thefigure}{\arabic{figure}}
\renewcommand{\thetable}{\arabic{table}}

\section{}\label{appA}
This paper studies solute dispersion in aqueous solutions of hydroxyethyl cellulose (HEC) and xanthan gum (XG). These polymers are ideal for microfluidic applications owing to their water solubility, pseudoplasticity, and stability. Such properties aid in flow regulation and help minimize unintended variations in dispersion. In microfluidic systems, HEC and XG solutions are commonly employed to control the transport and dispersion of solutes, such as drug molecules, through viscosity modulation. Therefore, accurate rheological modeling is essential for predicting and managing solute behavior in processes like mixing and separation. This section aims to establish appropriate viscoelastic models for these solutions to support subsequent analysis of solute diffusion and transport in microchannels.

\subsection{Rheological properties of HEC and XG aqueous solutions}\label{para_estimation}
Experimental studies~\citep{JDVS20,CCMK22} have shown that the frequency responses of aqueous solutions of HEC and XG adhere to a power-law relationship. The macroscopic viscoelastic phenomena observed in polymers, such as stress relaxation, originate from the motion of molecular segments. This classic concept is extensively detailed in \cite{Ferry1980}.
Specifically, the viscoelastic storage modulus and loss modulus exhibits a frequency-dependent behavior described by $G'(\omega)\sim\omega^{\alpha}, G''(\omega)\sim \omega^{\beta}$. However, traditional rheological models such as the Maxwell model and the Kelvin-Voigt model cannot adequately capture such power-law characteristics.

Although additional mechanical elements can be added in series or parallel to the basic Maxwell or Kelvin-Voigt elements~\citep{Tschoegl12} to introduce more relaxation modes and enable the modeling of power-law materials, this method often requires an excessive number of mechanical elements when dealing with many complex solutions. This not only increases model complexity but also raises computational demands, making it infeasible in practical modeling scenarios.
Furthermore, in models with a finite set of relaxation modes, the values of the fitted parameters are highly sensitive to the time scale of the experimental data used for fitting. As a result, the model parameters obtained through this process often lack clear physical significance~\citep{KPFB11}. The fractional constitutive model integrated with spring-pot mechanical elements is capable of providing a natural and quantitative description of the power-law behavior commonly observed in experiments with only a few constitutive parameters. Therefore, in this paper, we will use the fractional Maxwell viscoelastic model derived from the fractional stress-strain relationship to quantitatively describe the power-law rheological behavior exhibited by such solutions.

\subsection{Fractional Maxwell viscoelastic model}\label{F-Maxwell}
The classical linear viscoelastic Maxwell model has the following integral form
\begin{equation}\label{Maxwell-2}
  \sigma (t)=G_0\int_{-\infty}^{t}\mathrm{e}^{-\frac{t-t'}{\lambda}} \frac{\mathrm{d}\gamma(t')}{\mathrm{d} t'}{\mathrm{d} t'}.
\end{equation}
In this equation,  $G(t)=G_0\mathrm{e}^{-\frac{t}{\lambda}}$ is defined as the relaxation modulus, which describes the stress response to an instantaneous deformation jump. Notably, this integral form is a specific instance of the Boltzmann integral and is characterized by \textit{exponential decay}. It is widely used to characterize the linear viscoelastic deformation of complex materials. However, for materials exhibiting different forms of memory fading behavior, such as some polymer solutions with power-law decay behavior (e.g., the HEC solution and XG aqueous solution considered in this study), polymer gels~\citep{MAJP88,THEM08}, cells~\citep{FMBG01,BDID06}, etc., the relaxation modulus decays as $G(t)=St^{-\alpha}$, with the exponent $0<\alpha<1$. For such materials, the following form of stress response to an arbitrary shear history needs to be considered~\citep{SMBN95,ALCR18}
\begin{equation}\label{FMaxwell-1}
  \mathbf{\sigma}(t)=\frac{E\lambda^{\alpha}}{\Gamma(1-\alpha)}\int_{-\infty}^{t}(t-t')^{-\alpha} \frac{\mathrm{d}\gamma(t')}{\mathrm{d} t'}{\mathrm{d} t'},
\end{equation}
by reformulating the relaxation modulus as $G(t)=\frac{E}{\Gamma(1-\alpha)}\left(\frac{t}{\lambda}\right)^{-\alpha}$, where $E$ is a modulus and $\lambda$ is a characteristic time. In light of the definition of the fractional derivative operator (Caputo type)~\citep{Podlub99,Mainar22},
\begin{equation}\label{Caputo}
   \frac{\mathrm{d}^{\alpha}}{\mathrm{d}t^{\alpha}}f(t)
   =\frac{1}{\Gamma(1-\alpha)}\int_{-\infty}^{t}(t-t')^{-\alpha}\frac{\mathrm{d}f(t')}{\mathrm{d}t'}{\mathrm{d} t'},\;\;0<\alpha<1,
\end{equation}
the aforementioned equation (\ref{FMaxwell-1}) can be reformulated as follows:
\begin{equation}\label{FMaxwell-2}
  \mathbf{\sigma}(t)=E\lambda^{\alpha} \frac{\mathrm{d}^{\alpha}\mathbf{\gamma}(t)}{\mathrm{d} t^{\alpha}}.
\end{equation}
This formalism is commonly referred to as the Scott-Blair element or springpot~\citep{Scott-Blair47}. The proposed model can be physically implemented via hierarchical configurations of springs and dashpots, including ladder-like, tree-like, or fractal architectures. Microscopic and mesoscopic interpretations of such continuous viscoelastic structures can be found in~\cite{SHBA95}. In the limiting cases of $\alpha=1$ and $\alpha=0$,  equation (\ref{FMaxwell-2}) reduces to the constitutive laws for a viscous Newtonian fluid, $\mathbf{\sigma}(t)=\eta \frac{\mathrm{d}\mathbf{\gamma}}{\mathrm{d} t}$ with $\eta=E\lambda$, and a linear elastic solid, $\mathbf{\sigma}(t)=E \mathbf{\gamma}$, respectively. For intermediate values of $\alpha\in(0,1)$,  equation (\ref{FMaxwell-2})  generalizes these behaviors interpolate between the elastic and viscous extremes. Figure~\ref{fig19} illustrates the interpolation, showing the creep deformation of spring-pot elements with varying fractional exponents $\alpha$ under a step stress applied at $t=0$.

Analogous to classical integer-order viscoelastic models constructed via the series and parallel combination of springs and dashpots, spring-pot elements can be used to establish more complex constitutive models. Specifically, two spring-pots, characterized by the parameter pairs
($\alpha$, $E_1$, $\lambda_1$) and ($\beta$, $E_2$, $\lambda_2$),  respectively, are connected in series.  By assuming stress equivalence across the spring-pots ($\mathbf{\sigma}=\mathbf{\sigma_1}=\mathbf{\sigma_2}$) and strain additivity ($\mathbf{\gamma}=\mathbf{\gamma_1}+\mathbf{\gamma_2}$) , the constitutive equation of the fractional Maxwell model (FMM) can be systematically derived~\citep{SMBN95,TPMX03}.
\begin{equation}\label{FMaxwell}
   \left(1+\lambda^{\alpha-\beta}\frac{\mathrm{d}^{\alpha-\beta}}{\mathrm{d} t^{\alpha-\beta}}\right)\sigma(t)=E\lambda^{\alpha}\frac{\mathrm{d}^{\alpha}\gamma (t)}{\mathrm{d} t^{\alpha}},\;\;0\leq\beta\leq\alpha\leq1,
\end{equation}
where $\lambda=(E_1\lambda_1^\alpha/E_2\lambda_2^\beta)^{1/(\alpha-\beta)}$ and $E=E_1(\lambda_1/\lambda)^\alpha$. Without loss of generality, $\alpha>\beta$ is assumed here~\citep{SMBN95}.~\citet{Fried91} demonstrated that the model yields non-negative internal work and energy dissipation rate, fulfilling the laws of thermodynamics. It is important to note that the proposed model reduces to the classical integer-order Maxwell model when $\alpha=1$ and $\beta=0$. Conversely, when $\alpha=\beta=1$, the model simplifies to the classical Newtonian fluid model. This special-case analysis not only highlights the versatility of the model but also elucidates its connection with well-established constitutive models in the field of rheology.

In this paper, the experimental data of the viscoelastic storage modulus $G'(\omega)$ and loss modulus  $G''(\omega)$ of HEC and XG aqueous solutions are employed. These data are obtained through  small-amplitude oscillatory shear (SAOS) deformation tests in the frequency domain. Regarding the FMM, the complex modulus can be analytically derived by applying Fourier transform to the governing equation (\ref{FMaxwell})
\begin{equation}\label{G}
   G^*=\frac{E\lambda^{\alpha}(\mathrm{i}\omega)^{\alpha}*\lambda^{\beta}(\mathrm{i}\omega)^{\beta}}{\lambda^{\alpha}(\mathrm{i}\omega)^{\alpha}
   +\lambda^{\beta}(\mathrm{i}\omega)^{\beta}}.
\end{equation}
Subsequently, through the separation of the real and imaginary components of the derived equation, the expressions for the storage modulus and loss modulus can be obtained, as presented below:
\begin{equation}\label{G'}
  G'(\omega)=\frac{E\lambda^{\beta}\omega^{\beta}\cos(\pi\alpha/2)+E\lambda^{\alpha}\omega^{\alpha}\cos(\pi\beta/2)}
  {\lambda^{\alpha-\beta}\omega^{\alpha-\beta}+\lambda^{\beta-\alpha}\omega^{\beta-\alpha}+2\cos(\pi(\alpha-\beta)/2)},
\end{equation}
\begin{equation}\label{G''}
     G''(\omega)=\frac{E\lambda^{\beta}\omega^{\beta}\sin(\pi\alpha/2)+E\lambda^{\alpha}\omega^{\alpha}\sin(\pi\beta/2)}
  {\lambda^{\alpha-\beta}\omega^{\alpha-\beta}+\lambda^{\beta-\alpha}\omega^{\beta-\alpha}+2\cos(\pi(\alpha-\beta)/2)}.
\end{equation}
When $\alpha=1$ and $\beta=0$, expressions (\ref{G'}) to (\ref{G''}) simplify to those of the linear Maxwell model, which verifies the consistency between the fractional model and classical theory. The FMM concisely captures the broad spectrum of relaxation dynamics using only four parameters ($\alpha,\;\beta,\lambda,\;E$), providing a more parsimonious alternative to traditional models. In contrast, to accurately describe the power-law behavior of materials by using multiple Maxwell models, an impractically large number of discrete relaxation times is required~\citep{Tschoegl12}. 

\begin{figure}
  \centering
  \subfigure
  {\includegraphics[width=0.35\textwidth]{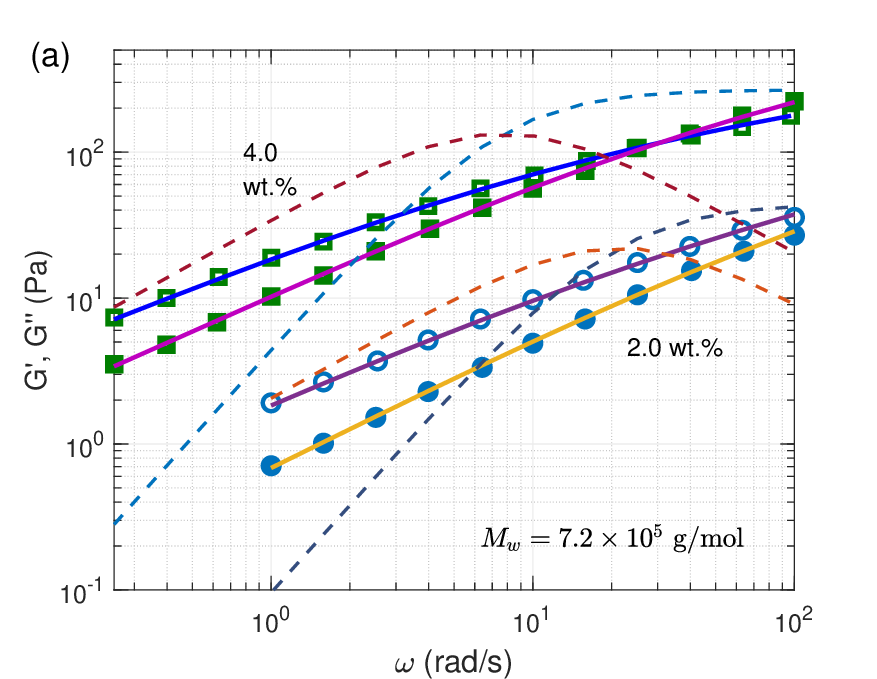}}
  \subfigure
  {\includegraphics[width=0.35\textwidth]{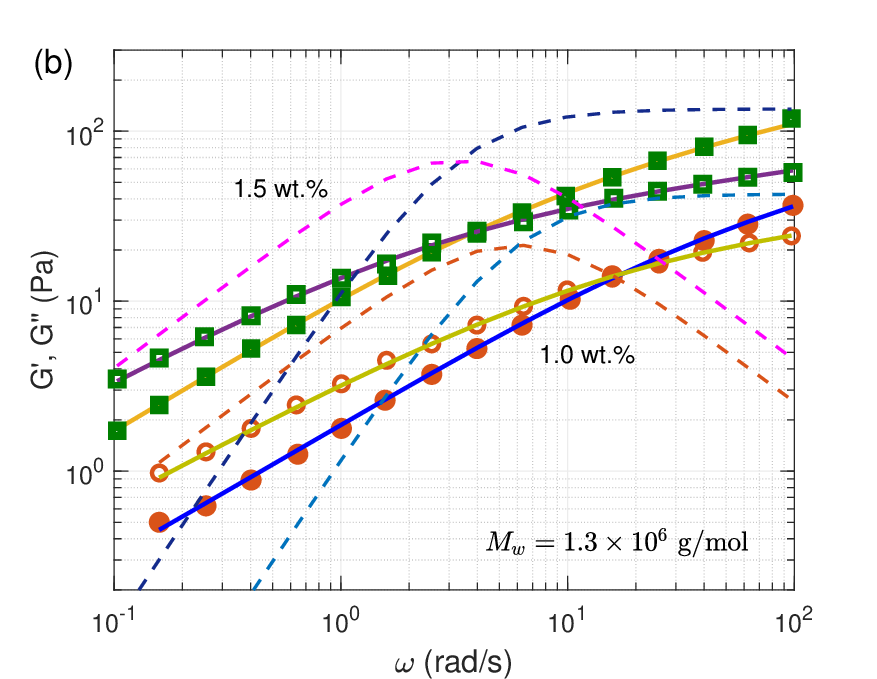}}
    \subfigure
  {\includegraphics[width=0.35\textwidth]{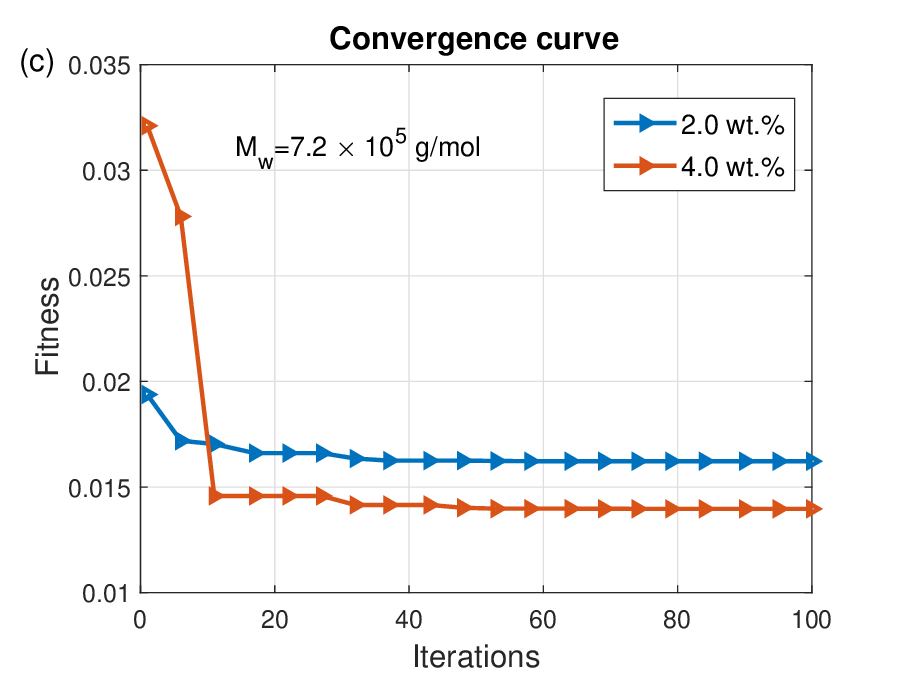}}
  \subfigure
  {\includegraphics[width=0.35\textwidth]{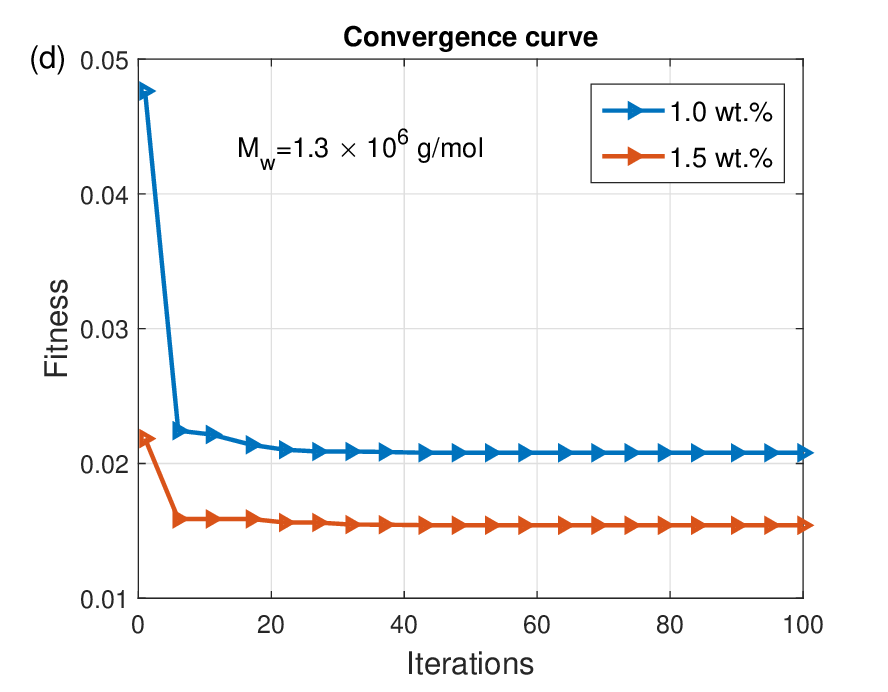}}
  \caption{The fitting results for the storage modulus $G'$ (closed symbols) and loss modulus $G''$ (open symbols) of HEC solutions in ~\cite{JDVS20} with fractional Maxwell model (solid lines) and classical Maxwell model (dashed lines) at $M_w=7.2\times10^5$ g/mol and $M_w=1.3\times10^6$ g/mol, respectively; (c,d) Iterative convergence curve of parameter inversion for HEC solutions with DCE method.\label{fig20}}
\end{figure}

\begin{table}
   \begin{center}
   \def~{\hphantom{0}}
   \begin{tabular}{lcccccccccccc}
   \hline
 $M_w$ &Concentration & &$\alpha$ & $\beta$  &  $E (Pa)$ & $\lambda (s)$  & &Fitness \\[3pt]
 $7.2\times10^5$ g/mol & 2.0 wt.\%  && 0.8417 & 0.4110 & 52.0171 & 0.0243 && 0.0162  \\
                       & 4.0 wt.\% & &0.7674 & 0.2908 & 238.3099 & 0.0532 && 0.0140 \\

$1.3\times10^6$ g/mol & 1.0 wt.\%&& 0.7398 & 0.2130 & 44.2188  & 0.0414 &&0.0208\\
                      & 1.5 wt.\%& & 0.7569 & 0.2345 & 66.9383  & 0.2518 &&0.0154 \\
   \hline
   \end{tabular}
 \caption{The parameter estimation results for HEC solutions.}\label{mytab1}
    \end{center}
\end{table}

\subsection{Parameter estimation on polymer solutions}\label{DCS}
This subsection focuses on fitting the experimental data for entangled HEC and XG aqueous solutions using the FMM and identifying the model parameters $\alpha$, $\beta$, $E$, and $\lambda$ via an enhanced differentiated creative search (DCS) algorithm~\citep{DSCN24}. The original DCS algorithm, known for its strong evolutionary and optimization capabilities, suffers from uneven solution space distribution due to random initialization. To address this, a logistic chaotic map is employed for population initialization to improve the algorithm's performance in the parameter identification process.
\begin{equation}
 P_{k,d} = L_d +  \chi\cdot \mathcal{L}(x_{k,d}, \vartheta) \cdot (U_d - L_d), \quad k=1,\dots,N_P; \quad d=1,\dots,D.
\end{equation}
Here $P_{k,d}$ represents the position coordinate of the $k$-th individual in the $d$-th dimension, $U_d$ and $L_d$ represent the lower bound and upper bound respectively for the $d$-th dimension, respectively, $N_p$ is the population size, $\chi$ denotes the chaotic disturbance strength coefficient, which regulates the search range of the chaotic mapping in the solution space, $\mathcal{L}(x_{i,d}, \vartheta)$ represents the logistic map function, whose expression is given by  
\begin{equation}
\mathcal{L}(x_{k,d}, \vartheta)=\vartheta x_{k,d} (1-x_{k,d}),
\end{equation}
where $\vartheta$ is the control parameter, typically set to $\vartheta=4$ to ensure the system operates in a full chaos state. 
\begin{figure}
  \centering
  \subfigure
  {\includegraphics[width=0.35\textwidth]{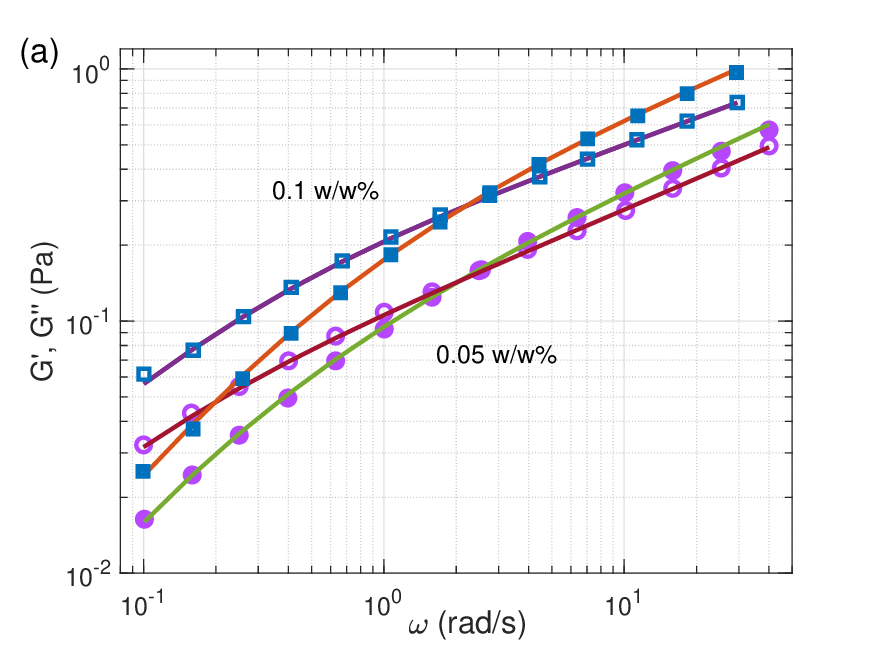}}
  \subfigure
  {\includegraphics[width=0.35\textwidth]{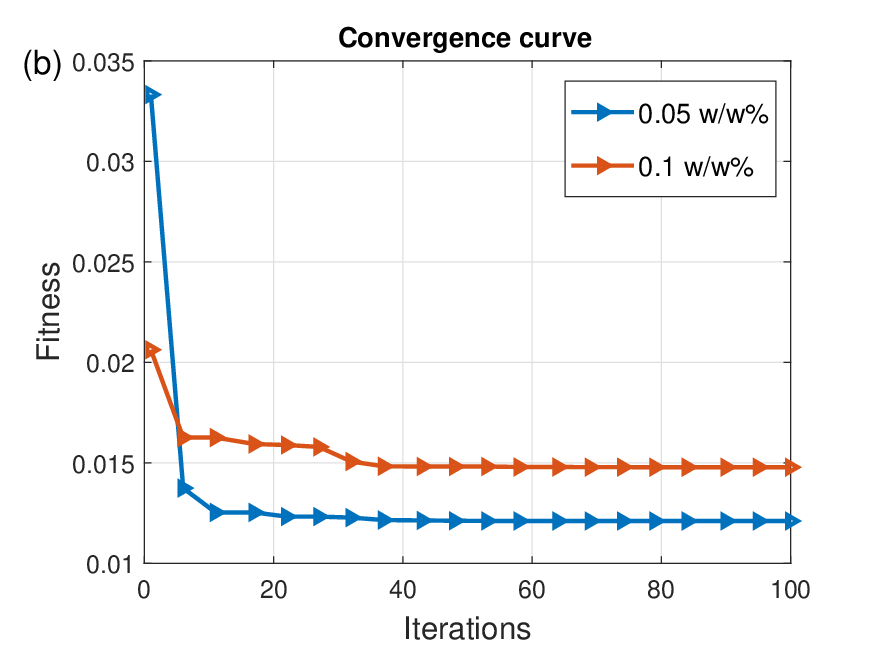}}
  \caption{(a) The fitting results for the storage modulus $G'$ (closed symbols) and loss modulus $G''$ (open symbols) of XG aqueous solutions in~\cite{CCMK22} with fractional Maxwell model; (b) Iterative convergence curve of parameter inversion for XG solution with DCE method.\label{fig21}}
\end{figure}
\begin{table}
   \begin{center}
   \def~{\hphantom{0}}
   \begin{tabular}{lccccccccccccccccc}
      \hline
   Concentration &&& $\alpha$ && $\beta$ & &  $E (Pa)$& & $\lambda (s)$ && & Fitness \\
   0.05 w/w\%  & &&0.9798 && 0.4169 && 0.0671 && 9.3457 &&&0.0121 \\
   0.1 w/w\%  &&&0.9314 && 0.3684 && 0.2399 && 3.3318 &&& 0.0148 \\
      \hline
   \end{tabular}
   \caption{The parameter estimation results for XG aqueous solutions.}\label{mytab2}
   \end{center}
 \end{table}
 
The fitness function and constraint conditions for this problem are given as
\begin{equation}
    Fitness(X)=\sqrt{\frac{\sum_{s=1}^{N_s}\left(G'_s-G'_{exp}\right)^2+\left(G''_s-G''_{exp}\right)^2}{N_s}}+cP,
\end{equation}
\begin{equation}
   0<\alpha\leq1,\;\;\beta-\alpha<0,\;\;0<E,\;\;0<\lambda.
\end{equation}
Here, $G'_s$ and $G''_s$ are the storage and loss moduli of FMM at the sample points, $G'_{exp}$ and $G''_{exp}$ designate the experimental data of storage and loss moduli of HEC solution or XG solution, $N_s$ represents the number of sample points, $P$ is the penalty coefficient given by a large number ($10^5$) and $c$ is the number of instances that do not meet the constraint conditions, ($\alpha$, $\beta$, $\lambda$, $E$) is the parameter vector to be estimated.

Using experimental data for the storage and loss moduli of HEC solutions with varying molecular weights ($M_w$) and concentrations at 5\% strain amplitude~\citep{JDVS20}, we determined the parameters of the FMM via the DCS method. The fitting results are presented in figure~\ref{fig20}(a,b), and estimated parameters are listed in table~\ref{mytab1}. For comparison, figure~\ref{fig20} also includes the fits obtained with the classical Maxwell model. It is evident that the FMM accurately captures the moduli across the different HEC solutions, whereas the classical Maxwell model exhibits significant deviations.
To evaluate the effectiveness of the DCS algorithm, convergence curves of the parameter inversion process are shown in figure~\ref{fig20}(c,d). The fitness value decreases rapidly within $5$–$10$ iterations and then stabilizes, demonstrating both high stability and fast convergence of the algorithm. The corresponding minimum fitness values are detailed in the last column of table~\ref{mytab1}. To further demonstrate the broad applicability of the FMM, the DCS algorithm was applied to fit the storage and loss moduli of XG aqueous solution at 10\% deformation~\citep{CCMK22}. As shown in figure~\ref{fig21}(a), the FMM predictions agree excellently with the experimental data. The estimated parameters, listed in table~\ref{mytab2}, are consistent with theoretical and literature values. Moreover, the convergence behavior in figure~\ref{fig21}(b) and the low fitness values in table~\ref{mytab2} confirm the algorithm’s robustness and stability, affirming both the reliability of the parameters and the effectiveness of the method.

In conclusion, t is imperative to highlight that the FMM not only excellently fits the two polymer solutions studied here but also shows strong versatility across diverse systems, including other polymers~\citep{FMRM18,KXKQ11,AJGM14,WBEB17}, biological solutions~\citep{YQJX18}, and viscoelastic interfaces~\citep{JAGH13}. Its ability to capture a wider range of relaxation times compared to classical and multi-mode models highlights the broad applicability and superiority of fractional calculus-based viscoelastic modeling. Furthermore, the capability of fractional calculus to describe complex multi-scale dynamics underscores its broader applicability, with significant potential in fields like turbulent flow modeling~\citep{Barros2024,Akhavan2023,Samiee2022,Keith2021}.

\section{}\label{appB}
The coefficients $C_1-C_4$ and parameter $A$ in equation (\ref{solution1}) are given as follows:
\begin{equation}
  A=\left(i^{1-\alpha}De^{1-\alpha}+i^{1-\beta}De^{1-\beta}\right)(Rei+Ha^2),\label{eq:A1}
\end{equation}
\begin{equation}
  B=-\frac{\kappa^2\zeta}{I_0(\kappa)}\left(i^{1-\alpha}De^{1-\alpha}+i^{1-\beta}De^{1-\beta}\right)\left(1+\kappa \delta\frac{\sinh\zeta}{\zeta}\right),
\end{equation}
\begin{equation}\label{C1}
  C_1=\frac{1}{2(A-\kappa^2)[\mathrm{I}_0(\sqrt{A})+\sqrt{A}\delta \mathrm{I}_1(\sqrt{A})]},
\end{equation}
\begin{equation}\label{C2}
 C_2=2B(\mathrm{I}_0(\kappa)+\delta \kappa \mathrm{I}_1(\kappa)),
\end{equation}
\begin{equation}\label{C3}
 C_3=\sqrt{A}B\pi \mathrm{I}_0(\sqrt{A})+AB\delta\pi \mathrm{I}_1(\sqrt{A}),
 \end{equation}
\begin{equation}\label{C4}
 C_4=B\kappa \pi \mathrm{I}_0(\sqrt{A})+\sqrt{A}B\delta\kappa\pi \mathrm{I}_1(\sqrt{A}).
 \end{equation}
Here, parameter $A$ is a comprehensive multi-physical coupling parameter that integrates viscoelastic effects ($De$, $\alpha$, $\beta$), oscillatory inertia ($Re$), and magnetic damping ($Ha^2$). Its magnitude directly governs the radial shape of the velocity profile, reflecting the combined modulation of viscoelasticity, inertia, and magnetic forces. Parameter $B$ acts as a correction coefficient for the effective EO driving strength, quantifying the EO force modulated by viscoelasticity and interfacial slip. Coefficients $C_1$ to $C_4$ incorporate the electrokinetic parameter $\kappa$, slip length $\delta$, and parameter $A$, ensuring that the velocity solution satisfies the wall boundary conditions.

The expression (\ref{solution1}) can be reformulated into the following equivalent form by means of the regularized confluent hypergeometric function
\begin{equation}\label{solutionu2}
  u_0(r)=N_1\mathrm{I}_0(\sqrt{A}r)-N_2\mathrm{I}_0(\kappa r),
\end{equation}
where
\begin{equation}
  N_1=\frac{2N_2\left[\mathrm{I}_0(\kappa)+\kappa\delta \mathrm{I}_1(\kappa)\right]}{2\mathrm{I}_0\left(\sqrt{A}\right)+A\delta
        _{\ 0}\!F_1\left(;2;\frac{A}{4}\right)},
\end{equation}
 \begin{equation}
  N_2=\frac{B}{A-\kappa^2}.
 \end{equation}
Here, the two Bessel function terms represent distinct flow modes: $\mathrm{I}_0(\sqrt{A}r)$ captures viscoelastic–magnetic–electroosmotic interactions, with $\sqrt{A}r$ governing its radial profile; $\mathrm{I}_0(\kappa r)$ is the pure AC electro-osmotic base flow, whose radial structure depends solely on $\kappa$. The coefficients $N_1$ and $N_2$ weight these two modes, incorporating $\kappa$, $\delta$, and all key coupling parameters to quantify the contribution of each mechanism to the overall velocity profile.
Moreover, this simple expression form will be used when solving the subsequent concentration distribution. $_{\ 0}\!F_1(a;b;z)$ is the regularized confluent hypergeometric function, which  is related with confluent hypergeometric function $_{\ 0}M_1(a;b;z)$~\citep{Brychkov08}
\begin{equation}\label{F1}
   _{\ 0}\!F_1(a;b;z)=\frac{_{\ 0}M_1(a;b;z)}{\Gamma(b)}.
\end{equation}
The confluent hypergeometric function  is one of the generalized hypergeometric function of order $m$, $n$, which is defined as follows
\begin{align}\label{F2}
  _{\ m}M_n(a;b;z)&=
  _{\ m}M_n(a_1,\ldots,a_j,\ldots,a_m;b_1,\ldots,b_k,\ldots b_n;z) \nonumber\\
 & =\sum_{q=0}^{\infty}\left(\frac{(a_1)_q,\ldots,(a_j)_q,\ldots,(a_m)_q}
  {(b_1)_q,\ldots,(b_k)_q,\ldots,(b_n)_q}\right)\left(\frac{z^q}{q!}\right),
\end{align}
where $(a)_q$ and $(b)_q$ are Pochhammer symbols, $a =(a_1,a_2,...,a_m)$ and $b = (b_1,b_2,...,b_n)$ are vectors of lengths $m$ and $n$, respectively. If  $a$ or $b$ is the empty vector, then the hypergeometric function $_{\ m}M_n(a;b;z)$ can be reduced as
\begin{align}\label{F3}
  _{\ 0}M_n(;b;z)=&\sum_{q=0}^{\infty}\left(\frac{1}{(b_1)_q,\ldots,(b_k)_q,\ldots,(b_n)_q}\right)\left(\frac{z^q}{q!}\right),\;\; \nonumber \\
     _{\ m}M_0(a;;z)=&\sum_{q=0}^{\infty}\left((a_1)_q,\ldots,(a_j)_q,\ldots,(a_m)_q\right)\left(\frac{z^q}{q!}\right),\;\;  \nonumber \\
       _{\ 0}M_0(;;z)=&\sum_{q=0}^{\infty}\left(\frac{z^q}{q!}\right)=\mathrm{e}^z.
\end{align}

\section{}\label{appC}
Suppose the solution of $g_1(r,t)$ takes the form $g_1(r,t)=\mathrm{Im}\{G(r) \mathrm{e}^{i\omega t}\}$, then upon substituting this into equations (\ref{g-K:g1}) and (\ref{g-conditions:b}) and omitting both the imaginary part and $\mathrm{e}^{i\omega t}$, the simplified equation and the corresponding boundary conditions are
\begin{equation}\label{g1-r}
   r^2 G^{''}(r)+rG^{'}(r)-i\omega r^2G(r)=r^2u_0(r)-r^2U_1,
\end{equation}
\begin{equation}\label{g1-r-condition}
 \frac{ \mathrm{d}G(0)}{\mathrm{d}r}=\frac{ \mathrm{d}G(1)}{\mathrm{d}r}=0.
\end{equation}
Equation (\ref{g1-r}) is a second-order linear non-homogeneous ordinary differential equation. Its solution can be expressed as the sum of the general solution $G_1(r)$ of the corresponding homogeneous equation and the particular solution $G_2(r)$ of the non-homogeneous equation, i.e., $G(r)=G_1(r)+G_2(r)$. The homogeneous equation that $G_1(r)$ satisfies and the non-homogeneous equation that $G_2(r)$ satisfies are presented as follows:
\begin{equation}\label{G1}
    r^2 G_1^{''}(r)+rG_1^{'}(r)-i\omega r^2G_1(r)=0,
\end{equation}
\begin{equation}\label{G2}
   r^2 G_2^{''}(r)+rG_2^{'}(r)-i\omega r^2G_2(r)=r^2u_0(r)-r^2U_1.
\end{equation}
Equation (\ref{G1})  is a modified Bessel equation of order $0$, and its solution can be directly written out
\begin{equation}\label{G1-solution}
   G_1(r)=A_1\mathrm{I}_0(\sqrt{i\omega}r)+A_2\mathrm{K}_0(\sqrt{i\omega}r),
\end{equation}
where  $\mathrm{I}_0$ and $\mathrm{K}_0$ are the modified Bessel function of the first kind and second kind zero order, constants $A_1$ and $A_2$ are decided by the relevant boundary conditions. In light of the composition of the right-hand side of equation (\ref{G2}), the particular solution assumes the following form according to the superposition principle of solutions
\begin{equation}\label{G2-solution}
   G_2(r)=A_3+A_4\mathrm{I}_0(\sqrt{A}r)+A_5\mathrm{I}_0(\kappa r).
\end{equation}
Firstly, the $A_3=U_1/i\omega$ can be determined by substituting $G_2=A_3$ into the following equation
\begin{equation}\label{B}
   r^2 G_2^{''}(r)+rG_2^{'}(r)-i\omega r^2G_2(r)=-r^2U_1.
\end{equation}
Then, bring $A_4\mathrm{I}_0(\sqrt{A}r)+A_5\mathrm{I}_0(\kappa r)$ into following the equation
\begin{equation}\label{D-E}
   r^2 G_2^{''}(r)+rG_2^{'}(r)-i\omega r^2G_2(r)=r^2u_0(r),
\end{equation}
we have
\begin{equation}\label{D1}
   A_4r^2 \left(\mathrm{I}_0(\sqrt{A}r)\right)^{''}+A_4r\left(\mathrm{I}_0(\sqrt{A}r)\right)^{'}-A_4i\omega r^2\mathrm{I}_0\left(\sqrt{A}r\right)=r^2N_1\mathrm{I}_0\left(\sqrt{A}r\right),
\end{equation}
\begin{equation}\label{E1}
   A_5r^2 \left(\mathrm{I}_0(\kappa r)\right)^{''}+A_5r\left(\mathrm{I}_0(\kappa r)\right)^{'}-A_5i\omega r^2\mathrm{I}_0\left(\kappa r\right)=-r^2N_2\mathrm{I}_0\left(\kappa r\right).
\end{equation}
Meanwhile, let  $A_4\mathrm{I}_0(\sqrt{A}r)$ and $A_5\mathrm{I}_0(\kappa r)$ be the solutions of the following homogeneous equation, respectively.
\begin{equation}\label{D2}
   A_4r^2 \left(\mathrm{I}_0(\sqrt{A}r)\right)^{''}+A_4r\left(\mathrm{I}_0(\sqrt{A}r)\right)^{'}-A_4A r^2\mathrm{I}_0\left(\sqrt{A}r\right)=0,
\end{equation}
\begin{equation}\label{E2}
   A_5r^2 \left(\mathrm{I}_0(\kappa r)\right)^{''}+A_5r\left(\mathrm{I}_0(\kappa r)\right)^{'}-A_5\kappa^2 r^2\mathrm{I}_0\left(\kappa r\right)=0.
\end{equation}
By solving the above four equations (\ref{D1})-(\ref{E2}), the constants $A_4$ and $A_5$ can be determined as
\begin{equation}\label{D-E}
   A_4=\frac{N_1}{A-i\omega},\;\;\;\;A_5=-\frac{N_2}{\kappa^2-i\omega}.
\end{equation}
Therefore, the solution of equation (\ref{g1-r}) can be expressed as the following form
\begin{equation}\label{G(r)}
   G(r)=A_1\mathrm{I}_0(\sqrt{i\omega}r)+A_2\mathrm{K}_0(\sqrt{i\omega}r)+A_3+A_4\mathrm{I}_0(\sqrt{A}r)+A_5\mathrm{I}_0(\kappa r).
\end{equation}
Based on the value of  $A_3$, $A_4$ and $A_5$ obtained above, and by means of the boundary conditions, the constants $A_1$ and $A_2$ can be deduced as
\begin{equation}\label{G(r)}
   A_1=-\frac{A_5\kappa \mathrm{I}_1(\kappa)+A_4\sqrt{A}\mathrm{I}_1(\sqrt{A})}{\sqrt{i\omega}\mathrm{I}_1(\sqrt{i \omega})},\;\;A_2=0.
\end{equation}
Ultimately, the solution of equation (\ref{g-K:g1}) can be presented as
\begin{align}\label{G(r,t)}
    g_1(r,t)&=\mathrm{Im}\left\{G(r)\mathrm{e}^{i\omega t}\right\} \\ \nonumber
               & =\mathrm{Im}\left\{\left[A_1\mathrm{I}_0(\sqrt{i\omega}r)+A_3+A_4\mathrm{I}_0(\sqrt{A}r)+A_5\mathrm{I}_0(\kappa r)\right]\mathrm{e}^{i\omega t}\right\}.
\end{align}

\bibliographystyle{elsarticle-harv}
\bibliography{references}

\begin{thebibliography}{125}
\expandafter\ifx\csname natexlab\endcsname\relax\def\natexlab#1{#1}\fi
\providecommand{\url}[1]{\texttt{#1}}
\providecommand{\href}[2]{#2}
\providecommand{\path}[1]{#1}
\providecommand{\DOIprefix}{doi:}
\providecommand{\ArXivprefix}{arXiv:}
\providecommand{\URLprefix}{URL: }
\providecommand{\Pubmedprefix}{pmid:}
\providecommand{\doi}[1]{\href{http://dx.doi.org/#1}{\path{#1}}}
\providecommand{\Pubmed}[1]{\href{pmid:#1}{\path{#1}}}
\providecommand{\bibinfo}[2]{#2}
\ifx\xfnm\relax \def\xfnm[#1]{\unskip,\space#1}\fi
\bibitem[{Ai et~al.(2022)Ai, Dai, Zhai, Chen and Huai}]{ADZCH22}
\bibinfo{author}{Ai, Y.D.}, \bibinfo{author}{Dai, H.C.}, \bibinfo{author}{Zhai,
  Y.W.}, \bibinfo{author}{Chen, B.}, \bibinfo{author}{Huai, W.X.},
  \bibinfo{year}{2022}.
\newblock \bibinfo{title}{The effect of solute release position on transient
  solute dispersion in floating wetlands: {A}n analytical study}.
\newblock \bibinfo{journal}{J. Clean. Prod.} \bibinfo{volume}{369},
  \bibinfo{pages}{133370}.
\bibitem[{Aime et~al.(2018)Aime, Cipelletti and Ramos}]{ALCR18}
\bibinfo{author}{Aime, S.}, \bibinfo{author}{Cipelletti, L.},
  \bibinfo{author}{Ramos, L.}, \bibinfo{year}{2018}.
\newblock \bibinfo{title}{Power law viscoelasticity of a fractal colloidal
  gel}.
\newblock \bibinfo{journal}{J. Rheol.} \bibinfo{volume}{62},
  \bibinfo{pages}{1429--1441}.
\bibitem[{Akhavan-Safaei and Zayernouri(2023)}]{Akhavan2023}
\bibinfo{author}{Akhavan-Safaei, A.}, \bibinfo{author}{Zayernouri, M.},
  \bibinfo{year}{2023}.
\newblock \bibinfo{title}{A non-local spectral transfer model and new scaling
  law for scalar turbulence}.
\newblock \bibinfo{journal}{J. Fluid Mech.} \bibinfo{volume}{956},
  \bibinfo{pages}{A26}.
\bibitem[{Arcos et~al.(2018)Arcos, M\'{e}ndez, Bautista and
  Bautista}]{Arcos2018}
\bibinfo{author}{Arcos, J.C.}, \bibinfo{author}{M\'{e}ndez, F.},
  \bibinfo{author}{Bautista, E.G.}, \bibinfo{author}{Bautista, O.},
  \bibinfo{year}{2018}.
\newblock \bibinfo{title}{Dispersion coefficient in an electro-osmotic flow of
  a viscoelastic fluid through a microchannel with a slowly varying wall zeta
  potential}.
\newblock \bibinfo{journal}{J. Fluid Mech.} \bibinfo{volume}{839},
  \bibinfo{pages}{348--386}.
\bibitem[{Aris(1956)}]{ARIS56}
\bibinfo{author}{Aris, R.}, \bibinfo{year}{1956}.
\newblock \bibinfo{title}{On the dispersion of a solute in a fluid flowing
  through a tube}.
\newblock \bibinfo{journal}{Proc. R. Soc. Lond. A} \bibinfo{volume}{235},
  \bibinfo{pages}{67--77}.
\bibitem[{Azari and Sadeghi(2022)}]{AzariS22}
\bibinfo{author}{Azari, M.}, \bibinfo{author}{Sadeghi, A.},
  \bibinfo{year}{2022}.
\newblock \bibinfo{title}{Unsteady convective-diffusive transport in
  semicircular microchannels with irreversible wall reaction}.
\newblock \bibinfo{journal}{J. Fluid Mech.} \bibinfo{volume}{949},
  \bibinfo{pages}{A1}.
\bibitem[{Balland et~al.(2006)Balland, Desprat, Icard, Feriol, Asnacios,
  Browaeys, Hanon and Gallet}]{BDID06}
\bibinfo{author}{Balland, M.}, \bibinfo{author}{Desprat, N.},
  \bibinfo{author}{Icard, D.}, \bibinfo{author}{Feriol, S.},
  \bibinfo{author}{Asnacios, A.}, \bibinfo{author}{Browaeys, J.},
  \bibinfo{author}{Hanon, S.}, \bibinfo{author}{Gallet, F.},
  \bibinfo{year}{2006}.
\newblock \bibinfo{title}{Power laws in microrheology experiments on living
  cells: {C}omparative analysis and modeling}.
\newblock \bibinfo{journal}{Phys. Rev. E} \bibinfo{volume}{74},
  \bibinfo{pages}{021911}.
\bibitem[{Banerjee et~al.(2021)Banerjee, Mehta, Pati and Biswas}]{BMPB21}
\bibinfo{author}{Banerjee, D.}, \bibinfo{author}{Mehta, S.K.},
  \bibinfo{author}{Pati, S.}, \bibinfo{author}{Biswas, P.},
  \bibinfo{year}{2021}.
\newblock \bibinfo{title}{Analytical solution to heat transfer for mixed
  electroosmotic and pressure-driven flow through a microchannel with
  slip-dependent zeta potential}.
\newblock \bibinfo{journal}{Int. J. Heat Mass Tran.} \bibinfo{volume}{181},
  \bibinfo{pages}{121989}.
\bibitem[{Banerjee et~al.(2023)Banerjee, Pati and Biswas}]{BPPB2023}
\bibinfo{author}{Banerjee, D.}, \bibinfo{author}{Pati, S.},
  \bibinfo{author}{Biswas, P.}, \bibinfo{year}{2023}.
\newblock \bibinfo{title}{Analytical study of pulsatile mixed electroosmotic
  and shear-driven flow in a microchannel with a slip-dependent zeta
  potential}.
\newblock \bibinfo{journal}{Appl. Math. Mech. -Engl. Ed.} \bibinfo{volume}{44},
  \bibinfo{pages}{1007--1022}.
\bibitem[{Barros and Escauriaza(2024)}]{Barros2024}
\bibinfo{author}{Barros, M.}, \bibinfo{author}{Escauriaza, C.},
  \bibinfo{year}{2024}.
\newblock \bibinfo{title}{Lagrangian and {E}ulerian perspectives of turbulent
  transport mechanisms in a lateral cavity}.
\newblock \bibinfo{journal}{J. Fluid Mech.} \bibinfo{volume}{984},
  \bibinfo{pages}{A1}.
\bibitem[{Bartels and Churchill(1942)}]{CFRV42}
\bibinfo{author}{Bartels, C.F.}, \bibinfo{author}{Churchill, R.V.},
  \bibinfo{year}{1942}.
\newblock \bibinfo{title}{Resolution of boundary problem by the use of a
  generalized convolution}.
\newblock \bibinfo{journal}{Bull. Am. Math. Soc.} \bibinfo{volume}{48},
  \bibinfo{pages}{276--282}.
\bibitem[{Barton(1983)}]{Barton83}
\bibinfo{author}{Barton, N.G.}, \bibinfo{year}{1983}.
\newblock \bibinfo{title}{The dispersion of solute from time-dependent releases
  in parallel flow}.
\newblock \bibinfo{journal}{J. Fluid Mech.} \bibinfo{volume}{136},
  \bibinfo{pages}{243--267}.
\bibitem[{Bird et~al.(2001)Bird, Stewart and Lightfoot}]{BSEN01}
\bibinfo{author}{Bird, R.B.}, \bibinfo{author}{Stewart, W.E.},
  \bibinfo{author}{Lightfoot, E.N.}, \bibinfo{year}{2001}.
\newblock \bibinfo{title}{Transport {P}henomena (second ed.)}.
\newblock \bibinfo{publisher}{New York, Wiley-Interscience Publication}.
\bibitem[{Brychkov(2008)}]{Brychkov08}
\bibinfo{author}{Brychkov, Y.A.}, \bibinfo{year}{2008}.
\newblock \bibinfo{title}{{H}andbook of {S}pecial {F}unctions}.
\newblock \bibinfo{publisher}{CRC Press, Boca Raton}.
\bibitem[{Cao and Kraume(2022)}]{CCMK22}
\bibinfo{author}{Cao, C.}, \bibinfo{author}{Kraume, M.}, \bibinfo{year}{2022}.
\newblock \bibinfo{title}{Improved mixing of viscoelastic fluids by the use of
  hysteresis effect}.
\newblock \bibinfo{journal}{Chem. Eng. Sci.} \bibinfo{volume}{260},
  \bibinfo{pages}{11787}.
\bibitem[{Chakraborty et~al.(2013)Chakraborty, Dey and
  Chakraborty}]{Chakraborty2013}
\bibinfo{author}{Chakraborty, R.}, \bibinfo{author}{Dey, R.},
  \bibinfo{author}{Chakraborty, S.}, \bibinfo{year}{2013}.
\newblock \bibinfo{title}{Thermal characteristics of
  elec-tromagnetohydrodynamic ﬂows in narrow channels with viscous
  dissipation and joule heating under constant wall heat ﬂux}.
\newblock \bibinfo{journal}{Int. J. Heat Mass Transfer} \bibinfo{volume}{67},
  \bibinfo{pages}{1151--1162}.
\bibitem[{Chatwin(1970)}]{Chatwin70}
\bibinfo{author}{Chatwin, P.C.}, \bibinfo{year}{1970}.
\newblock \bibinfo{title}{The approach to normality of the concentration
  distribution of a solute in a solvent flowing along a straight pipe}.
\newblock \bibinfo{journal}{J. Fluid Mech.} \bibinfo{volume}{43},
  \bibinfo{pages}{321--352}.
\bibitem[{Chaudhuri et~al.(2018)Chaudhuri, Mandal and
  Bandyopadhyay}]{Chaudhuri2018}
\bibinfo{author}{Chaudhuri, J.}, \bibinfo{author}{Mandal, T.K.},
  \bibinfo{author}{Bandyopadhyay, D.}, \bibinfo{year}{2018}.
\newblock \bibinfo{title}{Steady and oscillatory lorentz-force-induced
  transport and digitization of two-phase microflows}.
\newblock \bibinfo{journal}{Phys. Rev. Applied} \bibinfo{volume}{10},
  \bibinfo{pages}{034057}.
\bibitem[{Chen et~al.(2023)Chen, Li, Lv, Wei and Li}]{CJYC23}
\bibinfo{author}{Chen, Y.}, \bibinfo{author}{Li, J.}, \bibinfo{author}{Lv,
  Z.Y.}, \bibinfo{author}{Wei, Y.Q.}, \bibinfo{author}{Li, C.},
  \bibinfo{year}{2023}.
\newblock \bibinfo{title}{Mixing performance of viscoelastic fluids in an
  induced charge electroosmotic micromixer with a conductive cylinder}.
\newblock \bibinfo{journal}{J. Non-Newtonian Fluid Mech.}
  \bibinfo{volume}{317}, \bibinfo{pages}{105047}.
\bibitem[{Chowdhury et~al.(2025)Chowdhury, Pal and Gopmandal}]{CPGP25}
\bibinfo{author}{Chowdhury, S.}, \bibinfo{author}{Pal, S.K.},
  \bibinfo{author}{Gopmandal, P.P.}, \bibinfo{year}{2025}.
\newblock \bibinfo{title}{Dynamic electroosmotic flow and solute dispersion
  through a nanochannel filled with an electrolyte surrounded by a layer of a
  dielectric and immiscible liquid}.
\newblock \bibinfo{journal}{Soft Matter} \bibinfo{volume}{21},
  \bibinfo{pages}{1085--1112}.
\bibitem[{Churaev et~al.(2002)Churaev, Ralston, Sergeeva and Sobolev}]{CRSS02}
\bibinfo{author}{Churaev, N.V.}, \bibinfo{author}{Ralston, J.},
  \bibinfo{author}{Sergeeva, I.P.}, \bibinfo{author}{Sobolev, V.D.},
  \bibinfo{year}{2002}.
\newblock \bibinfo{title}{Electrokinetic properties of methylated quartz
  capillaries}.
\newblock \bibinfo{journal}{Adv. Colloid Interface Sci} \bibinfo{volume}{96},
  \bibinfo{pages}{265--278}.
\bibitem[{Colby et~al.(2007)Colby, Boris, Krause and Dou}]{CBKD07}
\bibinfo{author}{Colby, R.H.}, \bibinfo{author}{Boris, D.C.},
  \bibinfo{author}{Krause, W.E.}, \bibinfo{author}{Dou, S.},
  \bibinfo{year}{2007}.
\newblock \bibinfo{title}{Shear thinning of unentangled flexible polymer
  liquids}.
\newblock \bibinfo{journal}{Rheol. Acta} \bibinfo{volume}{46},
  \bibinfo{pages}{569--575}.
\bibitem[{Das et~al.(2024a)Das, Mondal, Poddar and Wang}]{DMPW24}
\bibinfo{author}{Das, C.}, \bibinfo{author}{Mondal, K.K.},
  \bibinfo{author}{Poddar, N.}, \bibinfo{author}{Wang, P.},
  \bibinfo{year}{2024}a.
\newblock \bibinfo{title}{Transient dispersion of a reactive solute in an
  oscillatory couette flow through an anisotropic porous medium}.
\newblock \bibinfo{journal}{Phys. Fluids} \bibinfo{volume}{36},
  \bibinfo{pages}{023610}.
\bibitem[{Das et~al.(2024b)Das, Poddar and Kairi}]{DPKR2024}
\bibinfo{author}{Das, D.}, \bibinfo{author}{Poddar, N.},
  \bibinfo{author}{Kairi, R.R.}, \bibinfo{year}{2024}b.
\newblock \bibinfo{title}{Modulating solute transport in magnetohydrodynamic
  pulsatile electroosmotic micro-channel flow: {R}ole of symmetric and
  asymmetric wall zeta potentials}.
\newblock \bibinfo{journal}{Phys. Fluids} \bibinfo{volume}{36},
  \bibinfo{pages}{092030}.
\bibitem[{Das et~al.(2025)Das, Vajravelu and Kairi}]{DVRK2025}
\bibinfo{author}{Das, D.}, \bibinfo{author}{Vajravelu, K.},
  \bibinfo{author}{Kairi, R.R.}, \bibinfo{year}{2025}.
\newblock \bibinfo{title}{Irreversibility estimation in electroosmotic {MHD}
  shear thinning nanofluid flow through a microchannel with slip-dependent zeta
  potentials}.
\newblock \bibinfo{journal}{Chinese J. Phys.} \bibinfo{volume}{95},
  \bibinfo{pages}{118--139}.
\bibitem[{Dehe et~al.(2021)Dehe, Rehm and Hardt}]{DRHS21}
\bibinfo{author}{Dehe, S.}, \bibinfo{author}{Rehm, I.S.},
  \bibinfo{author}{Hardt, S.}, \bibinfo{year}{2021}.
\newblock \bibinfo{title}{Hydrodynamic dispersion in {H}ele-{S}haw flows with
  in homogeneous wall boundary conditions}.
\newblock \bibinfo{journal}{J. Fluid Mech.} \bibinfo{volume}{925},
  \bibinfo{pages}{A11}.
\bibitem[{Deng et~al.(2024)Deng, Xiao and Liang}]{DXLC2024}
\bibinfo{author}{Deng, S.Y.}, \bibinfo{author}{Xiao, T.},
  \bibinfo{author}{Liang, C.X.}, \bibinfo{year}{2024}.
\newblock \bibinfo{title}{Analytical study of unsteady two-layer combined
  electroosmotic and pressure-driven flow through a cylindrical microchannel
  with slip-dependent zeta potential}.
\newblock \bibinfo{journal}{Chem. Eng. Sci.} \bibinfo{volume}{283},
  \bibinfo{pages}{119327}.
\bibitem[{Dentz et~al.(2018)Dentz, Icardi and Hidalgo}]{DIJJ18}
\bibinfo{author}{Dentz, M.}, \bibinfo{author}{Icardi, M.},
  \bibinfo{author}{Hidalgo, J.J.}, \bibinfo{year}{2018}.
\newblock \bibinfo{title}{Mechanisms of dispersion in a porous medium}.
\newblock \bibinfo{journal}{J. Fluid Mech.} \bibinfo{volume}{841},
  \bibinfo{pages}{851--882}.
\bibitem[{Dhellemmes et~al.(2024)Dhellemmes, Leclercq, Lichtenauer, Hochsmann,
  Leitner, Ebner, Michel~Martin, Neusu$\beta$ and Cottet}]{Dhellemmes24}
\bibinfo{author}{Dhellemmes, L.}, \bibinfo{author}{Leclercq, L.},
  \bibinfo{author}{Lichtenauer, L.}, \bibinfo{author}{Hochsmann, A.},
  \bibinfo{author}{Leitner, M.}, \bibinfo{author}{Ebner, A.},
  \bibinfo{author}{Michel~Martin, M.}, \bibinfo{author}{Neusu$\beta$, C.},
  \bibinfo{author}{Cottet, H.}, \bibinfo{year}{2024}.
\newblock \bibinfo{title}{Dual contributions of analyte adsorption and
  electroosmotic inhomogeneity to separation efficiency in capillary
  electrophoresis of proteins}.
\newblock \bibinfo{journal}{Anal. Chem.} \bibinfo{volume}{96},
  \bibinfo{pages}{11172--11180}.
\bibitem[{Dinic and Sharma(2020)}]{JDVS20}
\bibinfo{author}{Dinic, J.}, \bibinfo{author}{Sharma, V.},
  \bibinfo{year}{2020}.
\newblock \bibinfo{title}{Power laws dominate shear and extensional rheology
  response and capillarity-driven pinching dynamics of entangled hydroxyethyl
  cellulose ({HEC}) solutions}.
\newblock \bibinfo{journal}{Macromolecules} \bibinfo{volume}{53},
  \bibinfo{pages}{3424--3437}.
\bibitem[{Dong et~al.(2014)Dong, Zhang, Wang and Luo}]{DZWK14}
\bibinfo{author}{Dong, C.}, \bibinfo{author}{Zhang, J.S.},
  \bibinfo{author}{Wang, K.}, \bibinfo{author}{Luo, G.S.},
  \bibinfo{year}{2014}.
\newblock \bibinfo{title}{Micromixing performance of nanoparticle suspensions
  in a micro-sieve dispersion reactor}.
\newblock \bibinfo{journal}{Chem. Eng. J.} \bibinfo{volume}{253},
  \bibinfo{pages}{8--15}.
\bibitem[{Duankhan et~al.(2024)Duankhan, Sunat, Chiewchanwattana and
  Nasa-ngium}]{DSCN24}
\bibinfo{author}{Duankhan, P.}, \bibinfo{author}{Sunat, K.},
  \bibinfo{author}{Chiewchanwattana, S.}, \bibinfo{author}{Nasa-ngium, P.},
  \bibinfo{year}{2024}.
\newblock \bibinfo{title}{The {D}ifferentiated {C}reative {S}earch ({DCS}):
  Leveraging differentiated knowledge-acquisition and creative realism to
  address complex optimization problems}.
\newblock \bibinfo{journal}{Expert Syst. Appl.} \bibinfo{volume}{252},
  \bibinfo{pages}{123734}.
\bibitem[{Fabry et~al.(2001)Fabry, Maksym, Butler, Glogauer, Navajas and
  Fredberg}]{FMBG01}
\bibinfo{author}{Fabry, B.}, \bibinfo{author}{Maksym, G.N.},
  \bibinfo{author}{Butler, J.P.}, \bibinfo{author}{Glogauer, M.},
  \bibinfo{author}{Navajas, D.}, \bibinfo{author}{Fredberg, J.J.},
  \bibinfo{year}{2001}.
\newblock \bibinfo{title}{Scaling the microrheology of living cells}.
\newblock \bibinfo{journal}{Phys. Rev. Lett.} \bibinfo{volume}{87},
  \bibinfo{pages}{148102}.
\bibitem[{Ferr\'{a}s et~al.(2018)Ferr\'{a}s, Ford, Morgado, Rebelo, McKinley
  and N\'{o}brega}]{FMRM18}
\bibinfo{author}{Ferr\'{a}s, L.L.}, \bibinfo{author}{Ford, N.J.},
  \bibinfo{author}{Morgado, M.L.}, \bibinfo{author}{Rebelo, M.},
  \bibinfo{author}{McKinley, G.H.}, \bibinfo{author}{N\'{o}brega, J.M.},
  \bibinfo{year}{2018}.
\newblock \bibinfo{title}{Theoretical and numerical analysis of unsteady
  fractional viscoelastic flows in simple geometries}.
\newblock \bibinfo{journal}{Comput. Fluids} \bibinfo{volume}{174},
  \bibinfo{pages}{14--33}.
\bibitem[{Ferry(1980)}]{Ferry1980}
\bibinfo{author}{Ferry, J.D.}, \bibinfo{year}{1980}.
\newblock \bibinfo{title}{Viscoelastic {P}roperties of {P}olymers}.
\newblock \bibinfo{publisher}{New York, John Wiley \& Sons}.
\bibitem[{Friedrich(1991)}]{Fried91}
\bibinfo{author}{Friedrich, C.}, \bibinfo{year}{1991}.
\newblock \bibinfo{title}{Relaxation functions of rheological constitutive
  equations with fractional derivatives: Thermodynamical constraints}.
\newblock \bibinfo{journal}{In: Casas-Vázquez, J., Jou, D. (eds) Rheological
  Modelling: Thermodynamical and Statistical Approaches. Lecture Notes in
  Physics, vol 381. Springer, Berlin, Heidelberg} , \bibinfo{pages}{321--330}.
\bibitem[{Ghosal(2006)}]{GhosalS06}
\bibinfo{author}{Ghosal, S.}, \bibinfo{year}{2006}.
\newblock \bibinfo{title}{Electrokinetic flow and dispersion in capillary
  electrophoresis}.
\newblock \bibinfo{journal}{Annu. Rev. Fluid Mech.} \bibinfo{volume}{38},
  \bibinfo{pages}{309--338}.
\bibitem[{Ghosal and Chen(2012)}]{Ghosal12}
\bibinfo{author}{Ghosal, S.}, \bibinfo{author}{Chen, Z.}, \bibinfo{year}{2012}.
\newblock \bibinfo{title}{Electromigration dispersion in a capillary in the
  presence of electro-osmotic flow}.
\newblock \bibinfo{journal}{J. Fluid Mech.} \bibinfo{volume}{697},
  \bibinfo{pages}{436--454}.
\bibitem[{Gill and Sankarasubramanian(1970)}]{Gill1970}
\bibinfo{author}{Gill, W.N.}, \bibinfo{author}{Sankarasubramanian, R.},
  \bibinfo{year}{1970}.
\newblock \bibinfo{title}{Exact analysis of unsteady convective diffusion}.
\newblock \bibinfo{journal}{Proc. R. Soc. Lond. A} \bibinfo{volume}{316},
  \bibinfo{pages}{341--350}.
\bibitem[{Gill and Sankarasubramanian(1971)}]{Gill1971}
\bibinfo{author}{Gill, W.N.}, \bibinfo{author}{Sankarasubramanian, R.},
  \bibinfo{year}{1971}.
\newblock \bibinfo{title}{Dispersion of a non-uniform slug in time-dependent
  flow}.
\newblock \bibinfo{journal}{Proc. R. Soc. Lond. A} \bibinfo{volume}{322},
  \bibinfo{pages}{101--117}.
\bibitem[{Gill and Sankarasubramanian(1972)}]{Gill1972}
\bibinfo{author}{Gill, W.N.}, \bibinfo{author}{Sankarasubramanian, R.},
  \bibinfo{year}{1972}.
\newblock \bibinfo{title}{Dispersion of non-uniformly distributed time-variable
  continuous sources in time-dependent flow}.
\newblock \bibinfo{journal}{Proc. R. Soc. Lond. A} \bibinfo{volume}{327},
  \bibinfo{pages}{191--208}.
\bibitem[{Goyal et~al.(2024)Goyal, Datta and Chakraborty}]{GDCS24}
\bibinfo{author}{Goyal, V.}, \bibinfo{author}{Datta, S.},
  \bibinfo{author}{Chakraborty, S.}, \bibinfo{year}{2024}.
\newblock \bibinfo{title}{Generalizing electroosmotic-flow predictions over
  charge-modulated periodic topographies: tuneable far-field effects}.
\newblock \bibinfo{journal}{J. Fluid Mech.} \bibinfo{volume}{990},
  \bibinfo{pages}{A1}.
\bibitem[{Hegde and Harikrishnan(2021)}]{MGMW2024}
\bibinfo{author}{Hegde, A.S.}, \bibinfo{author}{Harikrishnan, A.R.},
  \bibinfo{year}{2021}.
\newblock \bibinfo{title}{Slip hydrodynamics of combined electroosmotic and
  pressure driven flows of power law fluids through narrow confinements}.
\newblock \bibinfo{journal}{Eur. J. Mech. B Fluids} \bibinfo{volume}{89},
  \bibinfo{pages}{525--550}.
\bibitem[{Hu and Zhu(2011)}]{KXKQ11}
\bibinfo{author}{Hu, K.X.}, \bibinfo{author}{Zhu, K.Q.}, \bibinfo{year}{2011}.
\newblock \bibinfo{title}{Anote on fractional {M}axwell model for {PMMA} and
  {PTFE}}.
\newblock \bibinfo{journal}{Polym. Test.} \bibinfo{volume}{30},
  \bibinfo{pages}{797--799}.
\bibitem[{Huang et~al.(2022)Huang, Kuo and Huang}]{HKKH22}
\bibinfo{author}{Huang, H.F.}, \bibinfo{author}{Kuo, J.E.},
  \bibinfo{author}{Huang, K.H.}, \bibinfo{year}{2022}.
\newblock \bibinfo{title}{Passive solute separation in {AC} electroosmosis
  including surface charge-coupled hydrodynamic slip effects}.
\newblock \bibinfo{journal}{Electrophoresis} \bibinfo{volume}{43},
  \bibinfo{pages}{571--580}.
\bibitem[{Huang et~al.(2024)Huang, Debnath, Roy, Wang, Jiang, Beg and
  Kuharat}]{HDRW24}
\bibinfo{author}{Huang, S.}, \bibinfo{author}{Debnath, S.},
  \bibinfo{author}{Roy, A.K.}, \bibinfo{author}{Wang, J.},
  \bibinfo{author}{Jiang, W.}, \bibinfo{author}{Beg, O.A.},
  \bibinfo{author}{Kuharat, S.}, \bibinfo{year}{2024}.
\newblock \bibinfo{title}{Transient dispersion of reactive solute transport in
  electrokinetic microchannel flow}.
\newblock \bibinfo{journal}{Phys. Fluids} \bibinfo{volume}{36},
  \bibinfo{pages}{052011}.
\bibitem[{Jaishankar and McKinley(2013)}]{JAGH13}
\bibinfo{author}{Jaishankar, A.}, \bibinfo{author}{McKinley, G.H.},
  \bibinfo{year}{2013}.
\newblock \bibinfo{title}{Power-law rheology in the bulk and at the interface:
  quasi-properties and fractional constitutive equations}.
\newblock \bibinfo{journal}{Proc. R. Soc. A} \bibinfo{volume}{469},
  \bibinfo{pages}{20120284}.
\bibitem[{Jaishankar and McKinley(2014)}]{AJGM14}
\bibinfo{author}{Jaishankar, A.}, \bibinfo{author}{McKinley, G.H.},
  \bibinfo{year}{2014}.
\newblock \bibinfo{title}{A fractional {K}-{BKZ} constitutive formulation for
  describing the nonlinear rheology of multiscale complex fluids}.
\newblock \bibinfo{journal}{J. Rheol.} \bibinfo{volume}{58},
  \bibinfo{pages}{1751--1788}.
\bibitem[{Jang and Lee(2000)}]{Jang2000}
\bibinfo{author}{Jang, J.}, \bibinfo{author}{Lee, S.S.}, \bibinfo{year}{2000}.
\newblock \bibinfo{title}{Theoretical and experimental study of mhd
  (magnetohydrodynamic) micropump}.
\newblock \bibinfo{journal}{Sensors Actuators} \bibinfo{volume}{80},
  \bibinfo{pages}{84--89}.
\bibitem[{Jimenez and Sullivan(1984)}]{CJPJ84}
\bibinfo{author}{Jimenez, C.}, \bibinfo{author}{Sullivan, P.J.},
  \bibinfo{year}{1984}.
\newblock \bibinfo{title}{Contaminant dispersion in some time-dependent laminar
  flows}.
\newblock \bibinfo{journal}{J. Fluid Mech.} \bibinfo{volume}{142},
  \bibinfo{pages}{57--77}.
\bibitem[{Joly et~al.(2006)Joly, Ybert, Trizac and Bocquet}]{Joly2006}
\bibinfo{author}{Joly, L.}, \bibinfo{author}{Ybert, C.},
  \bibinfo{author}{Trizac, E.}, \bibinfo{author}{Bocquet, L.},
  \bibinfo{year}{2006}.
\newblock \bibinfo{title}{Liquid friction on charged surfaces: from
  hydrodynamic slippage to electrokinetics}.
\newblock \bibinfo{journal}{J. Chem. Phys.} \bibinfo{volume}{125},
  \bibinfo{pages}{204716}.
\bibitem[{Keith et~al.(2021)Keith, Khristenko and Wohlmuth}]{Keith2021}
\bibinfo{author}{Keith, B.}, \bibinfo{author}{Khristenko, U.},
  \bibinfo{author}{Wohlmuth, B.}, \bibinfo{year}{2021}.
\newblock \bibinfo{title}{A fractional {PDE} model for turbulent velocity
  ﬁelds near solid walls}.
\newblock \bibinfo{journal}{J. Fluid Mech.} \bibinfo{volume}{916},
  \bibinfo{pages}{A21}.
\bibitem[{Kollmannsberger and Fabry(2011)}]{KPFB11}
\bibinfo{author}{Kollmannsberger, P.}, \bibinfo{author}{Fabry, B.},
  \bibinfo{year}{2011}.
\newblock \bibinfo{title}{Linear and nonlinear rheology of living cells}.
\newblock \bibinfo{journal}{Annu. Rev. Mater. Res.} \bibinfo{volume}{41},
  \bibinfo{pages}{75--97}.
\bibitem[{Kuntal et~al.(2025)Kuntal, Ghiya and Tiwari}]{KGTA2025}
\bibinfo{author}{Kuntal, Y.}, \bibinfo{author}{Ghiya, N.},
  \bibinfo{author}{Tiwari, A.}, \bibinfo{year}{2025}.
\newblock \bibinfo{title}{Solute dispersion in an electroosmotic flow of
  {C}arreau and {N}ewtonian fluids through a tube: analytical study}.
\newblock \bibinfo{journal}{Eur. Phys. J. Plus} \bibinfo{volume}{140},
  \bibinfo{pages}{221}.
\bibitem[{Larsen and Furst(2008)}]{THEM08}
\bibinfo{author}{Larsen, T.H.}, \bibinfo{author}{Furst, E.M.},
  \bibinfo{year}{2008}.
\newblock \bibinfo{title}{Microrheology of the liquid-solid transition during
  gelation}.
\newblock \bibinfo{journal}{Phys. Rev. Lett.} \bibinfo{volume}{100},
  \bibinfo{pages}{146001}.
\bibitem[{Li(2004)}]{LID2004}
\bibinfo{author}{Li, D.}, \bibinfo{year}{2004}.
\newblock \bibinfo{title}{{E}lectrokinetics in {M}icrofluidics}.
\newblock \bibinfo{publisher}{Academic Press}.
\bibitem[{Li and Jian(2019)}]{FQYJ19}
\bibinfo{author}{Li, F.Q.}, \bibinfo{author}{Jian, Y.J.}, \bibinfo{year}{2019}.
\newblock \bibinfo{title}{Solute dispersion generated by alternating current
  electric field through polyelectrolyte-grafted nanochannel with interfacial
  slip}.
\newblock \bibinfo{journal}{Int. J. Heat Mass Tran.} \bibinfo{volume}{141},
  \bibinfo{pages}{1066--1077}.
\bibitem[{Li and Jian(2017)}]{HLYJ17}
\bibinfo{author}{Li, H.C.}, \bibinfo{author}{Jian, Y.J.}, \bibinfo{year}{2017}.
\newblock \bibinfo{title}{Dispersion for periodic electro-osmotic flow of
  {M}axwell fluid through a microtube}.
\newblock \bibinfo{journal}{Int. J. Heat Mass Tran.} \bibinfo{volume}{115},
  \bibinfo{pages}{703--713}.
\bibitem[{Liao et~al.(2015)Liao, Dasgupta, Srinivasan and Liu}]{LDSL15}
\bibinfo{author}{Liao, H.Z.}, \bibinfo{author}{Dasgupta, P.K.},
  \bibinfo{author}{Srinivasan, K.}, \bibinfo{author}{Liu, Y.},
  \bibinfo{year}{2015}.
\newblock \bibinfo{title}{Mixing characteristics of mixers in flow analysis.
  {A}pplication to two-dimensional detection in ion chromatography}.
\newblock \bibinfo{journal}{Anal. Chem.} \bibinfo{volume}{87},
  \bibinfo{pages}{793--800}.
\bibitem[{Liu et~al.(2012)Liu, Jian, Chang and Yang}]{LJCY12}
\bibinfo{author}{Liu, Q.S.}, \bibinfo{author}{Jian, Y.J.},
  \bibinfo{author}{Chang, L.}, \bibinfo{author}{Yang, L.G.},
  \bibinfo{year}{2012}.
\newblock \bibinfo{title}{Alternating current ({AC}) electroosmotic flow of
  generalized {M}axwell fluids through a circular microtube}.
\newblock \bibinfo{journal}{Int. J. Phys. Sci.} \bibinfo{volume}{7},
  \bibinfo{pages}{5935--5941}.
\bibitem[{Liu et~al.(2024)Liu, Wu and Liu}]{LIUWL2024}
\bibinfo{author}{Liu, Y.B.}, \bibinfo{author}{Wu, Z.L.}, \bibinfo{author}{Liu,
  G.T.}, \bibinfo{year}{2024}.
\newblock \bibinfo{title}{Electrokinetic energy conversion efficiency in a
  nanochannel with slip-dependent zeta potential}.
\newblock \bibinfo{journal}{Phys. Scr.} \bibinfo{volume}{99},
  \bibinfo{pages}{025205}.
\bibitem[{Liu et~al.(2015)Liu, Jian, Liu and Li}]{LJLL15}
\bibinfo{author}{Liu, Y.P.}, \bibinfo{author}{Jian, Y.J.},
  \bibinfo{author}{Liu, Q.S.}, \bibinfo{author}{Li, F.Q.},
  \bibinfo{year}{2015}.
\newblock \bibinfo{title}{Alternating current magnetohydrodynamic
  electroosmotic flow of {M}axwell fluids between two micro-parallel plates}.
\newblock \bibinfo{journal}{J. Mol. Liq.} \bibinfo{volume}{211},
  \bibinfo{pages}{784--791}.
\bibitem[{Mainardi(2022)}]{Mainar22}
\bibinfo{author}{Mainardi, F.}, \bibinfo{year}{2022}.
\newblock \bibinfo{title}{Fractional {C}alculus and {W}aves in {L}inear
  {V}iscoelasticity, 2nd Ed}.
\newblock \bibinfo{publisher}{World Scientific}.
\bibitem[{Martin et~al.(1988)Martin, Adolf and Wilcoxon}]{MAJP88}
\bibinfo{author}{Martin, J.E.}, \bibinfo{author}{Adolf, D.},
  \bibinfo{author}{Wilcoxon, J.P.}, \bibinfo{year}{1988}.
\newblock \bibinfo{title}{Viscoelasticity of near-critical gels}.
\newblock \bibinfo{journal}{Phys. Rev. Lett.} \bibinfo{volume}{61},
  \bibinfo{pages}{2620--2623}.
\bibitem[{Masliyah and Bhattacharjee(2006)}]{MJBS2006}
\bibinfo{author}{Masliyah, J.H.}, \bibinfo{author}{Bhattacharjee, S.},
  \bibinfo{year}{2006}.
\newblock \bibinfo{title}{Electrokinetic and {C}olloid {T}ransport
  {P}henomena}.
\newblock \bibinfo{publisher}{John Wiley \& Sons}.
\bibitem[{Mazumder and Das(1992)}]{BSSK92}
\bibinfo{author}{Mazumder, B.S.}, \bibinfo{author}{Das, S.K.},
  \bibinfo{year}{1992}.
\newblock \bibinfo{title}{Effect of boundary reaction on solute dispersion in
  pulsatile flow through a tube}.
\newblock \bibinfo{journal}{J. Fluid Mech.} \bibinfo{volume}{239},
  \bibinfo{pages}{523--549}.
\bibitem[{Mederos et~al.(2020)Mederos, Arcos, Bautista and M\'{e}ndez}]{MABM20}
\bibinfo{author}{Mederos, G.}, \bibinfo{author}{Arcos, J.},
  \bibinfo{author}{Bautista, O.}, \bibinfo{author}{M\'{e}ndez, F.},
  \bibinfo{year}{2020}.
\newblock \bibinfo{title}{Hydrodynamics rheological impact of an oscillatory
  electroosmotic flow on a mass transfer process in a microcapillary with a
  reversible wall reaction}.
\newblock \bibinfo{journal}{Phys. Fluids} \bibinfo{volume}{32},
  \bibinfo{pages}{122003}.
\bibitem[{Mehta and Mondal(2024)}]{SKPK24}
\bibinfo{author}{Mehta, S.K.}, \bibinfo{author}{Mondal, P.K.},
  \bibinfo{year}{2024}.
\newblock \bibinfo{title}{Electroosmotic mixing of viscoplastic fluids in a
  microchannel}.
\newblock \bibinfo{journal}{Phys. Rev. Fluids} \bibinfo{volume}{9},
  \bibinfo{pages}{023301}.
\bibitem[{Mozafari et~al.(2025)Mozafari, Safarzadeh and Sadeghi}]{Mozafari2025}
\bibinfo{author}{Mozafari, S.}, \bibinfo{author}{Safarzadeh, H.},
  \bibinfo{author}{Sadeghi, A.}, \bibinfo{year}{2025}.
\newblock \bibinfo{title}{Analytical solutions for mass transfer and
  hydrodynamic dispersion by electroosmotic flow of viscoelastic fluids in
  heterogeneous microchannels}.
\newblock \bibinfo{journal}{Int. J. Heat Mass Tran.} \bibinfo{volume}{247},
  \bibinfo{pages}{127165}.
\bibitem[{Mukherjee et~al.(2020)Mukherjee, DasGupta and Chakraborty}]{SMSD20}
\bibinfo{author}{Mukherjee, S.}, \bibinfo{author}{DasGupta, S.},
  \bibinfo{author}{Chakraborty, S.}, \bibinfo{year}{2020}.
\newblock \bibinfo{title}{Temperature-gradient-induced massive augmentation of
  solute dispersion in viscoelastic micro-flows}.
\newblock \bibinfo{journal}{J. Fluid Mech.} \bibinfo{volume}{897},
  \bibinfo{pages}{A23}.
\bibitem[{Nghe et~al.(2011)Nghe, Terriac, Schneider, Li, Cloitre, B. and
  Tabeling}]{Nghe2011}
\bibinfo{author}{Nghe, P.}, \bibinfo{author}{Terriac, E.},
  \bibinfo{author}{Schneider, M.}, \bibinfo{author}{Li, Z.Z.},
  \bibinfo{author}{Cloitre, M.}, \bibinfo{author}{B., A.},
  \bibinfo{author}{Tabeling, P.}, \bibinfo{year}{2011}.
\newblock \bibinfo{title}{Microfluidics and complex fluids}.
\newblock \bibinfo{journal}{Lab Chip} \bibinfo{volume}{11},
  \bibinfo{pages}{788--794}.
\bibitem[{Peralta et~al.(2020)Peralta, Arcos, M\'{e}ndez and Bautista}]{PAMB20}
\bibinfo{author}{Peralta, M.}, \bibinfo{author}{Arcos, J.},
  \bibinfo{author}{M\'{e}ndez, F.}, \bibinfo{author}{Bautista, O.},
  \bibinfo{year}{2020}.
\newblock \bibinfo{title}{Mass transfer through a concentric-annulus
  microchannel driven by an oscillatory electroosmotic flow of a {M}axwell
  fluid}.
\newblock \bibinfo{journal}{J. Non-Newtonian Fluid Mech.}
  \bibinfo{volume}{279}, \bibinfo{pages}{104281}.
\bibitem[{Poddar et~al.(2024)Poddar, Saha, Mondal, Dhar and
  Mazumder}]{Poddar24}
\bibinfo{author}{Poddar, O.}, \bibinfo{author}{Saha, G.},
  \bibinfo{author}{Mondal, K.K.}, \bibinfo{author}{Dhar, S.},
  \bibinfo{author}{Mazumder, B.S.}, \bibinfo{year}{2024}.
\newblock \bibinfo{title}{Effect of phase exchange kinetics on taylor
  dispersion of chemically reactive solutes in an oscillatory
  magnetohydrodynamics flow between two parallel plates}.
\newblock \bibinfo{journal}{Phys. Fluids} \bibinfo{volume}{36},
  \bibinfo{pages}{053601}.
\bibitem[{Podlubny(1999)}]{Podlub99}
\bibinfo{author}{Podlubny, I.}, \bibinfo{year}{1999}.
\newblock \bibinfo{title}{Fractional {D}ifferential {E}quations}.
\newblock \bibinfo{publisher}{New York, Academic Press}.
\bibitem[{Ponalagusamy and Murugan(2023)}]{PRMD23}
\bibinfo{author}{Ponalagusamy, R.}, \bibinfo{author}{Murugan, D.},
  \bibinfo{year}{2023}.
\newblock \bibinfo{title}{Transport of a reactive solute in electroosmotic
  pulsatile flow of non-{N}ewtonian fluid through a circular conduit}.
\newblock \bibinfo{journal}{Chinese J. Phys.} \bibinfo{volume}{81},
  \bibinfo{pages}{243--269}.
\bibitem[{Rana and Murthy(2016a)}]{RanaM16}
\bibinfo{author}{Rana, J.}, \bibinfo{author}{Murthy, P.V.S.N.},
  \bibinfo{year}{2016}a.
\newblock \bibinfo{title}{Solute dispersion in pulsatile {C}asson fluid flow in
  a tube with wall absorption}.
\newblock \bibinfo{journal}{J. Fluid Mech.} \bibinfo{volume}{793},
  \bibinfo{pages}{877--914}.
\bibitem[{Rana and Murthy(2016b)}]{PVSN16}
\bibinfo{author}{Rana, J.}, \bibinfo{author}{Murthy, P.V.S.N.},
  \bibinfo{year}{2016}b.
\newblock \bibinfo{title}{Unsteady solute dispersion in {H}erschel-{B}ulkley
  fluid in a tube with wall absorption}.
\newblock \bibinfo{journal}{Phys. Fluids} \bibinfo{volume}{28},
  \bibinfo{pages}{111903}.
\bibitem[{Reshadi and Saidi(2019)}]{MRHS19}
\bibinfo{author}{Reshadi, M.}, \bibinfo{author}{Saidi, M.H.},
  \bibinfo{year}{2019}.
\newblock \bibinfo{title}{Tuning the dispersion of reactive solute by steady
  and oscillatory electroosmotic-{P}oiseuille flows in polyelectrolyte-grafted
  micro/nanotubes}.
\newblock \bibinfo{journal}{J. Fluid Mech.} \bibinfo{volume}{880},
  \bibinfo{pages}{73--112}.
\bibitem[{Rilwan et~al.(2024)Rilwan, Oni, Jha and Jibril}]{Rilwan2024}
\bibinfo{author}{Rilwan, U.S.}, \bibinfo{author}{Oni, M.O.},
  \bibinfo{author}{Jha, B.K.}, \bibinfo{author}{Jibril, H.M.},
  \bibinfo{year}{2024}.
\newblock \bibinfo{title}{Analysis of joule heating and viscous dissipation on
  electromagnetohydrodynamic flow with electroosmotic effect in a porous
  microchannel: {A} heat transfer miniature enhancement}.
\newblock \bibinfo{journal}{Heat Transfer} \bibinfo{volume}{53},
  \bibinfo{pages}{989--1013}.
\bibitem[{Rubinstein and Zaltzman(2013)}]{RIZB13}
\bibinfo{author}{Rubinstein, I.}, \bibinfo{author}{Zaltzman, B.},
  \bibinfo{year}{2013}.
\newblock \bibinfo{title}{Convective diffusive mixing in concentration
  polarization: from {T}aylor dispersion to surface convection}.
\newblock \bibinfo{journal}{J. Fluid Mech.} \bibinfo{volume}{728},
  \bibinfo{pages}{239--278}.
\bibitem[{Sadeghi et~al.(2020)Sadeghi, Saidi, Moosavi and Sadeghi}]{SSMS2020}
\bibinfo{author}{Sadeghi, M.}, \bibinfo{author}{Saidi, M.H.},
  \bibinfo{author}{Moosavi, A.}, \bibinfo{author}{Sadeghi, A.},
  \bibinfo{year}{2020}.
\newblock \bibinfo{title}{Unsteady solute dispersion by electrokinetic flow in
  a polyelectrolyte layer-grafted rectangular microchannel with wall
  absorption}.
\newblock \bibinfo{journal}{J. Fluid Mech.} \bibinfo{volume}{887},
  \bibinfo{pages}{A13}.
\bibitem[{Sadr et~al.(2004)Sadr, Yoda, Zheng and Conlisk}]{YZZC2004}
\bibinfo{author}{Sadr, R.}, \bibinfo{author}{Yoda, M.}, \bibinfo{author}{Zheng,
  Z.}, \bibinfo{author}{Conlisk, A.T.}, \bibinfo{year}{2004}.
\newblock \bibinfo{title}{An experimental study of electro-osmotic flow in
  rectangular microchannels}.
\newblock \bibinfo{journal}{J. Fluid Mech.} \bibinfo{volume}{506},
  \bibinfo{pages}{357--367}.
\bibitem[{Saha and Kundu(2022)}]{SSKB22}
\bibinfo{author}{Saha, S.}, \bibinfo{author}{Kundu, B.}, \bibinfo{year}{2022}.
\newblock \bibinfo{title}{Electroosmotic pressure-driven oscillatory flow and
  mass transport of {O}ldroyd-{B} fluid under high zeta potential and slippage
  conditions in microchannels}.
\newblock \bibinfo{journal}{Colloid. Surface. A} \bibinfo{volume}{647},
  \bibinfo{pages}{129070}.
\bibitem[{Saha and Kundu(2023)}]{Saha2023}
\bibinfo{author}{Saha, S.}, \bibinfo{author}{Kundu, B.}, \bibinfo{year}{2023}.
\newblock \bibinfo{title}{Multi-objective optimization of electrokinetic energy
  conversion efficiency and entropy generation for streaming potential driven
  electromagnetohydrodynamic flow of couple stress casson fluid in
  microchannels with slip-dependent zeta potentials}.
\newblock \bibinfo{journal}{Energy} \bibinfo{volume}{284},
  \bibinfo{pages}{129288}.
\bibitem[{Saha and Kundu(2025)}]{SSKB2025}
\bibinfo{author}{Saha, S.}, \bibinfo{author}{Kundu, B.}, \bibinfo{year}{2025}.
\newblock \bibinfo{title}{Highest electro-viscous energy and lowest
  irreversibility analysis for maxwell fluid in transient microchannel flow}.
\newblock \bibinfo{journal}{Appl. Therm. Eng.} \bibinfo{volume}{267},
  \bibinfo{pages}{125764}.
\bibitem[{Sahore et~al.(2018)Sahore, Doonan and Bailey}]{Sahore2018}
\bibinfo{author}{Sahore, V.}, \bibinfo{author}{Doonan, S.R.},
  \bibinfo{author}{Bailey, R.C.}, \bibinfo{year}{2018}.
\newblock \bibinfo{title}{Droplet microfluidics in thermoplastics: {D}evice
  fabrication, droplet generation, and content manipulation using integrated
  electric and magnetic fields}.
\newblock \bibinfo{journal}{Anal. Methods} \bibinfo{volume}{10},
  \bibinfo{pages}{4264--4274}.
\bibitem[{Samiee et~al.(2022)Samiee, Akhavan-Safaei and
  Zayernouri}]{Samiee2022}
\bibinfo{author}{Samiee, M.}, \bibinfo{author}{Akhavan-Safaei, A.},
  \bibinfo{author}{Zayernouri, M.}, \bibinfo{year}{2022}.
\newblock \bibinfo{title}{Tempered fractional {LES} modeling}.
\newblock \bibinfo{journal}{J. Fluid Mech.} \bibinfo{volume}{932},
  \bibinfo{pages}{A4}.
\bibitem[{Samuel et~al.(2025)Samuel, Chang, Ma and Santiago}]{CCMS25}
\bibinfo{author}{Samuel, C.J.}, \bibinfo{author}{Chang, R.},
  \bibinfo{author}{Ma, K.}, \bibinfo{author}{Santiago, J.G.},
  \bibinfo{year}{2025}.
\newblock \bibinfo{title}{Taylor dispersion for coupled electroosmotic and
  pressure-driven flows in all time regimes}.
\newblock \bibinfo{journal}{J. Fluid Mech.} \bibinfo{volume}{1011},
  \bibinfo{pages}{A36}.
\bibitem[{Sankarasubramanian and Gill(1972)}]{SANKAR1972}
\bibinfo{author}{Sankarasubramanian, R.}, \bibinfo{author}{Gill, W.N.},
  \bibinfo{year}{1972}.
\newblock \bibinfo{title}{Dispersion from a prescribed concentration
  distribution in time variable flow}.
\newblock \bibinfo{journal}{Proc. R. Soc. Lond. A} \bibinfo{volume}{329},
  \bibinfo{pages}{479--492}.
\bibitem[{Sankarasubramanian and Gill(1973)}]{SANKAR1973}
\bibinfo{author}{Sankarasubramanian, R.}, \bibinfo{author}{Gill, W.N.},
  \bibinfo{year}{1973}.
\newblock \bibinfo{title}{Unsteady convective diffusion with interphase mass
  transfer}.
\newblock \bibinfo{journal}{Proc. R. Soc. Lond. A} \bibinfo{volume}{333},
  \bibinfo{pages}{115--132}.
\bibitem[{Schiessel and Blumen(1995)}]{SHBA95}
\bibinfo{author}{Schiessel, H.}, \bibinfo{author}{Blumen, A.},
  \bibinfo{year}{1995}.
\newblock \bibinfo{title}{Mesoscopic pictures of the sol-gel transition: ladder
  models and fractal networks}.
\newblock \bibinfo{journal}{Macromolecules} \bibinfo{volume}{28},
  \bibinfo{pages}{4013--4019}.
\bibitem[{Schiessel et~al.(1995)Schiessel, Metzler, Blumen and
  Nonnenmacher}]{SMBN95}
\bibinfo{author}{Schiessel, H.}, \bibinfo{author}{Metzler, R.},
  \bibinfo{author}{Blumen, A.}, \bibinfo{author}{Nonnenmacher, T.F.},
  \bibinfo{year}{1995}.
\newblock \bibinfo{title}{Generalized viscoelastic models: their fractional
  equations with solutions}.
\newblock \bibinfo{journal}{J. Phys. A: Math. Gen.} \bibinfo{volume}{28},
  \bibinfo{pages}{6567--6584}.
\bibitem[{Schulz et~al.(2024)Schulz, G\"{a}rttner and Ray}]{SGNR24}
\bibinfo{author}{Schulz, R.}, \bibinfo{author}{G\"{a}rttner, S.},
  \bibinfo{author}{Ray, N.}, \bibinfo{year}{2024}.
\newblock \bibinfo{title}{Investigations of effective dispersion models for
  electroosmotic flow with rigid and free boundaries in a thin strip}.
\newblock \bibinfo{journal}{Math. Meth. Appl. Sci.} \bibinfo{volume}{47},
  \bibinfo{pages}{206--228}.
\bibitem[{Scott-Blair et~al.(1947)Scott-Blair, Veinoglou and
  Caffyn}]{Scott-Blair47}
\bibinfo{author}{Scott-Blair, G.W.}, \bibinfo{author}{Veinoglou, B.C.},
  \bibinfo{author}{Caffyn, J.E.}, \bibinfo{year}{1947}.
\newblock \bibinfo{title}{Limitations of the {N}ewtonian time scale in relation
  to non-equilibrium rheological states and a theory of quasi-properties}.
\newblock \bibinfo{journal}{Proc. R. Soc. Lond. A} \bibinfo{volume}{189},
  \bibinfo{pages}{69--87}.
\bibitem[{Sen et~al.(2025)Sen, Mondal and Kairi}]{SMRR2025}
\bibinfo{author}{Sen, R.}, \bibinfo{author}{Mondal, K.K.},
  \bibinfo{author}{Kairi, R.R.}, \bibinfo{year}{2025}.
\newblock \bibinfo{title}{Electroosmotic magnetohydrodynamics-driven solute
  dispersion in couple stress fluid flow through microchannel: Effect of
  transverse electric field}.
\newblock \bibinfo{journal}{Phys. Fluids} \bibinfo{volume}{37},
  \bibinfo{pages}{023138}.
\bibitem[{Shatrov and Gerbeth(2010)}]{SVGG10}
\bibinfo{author}{Shatrov, V.}, \bibinfo{author}{Gerbeth, G.},
  \bibinfo{year}{2010}.
\newblock \bibinfo{title}{Marginal turbulent magnetohydrodynamic flow in a
  square duct}.
\newblock \bibinfo{journal}{Phys. Fluids} \bibinfo{volume}{22},
  \bibinfo{pages}{084101}.
\bibitem[{Siva et~al.(2024)Siva, Dubey and Jangili}]{SDJS24}
\bibinfo{author}{Siva, T.}, \bibinfo{author}{Dubey, D.},
  \bibinfo{author}{Jangili, S.}, \bibinfo{year}{2024}.
\newblock \bibinfo{title}{Rotational flow dynamics of electroosmotic transport
  of couple stress fluid in a microfluidic channel under
  electromagnetohydrodynamic and slip-dependent zeta potential effects}.
\newblock \bibinfo{journal}{Phys. Fluids} \bibinfo{volume}{36},
  \bibinfo{pages}{092006}.
\bibitem[{Siva et~al.(2023)Siva, Jangili and Kumbhakar}]{SJBK2023}
\bibinfo{author}{Siva, T.}, \bibinfo{author}{Jangili, S.},
  \bibinfo{author}{Kumbhakar, B.}, \bibinfo{year}{2023}.
\newblock \bibinfo{title}{Entropy generation on {EMHD} transport of couple
  stress fluid with slip-dependent zeta potential under electrokinetic
  effects}.
\newblock \bibinfo{journal}{Int. J. Therm. Sci.} \bibinfo{volume}{191},
  \bibinfo{pages}{108339}.
\bibitem[{Soong et~al.(2010)Soong, Hwang and Wang}]{Soong10}
\bibinfo{author}{Soong, C.Y.}, \bibinfo{author}{Hwang, P.W.},
  \bibinfo{author}{Wang, J.C.}, \bibinfo{year}{2010}.
\newblock \bibinfo{title}{Analysis of pressure-driven electrokinetic flows in
  hydrophobic microchannels with slip-dependent zeta potential}.
\newblock \bibinfo{journal}{Microfluid. Nanofluid.} \bibinfo{volume}{9},
  \bibinfo{pages}{211--223}.
\bibitem[{Sparks et~al.(2003)Sparks, Smith, Straayer, Cripe, Schneider,
  Chimbayo, S. and N.}]{Sparks2003}
\bibinfo{author}{Sparks, D.}, \bibinfo{author}{Smith, R.},
  \bibinfo{author}{Straayer, M.}, \bibinfo{author}{Cripe, J.},
  \bibinfo{author}{Schneider, R.}, \bibinfo{author}{Chimbayo, A.},
  \bibinfo{author}{S., A.}, \bibinfo{author}{N., N.}, \bibinfo{year}{2003}.
\newblock \bibinfo{title}{Measurement of density and chemical concentration
  using a microfluidic chip}.
\newblock \bibinfo{journal}{Lab Chip} \bibinfo{volume}{3},
  \bibinfo{pages}{19--21}.
\bibitem[{Subudhi et~al.(2024)Subudhi, Jangili and Barik}]{SJSB24}
\bibinfo{author}{Subudhi, D.}, \bibinfo{author}{Jangili, S.},
  \bibinfo{author}{Barik, S.}, \bibinfo{year}{2024}.
\newblock \bibinfo{title}{Unsteady solute dispersion of electro-osmotic flow of
  micropolar fluid in a rectangular microchannel}.
\newblock \bibinfo{journal}{Phys. Fluids} \bibinfo{volume}{36},
  \bibinfo{pages}{073114}.
\bibitem[{Sujith et~al.(2025)Sujith, Mehta and Pati}]{Sujith2025}
\bibinfo{author}{Sujith, T.}, \bibinfo{author}{Mehta, S.K.},
  \bibinfo{author}{Pati, S.}, \bibinfo{year}{2025}.
\newblock \bibinfo{title}{Steric effect induced heat transfer characteristics
  of electromagnetohydrodynamic electroosmotic flow through a microchannel
  considering interfacial slip}.
\newblock \bibinfo{journal}{Phys. Fluids} \bibinfo{volume}{37},
  \bibinfo{pages}{052009}.
\bibitem[{Tan et~al.(2003)Tan, Pan and Xu}]{TPMX03}
\bibinfo{author}{Tan, W.C.}, \bibinfo{author}{Pan, W.X.}, \bibinfo{author}{Xu,
  M.Y.}, \bibinfo{year}{2003}.
\newblock \bibinfo{title}{A note on unsteady flows of a viscoelastic fluid with
  the fractional {M}axwell model between two parallel plates}.
\newblock \bibinfo{journal}{Int. J. Non-Linear Mech.} \bibinfo{volume}{38},
  \bibinfo{pages}{645--650}.
\bibitem[{Tandon and Kirby(2008)}]{VTBJ08}
\bibinfo{author}{Tandon, V.}, \bibinfo{author}{Kirby, B.J.},
  \bibinfo{year}{2008}.
\newblock \bibinfo{title}{Zeta potential and electroosmotic mobility in
  microfluidic devices fabricated from hydrophobic polymers: 2. {S}lip and
  interfacial water structure}.
\newblock \bibinfo{journal}{Electrophoresis} \bibinfo{volume}{29},
  \bibinfo{pages}{1102--1114}.
\bibitem[{Taylor(1953)}]{TAYLOR53}
\bibinfo{author}{Taylor, G.}, \bibinfo{year}{1953}.
\newblock \bibinfo{title}{Dispersion of soluble matter in solvent flowing
  slowly through a tube}.
\newblock \bibinfo{journal}{Proc. R. Soc. Lond. A} \bibinfo{volume}{219},
  \bibinfo{pages}{186--203}.
\bibitem[{Thompson and Troian(1997)}]{PASM97}
\bibinfo{author}{Thompson, P.A.}, \bibinfo{author}{Troian, S.M.},
  \bibinfo{year}{1997}.
\newblock \bibinfo{title}{A general boundary condition for liquid flow at solid
  surfaces}.
\newblock \bibinfo{journal}{Nature} \bibinfo{volume}{389},
  \bibinfo{pages}{360--362}.
\bibitem[{Tschoegl(1989)}]{Tschoegl12}
\bibinfo{author}{Tschoegl, N.W.}, \bibinfo{year}{1989}.
\newblock \bibinfo{title}{The {P}henomenological {T}heory of {L}inear
  {V}iscoelastic {B}ehavior}.
\newblock \bibinfo{publisher}{Berlin, Germany: Springer}.
\bibitem[{Vargas et~al.(2017)Vargas, Arcos, Bautista and
  M\'{e}ndez}]{Vargas2017}
\bibinfo{author}{Vargas, C.}, \bibinfo{author}{Arcos, J.},
  \bibinfo{author}{Bautista, O.}, \bibinfo{author}{M\'{e}ndez, F.},
  \bibinfo{year}{2017}.
\newblock \bibinfo{title}{Hydrodynamic dispersion in a combined
  magnetohydrodynamic-electroosmotic-driven flow through a microchannel with
  slowly varying wall zeta potentials}.
\newblock \bibinfo{journal}{Phys. Fluid} \bibinfo{volume}{29},
  \bibinfo{pages}{092002}.
\bibitem[{Vargas et~al.(2023)Vargas, M\'{e}ndez, Docoslis and
  Escobedo}]{VMDE23}
\bibinfo{author}{Vargas, C.}, \bibinfo{author}{M\'{e}ndez, F.},
  \bibinfo{author}{Docoslis, A.}, \bibinfo{author}{Escobedo, C.},
  \bibinfo{year}{2023}.
\newblock \bibinfo{title}{Colloid transport by an oscillatory electroosmotic
  flow between microelectrodes of axially variable shape}.
\newblock \bibinfo{journal}{Phys. Fluids} \bibinfo{volume}{35},
  \bibinfo{pages}{092014}.
\bibitem[{Vasista et~al.(2022)Vasista, Mehta and Pati}]{KSSP22}
\bibinfo{author}{Vasista, K.N.}, \bibinfo{author}{Mehta, S.K.},
  \bibinfo{author}{Pati, S.}, \bibinfo{year}{2022}.
\newblock \bibinfo{title}{Electroosmotic mixing in a microchannel with
  heterogeneous slip dependent zeta potential}.
\newblock \bibinfo{journal}{Chem. Eng. Process.} \bibinfo{volume}{176},
  \bibinfo{pages}{108940}.
\bibitem[{Vasista et~al.(2021)Vasista, Mehta, Pati and Sarkar}]{VMPS21}
\bibinfo{author}{Vasista, K.N.}, \bibinfo{author}{Mehta, S.K.},
  \bibinfo{author}{Pati, S.}, \bibinfo{author}{Sarkar, S.},
  \bibinfo{year}{2021}.
\newblock \bibinfo{title}{Electroosmotic flow of viscoelastic fluid through a
  microchannel with slip-dependent zeta potential}.
\newblock \bibinfo{journal}{Phys. Fluids} \bibinfo{volume}{33},
  \bibinfo{pages}{123110}.
\bibitem[{Vedel and Bruus(2012)}]{VEDEL2011}
\bibinfo{author}{Vedel, S.}, \bibinfo{author}{Bruus, H.}, \bibinfo{year}{2012}.
\newblock \bibinfo{title}{Transient {T}aylor-{A}ris dispersion for
  time-dependent flows in straight channels}.
\newblock \bibinfo{journal}{J. Fluid Mech.} \bibinfo{volume}{691},
  \bibinfo{pages}{95--122}.
\bibitem[{Wagner et~al.(2017)Wagner, Barbati, Engmann, Burbidge and
  McKinley}]{WBEB17}
\bibinfo{author}{Wagner, C.E.}, \bibinfo{author}{Barbati, A.C.},
  \bibinfo{author}{Engmann, J.}, \bibinfo{author}{Burbidge, A.S.},
  \bibinfo{author}{McKinley, G.H.}, \bibinfo{year}{2017}.
\newblock \bibinfo{title}{Quantifying the consistency and rheology of liquid
  foods using fractional calculus}.
\newblock \bibinfo{journal}{Food Hydrocoll.} \bibinfo{volume}{69},
  \bibinfo{pages}{242--254}.
\bibitem[{Wang and Kang(2010)}]{MWQK10}
\bibinfo{author}{Wang, M.}, \bibinfo{author}{Kang, Q.}, \bibinfo{year}{2010}.
\newblock \bibinfo{title}{Modeling electrokinetic flows in microchannels using
  coupled lattice {B}oltzmann methods}.
\newblock \bibinfo{journal}{J. Comput. Phys.} \bibinfo{volume}{229},
  \bibinfo{pages}{728--744}.
\bibitem[{Wang et~al.(2011)Wang, Huang and Yang}]{WHYC2011}
\bibinfo{author}{Wang, S.}, \bibinfo{author}{Huang, X.}, \bibinfo{author}{Yang,
  C.}, \bibinfo{year}{2011}.
\newblock \bibinfo{title}{Mixing enhancement for high viscous fluids in a
  microfluidic chamber}.
\newblock \bibinfo{journal}{Lab Chip} \bibinfo{volume}{11},
  \bibinfo{pages}{2081--2087}.
\bibitem[{Wang et~al.(2023)Wang, Xu and Qi}]{ZHHT23}
\bibinfo{author}{Wang, X.P.}, \bibinfo{author}{Xu, H.Y.}, \bibinfo{author}{Qi,
  H.T.}, \bibinfo{year}{2023}.
\newblock \bibinfo{title}{Mixing performance of an expansive mixer on
  viscoelastic solutions under alternating current electric field}.
\newblock \bibinfo{journal}{Phys. Fluids} \bibinfo{volume}{35},
  \bibinfo{pages}{103109}.
\bibitem[{Watson(1995)}]{GNWA1995}
\bibinfo{author}{Watson, G.N.}, \bibinfo{year}{1995}.
\newblock \bibinfo{title}{A {T}reatise on the {T}heory of {B}essel {F}unctions,
  2nd {E}dition}.
\newblock \bibinfo{publisher}{Cambridge University Press}.
\bibitem[{Wu and Chen(2014)}]{ZWGQ14}
\bibinfo{author}{Wu, Z.}, \bibinfo{author}{Chen, G.Q.}, \bibinfo{year}{2014}.
\newblock \bibinfo{title}{Approach to transverse uniformity of concentration
  distribution of a solute in a solvent flowing along a straight pipe}.
\newblock \bibinfo{journal}{J. Fluid Mech.} \bibinfo{volume}{740},
  \bibinfo{pages}{196--213}.
\bibitem[{Xie and Jian(2017)}]{ZYYJ17}
\bibinfo{author}{Xie, Z.Y.}, \bibinfo{author}{Jian, Y.J.},
  \bibinfo{year}{2017}.
\newblock \bibinfo{title}{Entropy generation of two-layer magnetohydrodynamic
  electroosmotic flow through microparallel channels}.
\newblock \bibinfo{journal}{Energy} \bibinfo{volume}{139},
  \bibinfo{pages}{1080--1093}.
\bibitem[{Yang et~al.(2019)Yang, Jian, Xie and Li}]{YJXL19}
\bibinfo{author}{Yang, C.}, \bibinfo{author}{Jian, Y.}, \bibinfo{author}{Xie,
  Z.}, \bibinfo{author}{Li, F.}, \bibinfo{year}{2019}.
\newblock \bibinfo{title}{Heat transfer characteristics of magnetohydrodynamic
  electroosmotic flow in a rectangular microchannel}.
\newblock \bibinfo{journal}{Eur. J. Mech. B Fluid} \bibinfo{volume}{74},
  \bibinfo{pages}{180--190}.
\bibitem[{Yang and Kwok(2002)}]{JYDY02}
\bibinfo{author}{Yang, J.}, \bibinfo{author}{Kwok, D.Y.}, \bibinfo{year}{2002}.
\newblock \bibinfo{title}{A new method to determine zeta potential and slip
  coefficient simultaneously}.
\newblock \bibinfo{journal}{J. Phys. Chem. B.} \bibinfo{volume}{106},
  \bibinfo{pages}{12851--12855}.
\bibitem[{Yang et~al.(2018)Yang, Qi and Jiang}]{YQJX18}
\bibinfo{author}{Yang, X.}, \bibinfo{author}{Qi, H.T.}, \bibinfo{author}{Jiang,
  X.Y.}, \bibinfo{year}{2018}.
\newblock \bibinfo{title}{Numerical analysis for electroosmotic flow of
  fractional {M}axwell fluids}.
\newblock \bibinfo{journal}{Appl. Math. Lett.} \bibinfo{volume}{78},
  \bibinfo{pages}{1--8}.
\bibitem[{Zhang et~al.(2025)Zhang, Zhou, Cui, Feng, Feng, Li, Hosokawa, Tian,
  Shen and Yalikun}]{ZZCFFL2025}
\bibinfo{author}{Zhang, T.}, \bibinfo{author}{Zhou, T.}, \bibinfo{author}{Cui,
  Q.}, \bibinfo{author}{Feng, X.}, \bibinfo{author}{Feng, S.},
  \bibinfo{author}{Li, M.and~Yang, Y.}, \bibinfo{author}{Hosokawa, Y.},
  \bibinfo{author}{Tian, G.}, \bibinfo{author}{Shen, A.Q.},
  \bibinfo{author}{Yalikun, Y.}, \bibinfo{year}{2025}.
\newblock \bibinfo{title}{Active microfluidic platforms for particle separation
  and integrated sensing applications}.
\newblock \bibinfo{journal}{ACS Sens.} \bibinfo{volume}{10},
  \bibinfo{pages}{5299--5313}.
\bibitem[{Zhu and Granick(2001)}]{YZSG01}
\bibinfo{author}{Zhu, Y.X.}, \bibinfo{author}{Granick, S.},
  \bibinfo{year}{2001}.
\newblock \bibinfo{title}{Rate-dependent slip of {N}ewtonian liquid at smooth
  surfaces}.
\newblock \bibinfo{journal}{Phys. Rev. Lett.} \bibinfo{volume}{87},
  \bibinfo{pages}{096105}.
\bibitem[{Zhu and Granick(2002)}]{YXGS02}
\bibinfo{author}{Zhu, Y.X.}, \bibinfo{author}{Granick, S.},
  \bibinfo{year}{2002}.
\newblock \bibinfo{title}{Limits of the hydrodynamic no-slip boundary
  condition}.
\newblock \bibinfo{journal}{Phys. Rev. Lett.} \bibinfo{volume}{88},
  \bibinfo{pages}{106102}.

\end{thebibliography}

\end{document}